\documentclass[nohyper,12pt,letterpaper]{JHEP3}
\usepackage{graphicx,epsfig,youngtab}

\newcommand{\myfig}[3]{\begin{figure}[ht]
\begin{center}
\leavevmode \epsfxsize=#2cm \epsfbox{#1}
\end{center}
\caption{#3} \label{fig:#1}
\end{figure}}

\author{David Bekker$^{1}$, Robert de Mello Koch$^{1,2}$ and Michael Stephanou$^{1}$\\
\qquad \\
$^{1}$ National Institute for Theoretical Physics,\\
Department of Physics and Centre for Theoretical Physics,\\
University of the Witwatersrand,\\
Wits, 2050,\\
South Africa\\
\qquad\\
$^{2}$Stellenbosch Institute for Advanced Studies,\\
Stellenbosch,\\
South Africa\\
\qquad\\
E-mail: \email{David.Bekker@students.wits.ac.za, robert@neo.phys.wits.ac.za, Michael.Stephanou@students.wits.ac.za}}

\abstract{
We develop techniques to compute the one-loop anomalous dimensions of operators in the ${\cal N}=4$ super
Yang-Mills theory that are dual to open strings ending on boundstates of sphere giant gravitons. Our results,
which are applicable to excitations involving an arbitrary number of open strings, generalize the single string
results of hep-th/0701067. The open strings we consider carry angular momentum on an S$^3$ embedded in the S$^5$
of the AdS$_5\times$S$^5$ background. The problem of computing the one loop anomalous dimensions is replaced with
the problem of diagonalizing an interacting Cuntz oscillator Hamiltonian. Our Cuntz oscillator dynamics illustrates
how the Chan-Paton factors for open strings propagating on multiple branes can arise dynamically.}

\preprint{WITS-CTP-033}

\title{Giant Gravitons - with Strings Attached (III)}


\def \Tr{\mbox{Tr\,}}
\def \threetwo{{}^3_2}
\def \twothree{{}^2_3}
\def \onetwo{{}^1_2}
\def \twoone{{}^2_1}

\def \onethree{{}^1_3}
\def \threeone{{}^3_1}

\def \oneplus{\mbox{\tiny{$1^+$}}}
\def \oneplustwo{{}^{\mbox{\tiny{$1^+$}}}_2}
\def \twooneplus{{}^2_{\mbox{\tiny{$1^+$}}}}
\def \w12{\mbox{\tiny{12}}}

\begin{document}

\section{Introduction and Summary}

The gauge theory/gravity correspondence\cite{Maldacena:1997re},\cite{Gubser:1998bc},\cite{Witten:1998qj} has provided powerful
clues into quantum gravity. For example, the correspondence claims the exact identity of maximally supersymmetric ${\cal N}=4$ Yang-Mills
quantum field theory with gauge group $SU(N)$ and type IIB string theory on the negatively curved AdS$_5\times$S$^5$ space with $N$
units of five form flux. Thus, we should be able to use the ${\cal N}=4$ super Yang-Mills theory as a definition of quantum gravity on
AdS$_5\times$S$^5$. For interesting recent progress in this direction,
see \cite{friends},\cite{Berenstein:2007wz}. The correspondence is however, not
yet understood well enough, for this to be possible. A detailed understanding of the gauge theory/gravity correspondence is frustrated
by the fact that it is a weak/strong coupling duality in the 't Hooft coupling. At weak 't Hooft coupling the field theory may be treated
perturbatively, but the spacetime of the dual quantum gravity is highly curved. In the opposite limit of strong 't Hooft coupling we
have to face the difficult problem of strongly coupled quantum field theory. The dual quantum gravity however, simplifies, because in this
limit the curvature of the spacetime is small. For this reason, most computations which can be carried out on
both sides of the correspondence (and hence clearly shed light on the correspondence) compute quantities that are protected by
symmetry - typically supersymmetry (see \cite{Aharony:1999ti} and references therein). The number of these tests and the agreement found
is impressive. However, computing and comparing protected quantities is not satisfying - to probe dynamical features of the
correspondence it would be nice to be able to compare quantities that are not protected by any symmetries. This is in general, a
formidable problem. In \cite{DBer}, the notion of an {\it almost BPS state} was introduced. These states are systematically small
deformations of states that are protected. For this reason, for almost BPS states, it is possible to reliably extrapolate from weak
to strong coupling. A good example of almost BPS states are the BMN loops\cite{Berenstein:2002jq}. By studying BMN loops it has been
possible to probe truly stringy aspects of the gauge theory/gravity correspondence (see \cite{Russo:2004kr} and references therein).

Giant gravitons, which are half-BPS states, have proved to be the source of many valuable quantities that are accessible on both sides
of the correspondence. Further, they are very interesting from a string theory point of view, since they are good examples of protected
non perturbative objects. Giant gravitons are spherical D3 branes extended in the sphere\cite{McGreevy:2000cw} or in the
AdS space \cite{Grisaru:2000zn},\cite{Hashimoto:2000zp},\cite{Das:2000fu} of the AdS$\times$S background. They are (classically) stable
due to the presence of the five form flux which produces a force that exactly balances their tension. The dual description of giant gravitons
is in terms of Schur polynomials in the Higgs fields\cite{Corley:2001zk},\cite{Balasubramanian:2001nh}.

Our interest in giant gravitons is related to the fact that excited giant gravitons provide a rich source of nearly BPS states.
Excitations of giant gravitons are obtained by attaching open strings to the giant. The gauge theory operator dual to an excited
sphere giant is known and the anomalous dimension of this operator reproduces the expected open string
spectrum\cite{Balasubramanian:2002sa}\footnote{See \cite{Berenstein:2006qk},\cite{Strings},\cite{Shahin} for further studies of non-BPS
excitations that have been interpreted as open strings attached to giant gravitons.}. This has been extended and the
operators dual to an arbitrary system of excited giant gravitons is now known\cite{Balasubramanian:2004nb}. The dual operators,
restricted Schur polynomials\footnote{We review the definition of the restricted Schur polynomial in Appendix E.}, beautifully
reproduce the restrictions imposed on excitations of the brane system by the Gauss law.
Further, these excited giant gravitons have recently been identified as the microstates of near-extremal black holes in
AdS$_5\times$S$^5$\cite{Balasubramanian:2007bs}. Although the evidence for identifying the restricted Schur polynomials as
the operators dual to excited giant gravitons is convincing, much remains to be done. For example, we do not yet understand
the detailed mechanism allowing Chan-Paton factors, expected for strings attached to a bound state of giant gravitons, to
emerge from the super Yang-Mills theory. In this article, our goal is to explore this issue, by providing techniques which
allow the computation of the anomalous dimensions of excited giant gravitons, to one loop. We will argue that the Chan-Paton
factors emerge from the symmetric group labels of the restricted Schur polynomials.

The computation of anomalous dimensions of operators in ${\cal N}=4$ super Yang-Mills theory has progressed considerably.
Much of the recent progress was sparked by a remarkable paper of Minahan and Zarembo\cite{Minahan:2002ve} which shows that
the spectrum of one loop anomalous dimensions of operators dual to closed string states, in a sub sector of the theory, gives
rise to an integrable $SO(6)$ spin chain. This result can be generalized to include the full set of local operators of the
theory\cite{Beisert:2003yb}. The integrable spin chain model describing the full planar one loop spectrum of anomalous dimensions
can be solved by Bethe-Ansatz techniques\cite{Beisert:2003yb}. Clearly, it is desirable to find a similar approach for operators dual
to open strings. A naive generalization is frustrated by the fact that, since the open string and giant can exchange momentum,
the number of sites of the open string lattice becomes a dynamical variable\footnote{An exception to this is an open string attached
to a maximal giant graviton\cite{Maxg}.}. This was circumvented in \cite{CuntzChain} by introducing a Cuntz oscillator chain.
Restricting to the $SU(2)$ sector, the spin chain is obtained by mapping one of the matrices, say $Z$, into a spin up
and the other, say $Y$, into a spin down. In contrast to this, the Cuntz chain uses the $Y$s to set up a lattice
which is populated by the $Z$s. Thus the number of sites in the Cuntz chain is fixed.

The power of the spin chain goes beyond the computation of anomalous dimensions. Indeed, the low energy description of the spin chain
relevant for closed string states appearing on the field theory side matches perfectly with the low energy limit of the string action
in AdS$_5\times$S$^5$\cite{Kr}. This is an important result because it shows how a string action can emerge from large $N$ gauge
theory. For the open string, the coherent state expectation value of the Cuntz chain Hamiltonian reproduces the open string action
for an open string attached to a sphere giant in AdS$_5\times$S$^5$\cite{CuntzChain},\cite{Berenstein:2006qk}, for an open string
attached to an AdS giant in AdS$_5\times$S$^5$\cite{adsgiant} and for an open string attached to a sphere giant in a deformed AdS$_5\times$S$^5$
background\cite{deMelloKoch:2005jg}. Recently\cite{Hofman:2007xp}, the worldsheet theory of an open string attached
to a maximal giant has been studied. Evidence that the system is integrable at two loops has been obtained.

The fact that the open string can exchange momentum with the giant is reflected in the fact that there are sources
and sinks (at the endpoints of the string) for the particles on the chain. The structure of these boundary
interactions is complicated: since the brane can exchange momentum with the string, the brane will in general
be deformed by these boundary interactions. The goal of this article is to determine this Cuntz chain Hamiltonian
for multiple strings attached to an arbitrary system of giant gravitons. In particular, this entails accounting for
back reaction on the giant graviton. To compute the Cuntz chain Hamiltonian, we need the two point functions of restricted Schur polynomials.
It is an involved combinatoric task to compute the two point functions of restricted Schur polynomials. The required technology
to compute these correlators, in the free field limit, has recently been developed in \cite{de Mello Koch:2007uu}\footnote{For some
earlier related work, see\cite{Corley:2002mj}.}. This was then extended to one loop, for operators dual to giants with a single
string attached\cite{de Mello Koch:2007uv}. In this article, we extend the existing technology, allowing the one loop
computation of correlators dual to giant graviton systems with an arbitrary number of strings attached. In the remainder of this
introduction, we will establish notation and give a sketch of the technology we develop.

To make our discussion concrete, we mostly consider the specific example of two strings attached to a bound state of two sphere
giants\footnote{In Appendix F we consider a boundstate of three sphere giants with two open strings attached.}.
Note however, that most of the formulas we derive (and certainly the techniques we develop) are applicable to the general
problem. Both the strings and the branes that we consider are distinguishable.
In this case there are a total of six possible states. For a bound
state of two sphere giant gravitons, we need to consider restricted Schur polynomials labeled by
Young diagrams with two columns each with $O(N)$ boxes. Denote the number
of boxes in the first column by $b_0+b_1$ and the number of boxes in the second column by $b_0$. It is natural to interpret the number
of boxes in each column as the momentum of each giant. We can use the state operator correspondence (see Appendices C.5 and D for further
discussion) to associate a Cuntz chain state with each restricted Schur polynomial. The Cuntz chain states have
six labels in total: the first two labels are $b_0$ and $b_1$ which determine
the momenta of the two giants; the next two labels are the branes on which the endpoints
of string one are attached and the final two labels are the branes on which the endpoints of string two are attached. We label the
strings by `1' and `2'. The brane corresponding to column 1 of the Young diagram is labeled `b' (for big brane) and the brane
corresponding to column 2 of the Young diagram is labeled `l' (for little brane). Since the second column of a Young diagram
can never contain more boxes that the first column, and since the radius of the giant graviton is determined by the square root
of its angular momentum, these are accurate labels. Consider a state with string 1 on big brane and
string 2 on little brane. The restricted Schur polynomial (written using the graphical notation of
\cite{de Mello Koch:2007uu},\cite{de Mello Koch:2007uv}) together with the corresponding Cuntz chain state are (in this case, $b_0=3$ and $b_1=4$)
$$\young({\,}{\,},{\,}{\,},{\,}{\,},{\,}{2},{\,},{\,},{\,},{1})\longleftrightarrow |3,4,bb,ll\rangle .$$
We will call states with strings stretching between branes ``stretched string states''. When labeling the Cuntz
chain state corresponding to a stretched string state, we will write the end point label corresponding to the upper index first. Thus,
$$\young({\,}{\,},{\,}{\,},{\,}{\,},{\,}{\onetwo},{\,},{\,},{\,},{\twoone})\longleftrightarrow |3,4,lb,bl\rangle .$$
The remaing four states are
$$\young({\,}{\,},{\,}{\,},{\,}{\,},{\,}{1},{\,},{\,},{\,},{2})\longleftrightarrow |3,4,ll,bb\rangle \qquad
  \young({\,}{\,},{\,}{\,},{\,}{\,},{\,}{\twoone},{\,},{\,},{\,},{\onetwo})\longleftrightarrow |3,4,bl,lb\rangle ,$$
$$\young({\,}{\,},{\,}{\,},{\,}{2},{\,}{1},{\,},{\,},{\,},{\,})\longleftrightarrow |2,6,ll,ll\rangle \qquad
  \young({\,}{\,},{\,}{\,},{\,}{\,},{\,}{\,},{\,},{\,},{2},{1})\longleftrightarrow |4,2,bb,bb\rangle .$$
The construction of the operators dual to excitations described by strings stretching between the branes requires the construction
of an ``intertwiner''\cite{de Mello Koch:2007uu}. One of the results of the present article, is to provide a general construction
of the intertwiner. This construction is given in Appendix A.
In the notation of \cite{de Mello Koch:2007uu}, we assume that when the restricted Schur polynomial is to be reduced, string 1 is
removed first and string 2 second. This implies that, when using the graphical notation, removing the box occupied by string 1 first
will always leave a valid Young diagram. This choice is arbitrary, but useful for explicit computation. Once we have the form of the
Hamiltonian, we can always change to a ``physical basis''.
To obtain operators dual to giant gravitons, we take $b_0$ to be $O(N)$ and $b_1$ to be $O(1)$.
We want to compute the matrix of anomalous dimensions to one loop and at large $N$.
To compute this matrix, we need to compute the two point functions of restricted Schur polynomials.
This is a hard problem: since the number of
fields in the giant graviton is $O(N)$, huge combinatoric factors pile up as the coefficient of non-planar diagrams and the usual the
planar approximation fails. We need to contract all of the fields in the giant gravitons exactly. The two open strings are described
by the words $W^{(1)}$ and $W^{(2)}$. The six Higgs fields $\phi^i$ $i=1,...,6$, of the ${\cal N}=4$ super Yang-Mills theory can be
grouped into the following complex combinations
$$ Z=\phi^1+i\phi^2,\qquad Y=\phi^3+i\phi^4,\qquad X=\phi^5+i\phi^6 .$$
The giant gravitons are built out of the $Z$ field; the open string words out of the $Z$ and $Y$ fields.
Thus, the open strings carry a component of angular momentum on the $S^3$ that the giant wraps, as well
a component parallel to the giant's angular momentum.
We will normalize things so that the action of ${\cal N}=4$ super Yang-Mills theory on $R\times S^3$ is
(we consider the Lorentzian theory and have set the radius of the $S^3$ to 1)
\begin{equation}
S={N\over 4\pi \lambda}\int dt \int_{S^3} {d\Omega_3\over 2\pi^2}
\left( {1\over 2}(D\phi^i)(D\phi^i)+{1\over 4}\big(\big[\phi^i,\phi^j\big]\big)^2
-{1\over 2}\phi^i\phi^i +\dots \right),
\end{equation}
With these conventions,
\begin{equation}
\langle Z^\dagger_{ij}(t)Z_{kl}(t)\rangle = {4\pi\lambda\over N}\delta_{il}\delta_{jk} =
\langle Y^\dagger_{ij}(t)Y_{kl}(t)\rangle ,
\end{equation}
The open string words can be labelled as
\begin{equation}
(W(\{n_1,n_2,\cdots , n_{L-1}\}))^i_j= (YZ^{n_1}YZ^{n_2} Y\cdots YZ^{n_{L-1}}Y)^i_j,
\end{equation}
where $\{n_1,n_2,\cdots , n_{L-1}\}$ are the Cuntz lattice occupation numbers.
The giant is built out of $Z$s; the first and last letters of the open string word $W$ are not $Z$s.
We will always use $L$ to denote the number of $Y$ fields in the
open string word and $J=n_1+n_2+\cdots +n_{L-1}$ to denote the number of $Z$ fields in the open string word. The number of fields in each word
is $J+L\approx L$ in the case that $J\ll L$ which we will assume in this article. For the words $W^{(1)},W^{(2)}$ to be dual to open strings,
we need to take $L\sim O(\sqrt{N})$. We do not know how to contract the open strings words exactly; when contracting the open string
words, only the planar diagrams are summed. To suppress the non-planar contributions we take ${L^2\over N}\ll 1$. To do this we consider
a double scaling limit in which the first limit takes $N\to\infty$ holding ${L^2\over N}$ fixed and the second limit takes the effective
genus counting parameter ${L^2\over N}$ to zero. Taking the limits in this way corresponds, in the dual string theory, to taking the string
coupling to zero, in the string theory constructed in a fixed giant graviton background. Since the two strings are distinguishable they
are represented by distinct words and hence, in the
large $N$ limit, we have
$$\langle W^{(i)}(W^{(j)})^\dagger\rangle\propto\delta^{ij}.$$
When computing a correlator of two restricted Schur polynomials, the fields
belonging to the giants in the two systems of excited giant gravitons are contracted amongst
each other, the fields in the first open string of each are contracted amongst each other and the fields in the second open string are
contracted amongst each other. We drop the contributions coming from contractions between $Z$s in the open strings and $Z$s
associated to the brane system, as well as contractions between $Z$s in different open string words. When computing two point functions
in free field theory, if the number of boxes in the representation $R$ is less than\footnote{When the number of operators in the Young diagram
is $O(N^2)$, the operator is dual to an LLM geometry\cite{LLM}.} $O(N^2)$ and the numbers of $Z$'s in the open
string is $O(1)$, the contractions between any $Z$s in the open string and the rest of the operator are suppressed in the large $N$
limit\cite{recent}. Contractions between $Z$s in different open string words are non planar and are hence subleading (clearly there are no
large combinatoric factors that modify this).

An important parameter of our excited giant graviton system is $N-b_0$. This parameter can scale as $O(N)$, $O(\sqrt{N})$ or $O(1)$.
In section 2, we will see that when $N-b_0$ is $O(1)$ the sphere giant boundary interaction is $O({1\over N})$, when $N-b_0$ is $O(\sqrt{N})$
the boundary interaction is $O({1\over\sqrt{N}})$ and when $N-b_0$ is $O(N)$, the boundary interaction is $O(1)$. Since we want to explore
the dynamics arising from the boundary interaction, we will assume that $N-b_0$ is $O(N)$.

The subspace of states reached by attaching two open strings to a giant graviton boundstate system is dynamically decoupled (from
subspaces obtained by attaching a different number of open strings) at large $N$.
It is possible to move out of this subspace by the process in which the word $W$ ``fragments'' thereby allowing $Y$s to populate more than
a single box in $R$. In the dual string theory this corresponds to a splitting of the original string into smaller strings, which are still
attached to the giant. This process was considered in \cite{de Mello Koch:2007uu}
and from that result we know that it does not contribute
in the large $N$ limit. One could also consider the process in which the open string detaches from the brane boundstate and is emitted as
a closed string state, so that it no longer occupies any box in $R$. This process (decay of the excited giant boundstate by gravitational
radiation) also does not contribute in the large $N$ limit\cite{Balasubramanian:2002sa},\cite{de Mello Koch:2007uu}.

Since the giant boundstate and the open string can exchange momentum, the value of $J$ is not a parameter that we can choose, but rather,
it is determined by the dynamics of the problem. Cases in which $J$ becomes large correspond to the situation in which a lot of momentum is
transferred from the giant to the open string, presumably signaling an instability. See \cite{Berenstein:2006qk} for a good physical
discussion of this instability. In cases where $J$ is large, back reaction is important
and the approximations we are employing are no longer valid. This will happen when $J$ becomes $O(\sqrt{N})$ since the assumption that we
can drop non-planar contributions when contracting the open string words breaks down. Essentially this is because as more and more $Z$s hop
onto the open string, it is starting to grow into a state which is eventually best described as a giant graviton itself. One can also no longer
neglect the contractions between any $Z$s in the open string and the rest of the operator, presumably because the composite system no longer
looks like a string plus giant (which can be separated nicely) but rather, it starts to look like one large deformed threebrane.
Thus, the fact that our approximation breaks down has a straight forward interpretation: We have set up our description by
assuming that the operator we study is dual to a threebrane with an open string attached. This implies that our operator can be
decomposed into a ``threebrane piece'' and a ``string piece''. These two pieces are treated very differently: when contracting the
threebrane piece, all contractions are summed; when contracting the string piece, only planar contractions are summed. Contractions between
the two pieces are dropped. When a large number of $Z$s hop onto the open string our operator is simply not
dual to a state that looks like a threebrane with an open string attached and our approximations are not valid. We are not
claiming that this operator can not be studied using large $N$ techniques - it may still be possible to set up a systematic
$1/N$ expansion. We are claiming that the diagrams we have summed do not give this approximation.

It is useful to decompose the potential for the scalars into D terms and F terms. The advantage of this
decomposition is that it is known that at one loop, the D term contributions cancel with the
gauge boson exchange and the scalar self energies\cite{Constable:2002hw}. Consequently we will only
consider the planar interactions arising from the F term. The F term interaction preserves the number of $Y$'s (the lattice is not dynamical)
and allows impurities (the $Z$s) to hop between neighboring sites. The bulk interactions are described by the Hamiltonian
\begin{equation}
H_{bulk} = 2\lambda\sum_{l=1}^L \hat{a}_l^\dagger \hat{a}_l -\lambda\sum_{l=1}^{L-1}(\hat{a}_l^\dagger \hat{a}_{l+1}
+\hat{a}_l \hat{a}^\dagger_{l+1}),
\label{bulk}
\end{equation}
where
\begin{equation}
\hat{a}_i \hat{a}_i^\dagger = I,\qquad \hat{a}^\dagger_i \hat{a}_i = I-|0\rangle\langle 0|.
\end{equation}
The interested reader is referred to \cite{Berenstein:2006qk} for the derivation of this result.
To obtain the full Hamiltonian, we need to include the boundary interactions arising from the string/brane
system interaction. This interaction introduces sources and sinks for the impurities at the boundaries
of the lattice. The boundary interaction allows $Z$s to hop from the string onto the giant, or from the giant
onto the string. Since the number of $Z$s gives the angular momentum of the system in the plane that the giant is orbiting
in, the boundary interaction allows the string and the brane to exchange angular momentum. We can classify the different
types of boundary interaction depending on whether momentum flows from the string to the brane or from the brane to the string.
Consider the interaction that allows a $Z$ to hop from the first or last site of either string onto the giant.
In this process the string loses momentum to the giant graviton. We call this
a ``hop off'' process because a $Z$ has hopped off the string. The opposite process in which a $Z$ hops off the
brane and onto the string is called a ``hop on'' process. In the ``hop on'' process the giant loses momentum to the string.
In addition to these momentum exchanging processes, there is also a boundary interaction in which a $Z$ belonging to the giant
``kisses" the first (or last) $Y$ in the open string word so that no momentum is exchanged. We call this the kissing interaction.

In Appendix C we will derive a set of identities that allow us to compute the term in the Hamiltonian describing the ``hop off" process.
These identities make extensive use of the technology for computing restricted characters which is developed in Appendix B.
We will now explain what we mean by a restricted character.
Let $R$ be an irrep of $S_n$ and let $R_1$ be an irrep of $S_{n-m}$ with $0<m<n$. If we restrict ourselves
to elements $\sigma\in S_{n-m}$, then $\Gamma_R(\sigma)$ will, in general, subduce a number of irreps of $S_{n-m}$. One of these
irreps is $R_1$. A restricted character $\chi_{R,R_1}(\sigma)$ is obtained by tracing the matrix representing $\sigma$ in irrep $R$,
$\Gamma_R(\sigma)$, over the $R_1$ subspace. If $\sigma\in S_{n-m}$ this simply gives the character of $\sigma$ in irrep $R_1$. Our
technology allows the computation of $\chi_{R,R_1}(\sigma)$ even when $\sigma\notin S_{n-m}$, in which case $\chi_{R,R_1}(\sigma)$
does not have an obvious group theory interpretation. The basic idea we exploit in constructing the ``hop off'' process is simple
to state: In a string-giant system, whenever a $Z$ field hops past the borders of the open string word $W$, the resulting
restricted Schur polynomial decomposes into a sum of two types of systems, one is a giant with a closed string and another is a string-giant
system where the giant is now bigger. In the large $N$ limit only this second type needs to be considered. Our identities express this
decomposition.

Since the Hamiltonian
must be Hermitian, we can obtain the ``hop on" term by daggering the ``hop off" term. Finally, we obtain the momentum conserving boundary
interaction by expressing the kiss as a hop on followed by a hop off. This determines the complete Cuntz oscillator chain Hamiltonian
needed for a one loop computation of the anomalous dimensions of operators dual to excited giant graviton bound states. This derivation
of the Cuntz chain Hamiltonian, which is the main technical result of this article, is given in section 2.

The resulting Hamiltonian clearly reflects the worldsheet structure of the open strings that are interacting. This explains how the
Chan-Paton factors associated with strings in a multi-brane system dynamically emerge from Yang-Mills theory: they emerge from the
symmetric group labels of the restricted Schur polynomials. Our Hamiltonian
treats string 1 and string 2 differently. This is not at all surprising, since when we built our operator
we treated the two strings differently. In section 3 we describe a new ``physical basis'' singled out by the requirement that the two
strings enter on an equal footing. In section 4 we present our conclusions.

Strings stretching between giants in AdS can be realized as solutions to the Born-Infeld action describing the world volume dynamics
of these branes\cite{Shahin}. In this work the Gauss law is enforced by the construction of consistent solutions to the equations
of motion on a compact space. In the work \cite{Balasubramanian:2004nb} the one loop anomalous dimension of operators representing
a string attached to a two brane bound state was considered. One of the branes was taken to be a maximal giant to simplify the
computation. For two coincident branes, the one-loop anomalous dimension for an open string is twice the answer for a single brane.
This is the first, to the best of our knowledge, hint of the dynamical emergence of Chan-Paton factors for open strings
on coincident branes. The demonstration of \cite{Balasubramanian:2004nb} identifies the extra factor of 2 with the trace over the
indices of the enhanced gauge group associated to coincident branes. Our demonstration proves this identification: we can follow the
Chan-Paton indices in the tree level transitions of two open strings. Further, using the technology we develop, it is straight forward
(but technically involved) to generalize this result to a bound state of $m$ branes, where we expect a $U(m)$ gauge theory to emerge.
As an example, in Appendix F we consider a boundstate of three sphere giants. In this case, a $U(3)$ gauge theory emerges.

\section{Cuntz Chain Hamiltonian}

In this section we will derive the form of the terms in the Hamiltonian describing the string boundary interactions. This will
allow us to compute the complete Cuntz chain Hamiltonian, since the bulk Hamiltonian has already been given in (\ref{bulk}).

\subsection{Hop Off Interaction}

We start by deriving the hop off interaction. The F term vertex allows a $Z$ and a $Y$ to change position within a word.
The hopping interaction corresponds to the situation in which a $Z$ hops past the $Y$ marking the end point of the string,
i.e. a $Z$ hops off the string and onto the giant. Concretely, when acting on either open string, this hop takes
$$ W(\{n_1,n_2,\cdots,n_{L-1}\})\to ZW(\{n_1-1,n_2,\cdots,n_{L-1}\})\quad {\rm or}$$
$$ W(\{n_1,n_2,\cdots,n_{L-1}\})\to W(\{n_1,n_2,\cdots,n_{L-1}-1\})Z .$$
To determine the corresponding term in the interaction Hamiltonian, we need to be able to express
objects like $\chi_{R,R''}^{(2)}(Z,ZW^{(1)},W^{(2)})$ in terms of $\chi_{S,S''}^{(2)}(Z,W^{(1)},W^{(2)})$ where $S$ is a Young diagram
with one more box than $R$\footnote{The number of primes on the label of the restricted Schur polynomial indicates how many boxes are
dropped, i.e. $R''$ is obtained by dropping two boxes from $R$.}. This is easily achieved
by inverting the identities derived in Appendix C. To get the hop off interaction in
the Hamiltonian, we rewrite the identities in terms of normalized Cuntz chain states.

{\vskip 1.0cm}

\noindent
{\it $+1\to 1$ Hop off Interaction:} This term in the Hamiltonian describes the hop off process in which a $Z$ hops out
of the first site of string 1. We write $+1\to 1$ to indicate that the string before the hop has one extra $Z$ in its first
site.
\begin{equation}
 H_{+1\to 1}
\left[\matrix{|b_0-1,b_1,bb,ll\rangle\cr |b_0-1,b_1,ll,bb\rangle\cr |b_0-1,b_1,bl,lb\rangle\cr
              |b_0-1,b_1,lb,bl\rangle\cr |b_0-2,b_1+2,ll,ll\rangle\cr |b_0,b_1-2,bb,bb\rangle}\right]
=-\lambda\sqrt{1-{b_0\over N}}M_1
\left[\matrix{|b_0-1,b_1+1,bb,ll\rangle\cr |b_0,b_1-1,ll,bb\rangle\cr |b_0-1,b_1+1,bl,lb\rangle\cr
              |b_0,b_1-1,lb,bl\rangle\cr |b_0-1,b_1+1,ll,ll\rangle\cr |b_0,b_1-1,bb,bb\rangle}\right],
\end{equation}
where
$$M_1=\left[\matrix{
-(b_1)_1^2 &{1\over b_1(b_1+1)^2} &0 &{(b_1)_0\over b_1+1} &{(b_1)_1\over b_1+1} &-{(b_1)_1\over b_1 (b_1+1)}\cr
-{1\over (b_1+2)(b_1+1)^2} &-(b_1)_1^2 &-{(b_1)_2\over b_1+1} &0 &-{(b_1)_1\over (b_1+1)(b_1+2)} &-{(b_1)_1\over b_1+1}\cr
-{(b_1)_1\over (b_1+1)(b_1+2)} &{(b_1)_1\over b_1+1} &-(b_1)_1(b_1)_2 &0 &-{b_1\over (b_1+1)^2} &{1\over (b_1+1)^2}\cr
-{(b_1)_1\over b_1+1} &-{(b_1)_1\over b_1(b_1+1)} &0 &-(b_1)_0(b_1)_1 &{1\over (b_1+1)^2} &{b_1+2\over (b_1+1)^2}\cr
-{(b_1)_2\over b_1+1} &0 &{1\over b_1+2} &0 &-(b_1)_1(b_1)_2 &0\cr
0 &{(b_1)_0\over b_1+1} &0 &-{1\over b_1} &0 &-(b_1)_0(b_1)_1
}\right],$$
and
$$ (b_1)_n={\sqrt{b_1+n-1}\sqrt{b_1+n+1}\over b_1+n}.$$
The term in the Hamiltonian describing the process in which the $Z$ hops out of the last site of string 1 is described by
swapping the labels of the endpoints of the open strings. Concretely, it is given by
\begin{equation}
H_{1+\to 1}
\left[\matrix{|b_0-1,b_1,bb,ll\rangle\cr |b_0-1,b_1,ll,bb\rangle\cr |b_0-1,b_1,lb,bl\rangle\cr
              |b_0-1,b_1,bl,lb\rangle\cr |b_0-2,b_1+2,ll,ll\rangle\cr |b_0,b_1-2,bb,bb\rangle}\right]
=-\lambda\sqrt{1-{b_0\over N}}M_1
\left[\matrix{|b_0-1,b_1+1,bb,ll\rangle\cr |b_0,b_1-1,ll,bb\rangle\cr |b_0-1,b_1+1,lb,bl\rangle\cr
              |b_0,b_1-1,bl,lb\rangle\cr |b_0-1,b_1+1,ll,ll\rangle\cr |b_0,b_1-1,bb,bb\rangle}\right],
\end{equation}
where $M_1$ is the matrix given above. We write $1+\to 1$ to indicate that the string before the hop has one extra $Z$ in its last
site.

{\vskip 1.0cm}

\noindent
{\it $+2\to 2$ Hop off Interaction:} This term in the Hamiltonian describes the hop off process in which a $Z$ hops out
of the first site of string 2.
\begin{equation}
 H_{+2\to 2}
\left[\matrix{|b_0-2,b_1+1,bb,ll\rangle\cr |b_0-1,b_1-1,ll,bb\rangle\cr |b_0-2,b_1+1,bl,lb\rangle\cr
              |b_0-1,b_1-1,lb,bl\rangle\cr |b_0-2,b_1+1,ll,ll\rangle\cr |b_0-1,b_1-1,bb,bb\rangle}\right]
=-\lambda\sqrt{1-{b_0\over N}}M_2
\left[\matrix{|b_0-1,b_1,bb,ll\rangle\cr |b_0-1,b_1,ll,bb\rangle\cr |b_0-1,b_1,bl,lb\rangle\cr
              |b_0-1,b_1,lb,bl\rangle\cr |b_0-1,b_1,ll,ll\rangle\cr |b_0-1,b_1,bb,bb\rangle}\right],
\end{equation}
where
$$M_2=\left[\matrix{
-(b_1)_1^2 &-{1\over (b_1+2)(b_1+1)^2} &-{(b_1)_1\over (b_1+1)(b_1+2)} &-{(b_1)_1\over b_1+1} &0 &-{(b_1)_2\over b_1+1}\cr
{1\over b_1(b_1+1)^2} &-(b_1)_1^2 &{(b_1)_1\over b_1+1} &-{(b_1)_1\over (b_1+1)b_1} &{(b_1)_0\over b_1+1} &0\cr
0 &-{(b_1)_2\over b_1+1} &-(b_1)_1(b_1)_2 &0 &0 &{1\over b_1+2}\cr
{(b_1)_0\over b_1+1} &0 &0 &-(b_1)_0(b_1)_1 &-{1\over b_1} &0\cr
-{(b_1)_1\over b_1(b_1+1)} &-{(b_1)_1\over b_1+1} &{1\over (b_1+1)^2} &{b_1+2\over (b_1+1)^2} &-(b_1)_1(b_1)_0 &0\cr
{(b_1)_1\over b_1+1} &-{(b_1)_1\over (b_1+1)(b_1+2)} &-{b_1\over (b_1+1)^2} &{1\over (b_1+1)^2} &0 &-(b_1)_2(b_1)_1
}\right].$$
Notice that these interactions (as is the case for all of the boundary interactions) are highly suppressed for a maximal
giant\cite{Maxg}. The term in the Hamiltonian describing the process in which the $Z$ hops out of the last site of string 2 is described by
swapping the labels of the endpoints of the open strings.

The function $(b_1)_n$ also appears in the Hamiltonian relevant for a single string attached to a giant\cite{de Mello Koch:2007uv}.
Notice that $(b_1)_n$ vanishes when $b_1=1-n$, but tends to 1 very rapidly as $b_1$ is increased from this value.
The diagonal terms in the Hamiltonian with a $(b_1)_1$ factor will thus vanish when $b_1=0$.
The radius of each giant is determined by their momentum.
Since $b_1$ is the difference in momentum of the two giants, $b_1=0$ corresponds to coincident giants.
Thus, $(b_1)_n$ is switching off short distance interactions.
The hop off Hamiltonian does not generate illegal Young diagrams from legal ones precisely because these interactions are switched off.

It may seem puzzling that the boundary interaction has the universal strength $\sqrt{1-{b_0\over N}}$ regardless of which end the $Z$
peels off. Indeed, any local boundary interaction should only know about the boundary which is participating. Since the string end points
are on branes of different sizes, one would expect two different strengths for the two endpoints. This universal strength is an artifact
of the limit we consider. We take $b_0$ to be $O(N)$ and $b_1$ to be $O(1)$. The strength $\sqrt{1-{b_0\over N}}$ arises from the
normalization of the Cuntz oscillator states for the string endpoint attached to the smaller giant graviton (see appendix D for these
normalizations). Similarly, the strength $\sqrt{1-{b_0+b_1\over N}}$ arises from the normalization of the Cuntz oscillator states for the
string endpoint attached to the larger giant graviton. In the limit we consider
$$ \sqrt{1-{b_0+b_1\over N}}-\sqrt{1-{b_0\over N}}=O\Big({1\over N}\Big). $$
Physically, taking $b_0=O(N)$ and $b_1=O(1)$ implies that the two branes are very nearly coincident.

Finally, note that the structure of the hop on and hop off interactions, clearly reflect the fact that the open strings attached
to the giants are orientable.

\subsection{Hop On Interaction}

Since ${\cal N}=4$ super Yang-Mills theory is a unitary conformal field theory, we know that the spectrum of anomalous dimensions
of the theory is real. This implies that the energy spectrum of our Cuntz chain Hamiltonian must be real and hence the Hamiltonian
must be Hermitian. Thus, the hop on term in the Hamiltonian is given by the Hermitian conjugate of the hop off term.

To give an example, we will now derive the term in the Hamiltonian describing the process in which a $Z$ from the brane hops into
the first site of string 1. Let $|\psi\rangle$ denote the state with a brane of momentum $P_{\rm brane}=P$ and a string of
momentum $P_{\rm string}=p$ and $|\phi\rangle$ denote the state with $P_{\rm brane}=P+1$ and $P_{\rm string}=p-1$. Then,
\begin{equation}
H_{+1\to 1}|\psi\rangle =-\lambda\sqrt{1-{b_0\over N}}M_1|\phi\rangle ,
\end{equation}
and
$$ \langle\phi' |H_{+1\to 1}|\psi\rangle =-\lambda\sqrt{1-{b_0\over N}}\langle\phi' |M_1|\phi\rangle =
-\lambda\sqrt{1-{b_0\over N}} (M_1)_{\phi'\phi}.$$
Daggering we find (keep in mind that $M_1$ is real)
\begin{eqnarray}
\langle\psi |H_{1\to +1}|\phi'\rangle &=&(\langle\phi' |H_{+1\to 1}|\psi\rangle)^\dagger\nonumber\\
&=&-\lambda\sqrt{1-{b_0\over N}}\langle\phi |(M_1)^T|\phi'\rangle \nonumber\\
&=&-\lambda\sqrt{1-{b_0\over N}} \left( (M_1{})^T\right)_{\phi\phi'}.\nonumber
\end{eqnarray}
Thus we obtain
\begin{equation}
 H_{1\to +1} |\phi\rangle=-\lambda\sqrt{1-{b_0\over N}}N_1|\psi\rangle ,
\end{equation}
with $N_1=(M_1)^T$.

\subsection{Kissing Interaction}
\myfig{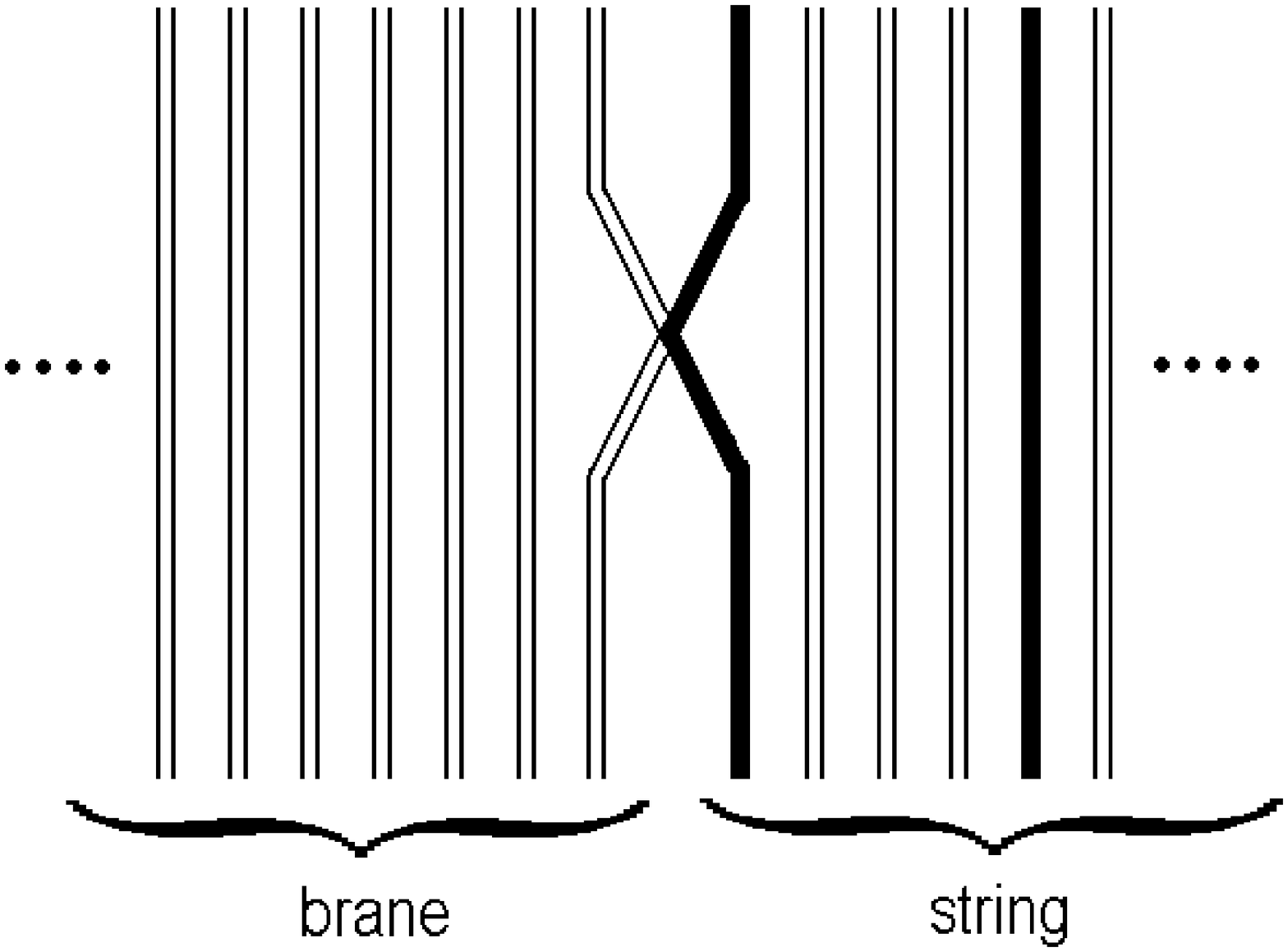}{16.0}{The Feynman diagram on the left shows the kissing interaction. The white ribbons are $Z$ fields, the
black ribbons are $Y$ fields. The interacting black ribbon shown marks the beginning of the string; there are 3 $Z$s in the
first site of the string. The Feynman diagram on the right shows a hop on interaction followed by a hop off interaction. If you
shrink the composite hop on/hop off interaction to a point, you recover the kissing interaction.}
The kissing interaction corresponds to the Feynman diagram shown on the left in Figure \ref{fig:kissing.ps}. Notice that the number of $Z$ fields
in the giant is unchanged by this process so that the string and brane do not exchange momentum by this process.
As far as the combinatorics goes, we can model the kissing interaction as a hop on (the string) followed by a hop off. We know both the
hop on and hop off terms so the kissing interaction follows. This is illustrated by the Feynman diagram shown on the right in Figure 1.
The kissing interaction must be included for both endpoints of both strings.

A straight forward computation easily gives
\begin{equation}
 H_{\rm kissing}=\lambda \left(1-{b_0\over N}\right){\bf 1},
\end{equation}
for each endpoint of either string. In this formula ${\bf 1}$ is the identity operator.

The fact that the kissing interaction comes out proportional to the identity operator is a non-trivial check of our hop on and hop off
interactions. Indeed, the contraction of the F term vertex which leads to the kissing interaction removes an adjacent $Z$ and $Y$ and
then replaces them in the same order. Thus, the kissing interaction had to come out proportional to the identity. The careful reader
may worry that this is not in fact true - indeed, the restricted Schur polynomial includes terms in which the open string word is
traced and terms in which the two open string words are multiplied. For these terms there is no $Z$ next to the word to ``do the kissing''.
Precisely these terms were considered in Appendix C.5. They do not contribute at large $N$.

\subsection{Validity of the Cuntz Chain Hamiltonian}

We have made a number of approximations. When contracting the open string words, only the planar diagrams have been summed.
The non-planar contributions can only be neglected if ${L^2\over N}\ll 1$. Contributions coming from contractions between $Z$s
in the open strings and $Z$s associated to the brane system have been dropped. When computing two point functions
in free field theory, if the number of boxes in the representation $R$ is less than $O(N^2)$ and the numbers of $Z$'s in the open
string is $O(1)$, the contractions between any $Z$s in the open string and the rest of the operator are suppressed in the large $N$
limit\cite{recent}. Contractions between $Z$s in different open string words have been dropped because they are non planar and are
hence subleading. No large combinatoric factors modify this. Finally, when $J$ is large, back reaction is important
and the approximations we are employing are no longer valid. When $J$ becomes $O(\sqrt{N})$ the assumption that we
can drop non-planar contributions when contracting the open string words breaks down.

\section{Interpretation}

The operators we are studying are dual to giant gravitons with open strings attached. Since the giant gravitons have finite volume,
the Gauss Law implies that the total charge on each giant must vanish - there must be the same number of strings leaving each brane as
there are arriving on each brane. These operators do indeed satisfy these non-trivial constraints\cite{Balasubramanian:2004nb}, providing
convincing evidence for the proposed duality. The low energy dynamics of the open strings attached to the giant gravitons is a Yang-Mills
theory. This new {\it emergent} $3+1$ dimensional Yang-Mills theory is not described as a local field theory on the $S^3$ on which the
original Yang-Mills theory is defined - it is local on a new space, the world volume of the giant
gravitons\cite{Balasubramanian:2004nb},\cite{Balasubramanian:2001dx}. This new space emerges from the matrix degrees of freedom
participating in the Yang-Mills theory. Reconstructing this emergent gauge theory may provide a simpler toy model that will give us important
clues into reconstructing the full AdS$_5\times$S$^5$ quantum gravity. In this section, our goal is to make contact with this emergent
Yang-Mills dynamics.

\subsection{Dynamical Emergence of Chan-Paton Factors}

Return to the $H_{+1\to 1}$ hop off interaction obtained in section 2.1. Recall that this corresponds to the interaction in which a $Z$
hops out of the first site of string 1. If we expand the matrix $M_1$ for large $b_1$, we find
\begin{equation}
M_1= \sum_{n=0}^{\infty} M_1(n)b_1^{-n}.
\end{equation}
The leading order $M_1(0)$ is simply $-{\bf 1}$ with ${\bf 1}$ the $6\times 6$ identity matrix. The $Z$ simply hops off
the string and onto the brane without much rearranging of the system. This is the dominant process.
Next, consider the term of order $b_1^{-1}$. It is simple to compute
\begin{equation}
 M_1(1)=\left[\matrix{
0  &0  &0   &1  &1  &0 \cr
0  &0  &-1  &0  &0  &-1\cr
0  &1  &0   &0  &-1 &0 \cr
-1 &0  &0   &0  &0  &1 \cr
-1 &0  &1   &0  &0  &0 \cr
0  &1  &0   &-1 &0  &0
}\right].
\end{equation}
The radius of the giant graviton $R_g$ is related to its momentum $P$ by $R_g=\sqrt{P\over N}$. The giant orbits with a radius
$R =\sqrt{1-R_g^2}.$ For the two giants in the bound state we are considering we have $P_1=b_0$ and $P_2=b_0+b_1$. Using the
fact that $b_0=O(N)$ and $b_1=O(1)$ it is simple to verify that both the difference in the radii of the two giants and the difference
in the radii of their orbits is proportional to $b_1$. Thus, a $b_1^{-1}$ dependence indicates a potential with an inverse distance
dependence which is the correct dependence for massless particles moving in $3+1$ dimensions.
In Figure \ref{fig:interactions1.ps} we have represented the transitions implied by $M_1(1)$ graphically.
Transitions between any two adjacent Young diagrams are allowed.
\myfig{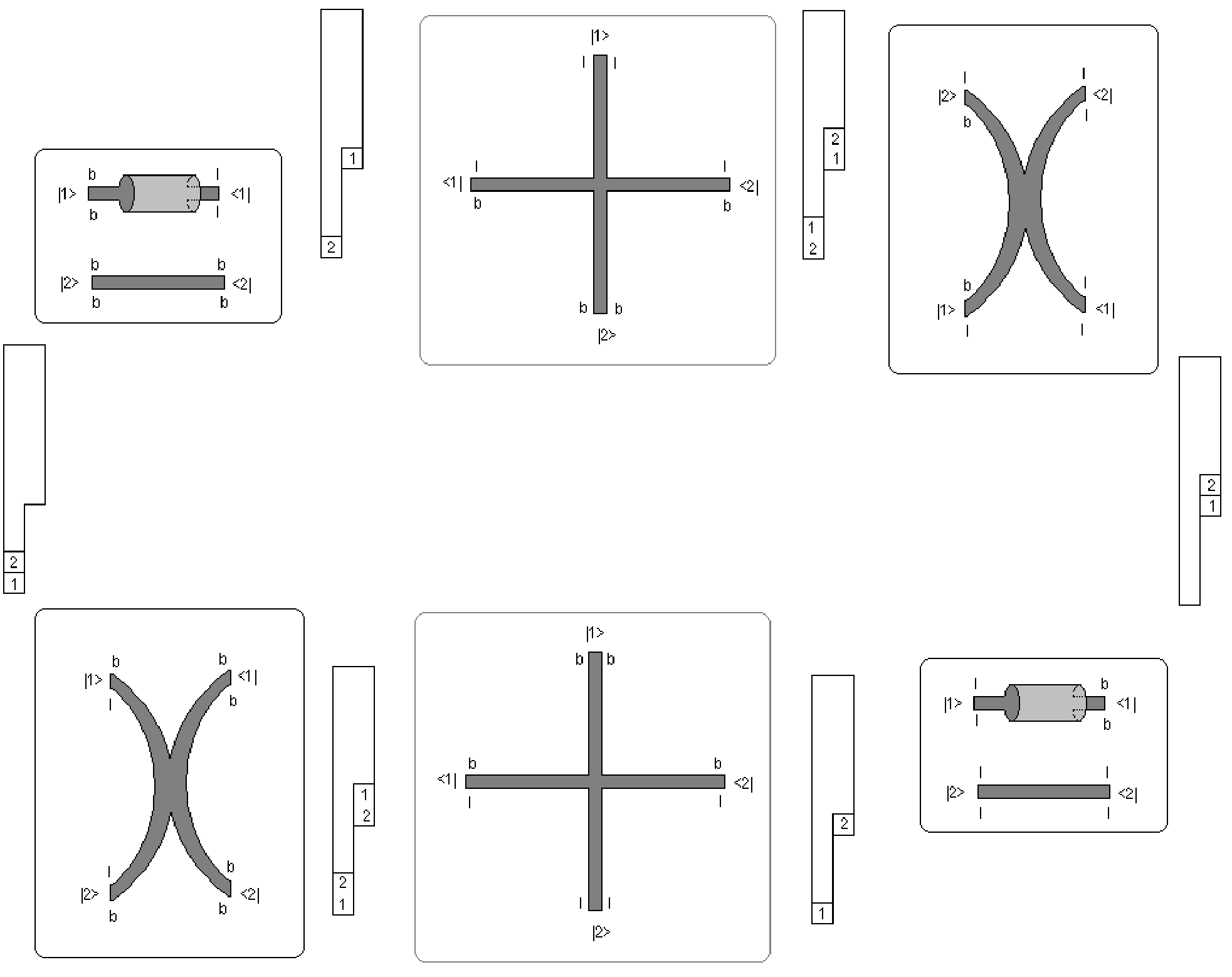}{18.0}{The order $b_1^{-1}$ terms in the hop off interaction. This interaction allows a transition between
the operators described by any two adjacent Young diagrams. The figures between the Young diagram show the open string diagram
relevant for the clockwise transition. The kets are associated to the open string states before the transition; the bras to the states
after the transition. The end point labels `b' and `l' are for big brane and little brane.}

As an example, consider the transition
$$\young({\,}{\,},{\,}{\,},{\,}{\,},{\,}{1},{\,},{\,},{\,},{2})\to \young({\,}{\,},{\,}{\,},{\,}{\,},{\,}{\twoone},{\,},{\,},{\,},{\,},{\onetwo}).$$
The upper label of string 1 has moved. In all of the transitions shown, the upper index of string 1 always moves, so that it is natural
to associate the upper index of string 1 with the first site of string one, and to look for an interpretation of this interaction in
terms of open string processes that involve the upper index of string 1. The figures between the Young diagram show that there is indeed a
natural interpretation for these transitions. {\it It is clear that our Cuntz oscillator dynamics illustrates how the Chan-Paton factors
for open strings propagating on multiple branes arise dynamically.} Drawing all possible ribbon diagrams correctly accounts for both
$M_1(0)$ and $M_1(1)$.

\subsection{Physical Basis}

Although the interpretation of the $b_1^{-1}$ terms is encouraging, there are extra higher order corrections
($M_1(2)b_1^{-2}$, $M_1(3)b_1^{-3}$ and higher orders) that do not seem to have a natural open string interpretation. In addition to this,
the interaction we have obtained depends on the open string
words describing each open string, the Young diagram describing the brane bound state system as well as the order in which the strings
were attached. This dependence on the order in which the strings are attached is not physically sensible.

It is natural to expect that the resolution to these two puzzles is connected. Recall that when constructing the restricted Schur polynomial
we have assumed that when computing reductions, string 1 is removed first and string 2
second. This arbitrary choice defines a basis for the Cuntz oscillator chain.
We interpret the unphysical features of our interactions, described in the previous
paragraph, as reflecting a property of the basis it is written in and not as an inherent problem with the interaction.
In this section we will define a new physical basis, singled out by the requirement that the boundary interaction does
not depend on the order in which the open strings are attached.

A few comments are in order. A basis for the ${1\over 2}$ BPS states (giants with no open strings attached) is provided by the
taking traces of $Z$ or by taking subdeterminants or by the Schur polynomials. These are three perfectly acceptable bases, since
using any single one of these bases we can generate, by taking linear combinations of the elements of the basis considered,
 a member from every ${1\over 2}$ BPS multiplet\cite{Corley:2001zk}. From a physical point of view, these different bases are not
on an equal footing: the Schur polynomial is the most useful. Indeed, the Schur polynomials diagonalize the matrix of two point
correlators (Zamolodchikov metric) so that they can be put into correspondence with the (orthogonal) states of a Fock space.
In the same way, the basis for excited giants gravitons we have been considering is a perfectly acceptable basis. However, it
is the operators in the physical basis (defined below) that have a good physical interpretation.

Denote our two strings by string $A$ and string $B$. The state obtained by attaching string $A$ first will be denoted by
$|b_0,b_1,x_A y_A, x_B y_B \rangle $, where $x_A y_A$ are the endpoints of string $A$ and  $x_B y_B$ are the endpoints of
string $B$. The state obtained by attaching string $B$ first will be denoted by $|b_0,b_1,x_B y_B, x_A y_A \rangle\!\rangle $.
In each subspace of sharp giant graviton momentum (definite $b_0$ and $b_1$), we can write the following relation between
these two sets of states
\begin{equation}
 \left[\matrix{
|b_0,b_1,bb,ll \rangle\cr
|b_0,b_1,ll,bb \rangle\cr
|b_0,b_1,bl,lb \rangle\cr
|b_0,b_1,lb,bl \rangle\cr
|b_0,b_1,ll,ll \rangle\cr
|b_0,b_1,bb,bb \rangle}\right]=PT
\left[\matrix{
|b_0,b_1,bb,ll \rangle\!\rangle\cr
|b_0,b_1,ll,bb \rangle\!\rangle\cr
|b_0,b_1,bl,lb \rangle\!\rangle\cr
|b_0,b_1,lb,bl \rangle\!\rangle\cr
|b_0,b_1,ll,ll \rangle\!\rangle\cr
|b_0,b_1,bb,bb \rangle\!\rangle
}\right],
\end{equation}
where
$$
P=\left[\matrix{
0 &1 &0 &0 &0 &0\cr
1 &0 &0 &0 &0 &0\cr
0 &0 &0 &1 &0 &0\cr
0 &0 &1 &0 &0 &0\cr
0 &0 &0 &0 &1 &0\cr
0 &0 &0 &0 &0 &1}\right]\qquad {\rm and}$$
{\small
$$ T=\left[\matrix{
\left(1-{1\over (b_1+1)^2}\right)           & {1\over (b_1+1)^2}                         &-{1\over (b_1+1)}\sqrt{1-{1\over (b_1+1)^2}} &-{1\over (b_1+1)}\sqrt{1-{1\over (b_1+1)^2}} &0 &0\cr
{1\over (b_1+1)^2}                          & \left(1-{1\over (b_1+1)^2}\right)          &{1\over (b_1+1)}\sqrt{1-{1\over (b_1+1)^2}}  &{1\over (b_1+1)}\sqrt{1-{1\over (b_1+1)^2}}  &0 &0\cr
{1\over (b_1+1)}\sqrt{1-{1\over (b_1+1)^2}}  &-{1\over (b_1+1)}\sqrt{1-{1\over (b_1+1)^2}} &\left(1-{1\over (b_1+1)^2}\right)          &-{1\over (b_1+1)^2}                          &0 &0\cr
{1\over (b_1+1)}\sqrt{1-{1\over (b_1+1)^2}}  &-{1\over (b_1+1)}\sqrt{1-{1\over (b_1+1)^2}} &-{1\over (b_1+1)^2}                          &\left(1-{1\over (b_1+1)^2}\right)           &0 &0\cr
0                                          &0                                          &0                                          &0                                          &1 &0\cr
0                                          &0                                          &0                                          &0                                          &0 &1}
\right].$$
}
The matrix $T$ is determined by the subgroup swap rule of \cite{de Mello Koch:2007uu}. It is satisfying that $PT\times PT=1$.
It is straight forward to check that
\begin{equation}
 H_{+1\to 1}=A_{2\to 1}\, PT\, H_{+2\to 2}\, A_{1\to 2}\, PT,
\end{equation}
where
$$ \left[\matrix{
|b_0-2,b_1+2,bb,ll \rangle\cr
|b_0-1,b_1,ll,bb \rangle\cr
|b_0-2,b_1+2,bl,lb \rangle\cr
|b_0-1,b_1,lb,bl \rangle\cr
|b_0-2,b_1+2,ll,ll \rangle\cr
|b_0-1,b_1,bb,bb \rangle}\right]=A_{2\to 1}
\left[\matrix{
|b_0-1,b_1,bb,ll \rangle\cr
|b_0-1,b_1,ll,bb \rangle\cr
|b_0-1,b_1,bl,lb \rangle\cr
|b_0-1,b_1,lb,bl \rangle\cr
|b_0-1,b_1,ll,ll \rangle\cr
|b_0-1,b_1,bb,bb \rangle
}\right],\qquad {\rm and}$$
$$ \left[\matrix{
|b_0-2,b_1+1,bb,ll \rangle\cr
|b_0-1,b_1-1,ll,bb \rangle\cr
|b_0-2,b_1+1,bl,lb \rangle\cr
|b_0-1,b_1-1,lb,bl \rangle\cr
|b_0-2,b_1+1,ll,ll \rangle\cr
|b_0-1,b_1-1,bb,bb \rangle}\right]=A_{1\to 2}
\left[\matrix{
|b_0-2,b_1+1,bb,ll \rangle\cr
|b_0-2,b_1+1,ll,bb \rangle\cr
|b_0-2,b_1+1,bl,lb \rangle\cr
|b_0-2,b_1+1,lb,bl \rangle\cr
|b_0-3,b_1+3,ll,ll \rangle\cr
|b_0-1,b_1-1,bb,bb \rangle
}\right].$$
Denote the similarity transformation which takes us to the physical basis by $S$. In this basis, we denote
$H_{+1\to 1}$ by $\hat{H}_{+1\to 1}$ and $H_{+2\to 2}$ by $\hat{H}_{+2\to 2}$. Clearly
$$\hat{H}_{+1\to 1} = SH_{+1\to 1}S^{-1},\qquad \hat{H}_{+2\to 2}=S H_{+2\to 2}S^{-1}.$$
The transformation $S$ is now determined by the requirement
\begin{equation}
\hat{H}_{+1\to 1}=P\hat{H}_{+2\to 2}P .
\end{equation}
We have not yet been able to solve this equation for $S$. Due to the presence of $A_{1\to 2}$ and $A_{2\to 1}$ in the
relation between $H_{+1\to 1}$ and $H_{+2\to 2}$, it seems that $S$ must mix subspaces of different giant momenta
$(b_0,b_1)$. In this case the physical basis will not have sharp giant momentum and hence the resulting states will not have a
definite radius. This is not too surprising: the open strings will pull ``dimples'' out of the giant graviton's world volume
so that the giant with an open string attached does not have a definite radius.
We leave the interesting question of determining the transformation $S$ for the future.

\section{Discussion}

A bound state of giant gravitons can be excited by attaching open strings. The problem of computing the anomalous
dimensions of these operators can be replaced with the problem of diagonalizing a Cuntz oscillator Hamiltonian. In
this article we have developed the technology needed to construct this Cuntz oscillator Hamiltonian to one loop.
Firstly, we have given an algorithmic construction of the operators dual to excitations described by open strings which
stretch between the branes. This involved giving an explicit construction of the intertwiner which is used to construct
the relevant restricted Schur polynomial. Secondly, we have developed methods that allow an efficient evaluation
of any restricted character. Our method expresses the restricted character graphically as a sum of strand diagrams.
Finally, we have explained how to derive the boundary interaction terms from identities satisfied by the restricted
Schur polynomials. Since the excited giant graviton operators are small excitations of
BPS states, we expect that our results can be extrapolated to strong
coupling and hence can be compared with results from the dual string theory. The form of our Cuntz
oscillator Hamiltonian provides evidence that the excitations of the giant gravitons have the detailed interactions of
an emergent gauge theory. In particular, we have demonstrated the dynamical emergence of the Chan-Paton factors of the
open strings. We have also started to clarify the dictionary
relating the states of the Cuntz oscillator chain to the states of string field theory on D-branes in AdS$_5\times$S$^5$.
Although we have mainly considered a bound state of two sphere giants with two open strings attached, our methods are
applicable to an arbitrary bound state of giant gravitons with any number of open strings attached.

Our result is a generalization of the spin chains considered so far in the literature: usually the spin chain gives a description of
closed strings. Our Cuntz oscillator describes the dynamics of an open string interacting with a giant graviton. Both the state
of the string (described by the Cuntz chain occupation numbers) and the state of the giant graviton (the shape of the Young
diagram) are dynamical in our approach.

It is worth emphasizing that the new emergent gauge symmetry is distinct from the original gauge symmetry of the theory\cite{Balasubramanian:2004nb}.
The excited giant graviton operators\cite{Balasubramanian:2004nb} are obtained by taking a trace over the indices of the
symmetric group matrix $\Gamma_R(\sigma)$ appearing in the sum
$$
{1\over (n-k)!} \sum_{\sigma\in S_n} \Gamma_R(\sigma)\Tr(\sigma Z^{\otimes n-k}W^{(1)}\cdots W^{(k)}),\qquad {\rm where}
$$
$$
\Tr (\sigma Z^{\otimes n-k}W^{(1)}\cdots W^{(k)})= Z^{i_1}_{i_{\sigma (1)}}Z^{i_2}_{i_{\sigma (2)}}\cdots
Z^{i_{n-k}}_{i_{\sigma (n-k)}}(W^{(1)})^{i_{n-k+1}}_{i_{\sigma (n-k+1)}}\cdots
(W^{(k)})^{i_{n}}_{i_{\sigma (n)}}.
$$
The color indices of the original super Yang-Mills theory are all traced: every term in the above sum is a color singlet
with respect to the gauge symmetry of the original Yang-Mills theory. The color indices of the new gauge theory
arise from the labeling of the partial trace over $\Gamma_R(\sigma)$. In some sense we are ``substituting'' symmetric group indices
for the original gauge theory indices. We call this mechanism ``{\it color substitution}''.

There are a number of directions in which this work can be extended. For Young diagrams with $m$ columns we expect an
emergent Yang-Mills theory with gauge group $U(m)$. It would be nice to repeat the calculations we performed here in that
setting\footnote{For the $m=3$ case, see Appendix F.}.
Another interesting calculation would involve studying the dynamics of two giant gravitons with strings stretched between them. In
general, the boundary terms will certainly have different values at each boundary (as anticipated in
\cite{Berenstein:2006qk}) in which case there will be a net flow of $Z$s from one brane to the other.
This flow of $Z$'s will produce a force between the two giants, conjectured to be an attractive force in\cite{Berenstein:2006qk}.

A very concrete application of our methods is the construction of the gauge theory operator dual to the fat
magnon\cite{Hirano:2006ti}\footnote{The fat magnon in the plane wave background is the hedgehog of \cite{Shahin}}.
The fat magnon is a bound state of a giant graviton and giant magnons (fundamental strings). Essentially, due to the background
five form flux, the giant magnon becomes fat by the Myers effect\cite{Myers:1999ps}. The fat magnon has the same anomalous
dimension as the giant magnon. It would be nice to explicitely recover this anomalous dimension using our technology\footnote{We
would like to thank Shahin Sheikh-Jabbari for suggesting this to us.}.

Finally, there is now a proposal
for gauge theory operators dual to brane-anti-brane states\cite{Kimura:2007wy}. This proposal was made, at the level of the free field
theory, by identifying the operators that diagonalize the two point functions of operators built from $Z$ and $Z^\dagger$. Since these
states are non-supersymmetric, corrections when the coupling is turned on are expected to be important for the physics. It would be
interesting to extend the technology developed in this article to this non-supersymmetric setting.

$$ $$

\noindent
{\it Acknowledgements:}
We would like to thank Rajsekhar Bhattacharyya, Norman Ives, Sanjaye Ramgoolam, Jo\~ao Rodrigues,
Shahin Sheikh-Jabbari and Alex Welte for pleasant discussions and/or helpful correspondence.
This work is based upon research supported by the South African Research Chairs
Initiative of the Department of Science and Technology and National Research Foundation.
Any opinion, findings and conclusions or recommendations expressed in this material
are those of the authors and therefore the NRF and DST do not accept any liability
with regard thereto. This work is also supported by NRF grant number Gun 2047219.

\appendix

$$ $$

\noindent
{\it Navigating the Appendices:} In the appendices we will freely make use of results obtained in
the previous two articles in this series\cite{de Mello Koch:2007uu,de Mello Koch:2007uv}. A reader
wishing to master the details of our analysis will need to review this background. We will now explain
which results are used when. For a discussion of intertwiners see sections 2.2 and section C.1 of
\cite{de Mello Koch:2007uu}. In appendix B we make frequent use of the subgroup swap rule which
is derived in appendix D of \cite{de Mello Koch:2007uu}. This is perhaps the most technical
result from \cite{de Mello Koch:2007uu,de Mello Koch:2007uv} that is used in this article. For
this reason, we have reviewed a concrete example in the first section of appendix B. We also use
the character identity given at the end of section D.1 of \cite{de Mello Koch:2007uu}. The strategy
for deriving the hopping identities of appendix C was given in \cite{de Mello Koch:2007uv}.

\section{Intertwiners}

Intertwiners are used to construct operators dual to states with open strings stretching between giant gravitons.
In this appendix we provide a general discussion of intertwiners and their construction.

\subsection{Strings stretching between two branes}

The Gauss Law is a strict constraint on the allowed excited brane configurations\cite{Balasubramanian:2004nb}:
since the branes we consider have
a compact world volume, the total charge on any given brane must vanish. This implies that to construct a state with
strings stretching between two branes, we need at least two strings in the brane plus string system. Thus, in constructing
the restricted Schur polynomial, we will need to remove at least two boxes. For concreteness, consider the case of two
sphere giants, so that our restricted Schur polynomial is built with the Young diagram $R$ that has two columns and each
column has $O(N)$ boxes. $R$ has a total of $n=O(N)$ boxes. Denote the two boxes to be removed
in constructing the restricted Schur polynomial\footnote{See appendix E for a quick review of restricted Schur polynomials
and \cite{Balasubramanian:2004nb,de Mello Koch:2007uu,de Mello Koch:2007uv} for a detailed discussion.} by box 1 and box 2. To
attach strings stretching between these two giants, the two boxes must belong to different columns. Assume that box 1
belongs to column 1 and box 2 to column 2. After restricting $S_n$ to an $S_{n-1}$ subgroup, representation $R$ subduces
irrep $R'$ (whose Young diagram is obtained by removing box 1 from $R$) and irrep $S'$ (whose Young diagram is obtained by
removing box 2 from $R$). If we now further restrict to an $S_{n-2}$ subgroup, one of the irreps subduced by $R'$ is $R''$
(whose Young diagram is obtained by removing box 2 from $R'$) and one of the irreps subduced by $S'$ is $S''$ (whose
Young diagram is obtained by removing box 1 from $S'$). Note that $R''$ and $S''$ have the same Young diagram (and hence the
same dimension) but act on distinct states in the carrier space of $R$. The two possible intertwiners we can define map
between the states belonging to $R''$ and the states belonging to $S''$.

The precise form of the intertwiners depends on the basis used for the $S_{n-2}$ irreps $\Gamma_{R''}(\sigma)$ and
$\Gamma_{S''}(\sigma)$. In writing down the intertwiner, we assume that $\Gamma_{R''}(\sigma)$ and $\Gamma_{S''}(\sigma)$
represent $\sigma$ with the same matrix. With this assumption, it is possible to put the elements of the basis of the carrier
space of $R''$ into one to one correspondence with the elements of the basis of the carrier space of $S''$:
$|i,R''\rangle \leftrightarrow |i,S''\rangle$. We will use this correspondence below. In a suitable basis, we have
$$
\Gamma_R (\sigma )=\left[
\matrix{\Gamma_{R''}(\sigma ) &0                    &\cdots\cr
        0                    &\Gamma_{S''}(\sigma) &\cdots\cr
        \cdots               &\cdots               &\cdots
}\right],
$$
for $\sigma\in S_{n-2}$. In constructing the restricted Schur polynomial, we also consider more general $\sigma\in S_n$.
In this case, if $\sigma\notin S_{n-2}$, $\Gamma_R(\sigma )$ will not be block diagonal. Even in this more general case,
we will use the labels of the $S_{n-2}$ subduced subspaces to label the carrier space of irrep $R$. Denote the projection
operator that projects from the carrier space of $R$ to the $R''$ subspace by $P_{R\to R'\to R''}$, and the projection
operator that projects from the carrier space of $R$ to the $S''$ subspace by $P_{R\to S'\to S''}$. Clearly, the intertwiner
which maps from $S''$ to $R''$ must take the form
\begin{equation}
I_{R'',S''}=P_{R\to R'\to R''}OP_{R\to S'\to S''}=\left[
\matrix{0      &M      &\cdots\cr
        0      &0      &\cdots\cr
        \cdots &\cdots &\cdots
}\right].
\label{intertwiner}
\end{equation}
The second possible intertwiner that we can construct is given by
$$
I_{S'',R''}=P_{R\to S'\to S''}OP_{R\to R'\to R''}=\left[
\matrix{0      &0      &\cdots\cr
        M      &0      &\cdots\cr
        \cdots &\cdots &\cdots
}\right].
$$
We want to find a unique specification for $O$ so that $M$ is simply the identity matrix. For $\sigma\in S_{n-2}$ we have
$$\Gamma_R(\sigma )I_{R'',S''}=\left[
\matrix{0      &\Gamma_{R''}(\sigma)M      &\cdots\cr
        0      &0      &\cdots\cr
        \cdots &\cdots &\cdots
}\right]$$
and
$$ I_{R'',S''}\Gamma_R(\sigma )=\left[
\matrix{0      &M\Gamma_{S''}(\sigma)      &\cdots\cr
        0      &0      &\cdots\cr
        \cdots &\cdots &\cdots
}\right].$$
Now, by assumption, $\Gamma_{R''}(\sigma)=\Gamma_{S''}(\sigma)$ since we have $\sigma\in S_{n-2}$. Thus,
\begin{equation}
\big[ \Gamma_R(\sigma ),I_{R'',S''}\big]=\left[
\matrix{0      &\big[\Gamma_{R''}(\sigma),M\big]      &\cdots\cr
        0      &0      &\cdots\cr
        \cdots &\cdots &\cdots
}\right].
\label{offdiag}
\end{equation}
Applying Schur's Lemma (for irrep $R''$) to the right hand side implies that $M$ is the identity matrix if and only if
$\big[ \Gamma_R(\sigma ),I_{R'',S''}\big]=0$ for all $\sigma\in S_{n-2}$.  Clearly, for $\sigma\in S_{n-2}$ we have
$\big[ \Gamma_R(\sigma ),P_{R\to R'\to R''}\big]=\big[ \Gamma_R(\sigma ),P_{R\to S'\to S''}\big]=0$ so that
$$0=\big[ \Gamma_R(\sigma ),I_{R'',S''}\big]=P_{R\to R'\to R''}\big[ \Gamma_R(\sigma ),O\big] P_{R\to S'\to S''}.$$
Thus, we will require
\begin{equation}
\big[ \Gamma_R(\sigma ),O\big]=0,\qquad \forall \sigma\in S_{n-2}.
\label{determineO}
\end{equation}
If we specify a condition that determines the normalization of the intertwiner, then this normalization condition and
(\ref{determineO}) provide the specification for $O$ that we were looking for. The normalization of the intertwiner is
fixed by demanding that
$$ \Tr (M)={\rm dim}_{R''},$$
with ${\rm dim}_{R''}$ the dimension of irrep $R''$. This provides a unique definition of the intertwiner.

For the example we are considering here, imagine that the $S_{n-1}$ subgroup is obtained as
$$ {\cal G}=\{\sigma\in S_n | \sigma (n)=n\},$$
and further that the $S_{n-2}$ subgroup is obtained as
$$ {\cal H}=\{\sigma\in {\cal G} | \sigma (n-1)=n-1\} .$$
Then the intertwiner is given by
$$
I_{R'',S''}={\cal N} P_{R\to R'\to R''}\Gamma_R (n,n-1) P_{R\to S'\to S''},
$$
with
$${\cal N}^{-1}={\Tr_{R'',S''}(\Gamma_R (n,n-1) )\over {\rm dim}_{R''}}\equiv\sum_{i=1}^{{\rm dim}_{R''}}
{\langle R'',i |\Gamma_R(n,n-1)|S'',i\rangle\over {\rm dim}_{R''} }. $$
This last equation makes use of the correspondence between the bases of the carrier spaces $R''$ and $S''$.
Using the technology developed in the next appendix, we find
$${\Tr_{R'',S''}(\Gamma_R (n,n-1) )\over {\rm dim}_{R''}}=\sqrt{1-{1\over (c_1-c_2)^2}},$$
where $c_1$ and $c_2$ are the weights associated with box 1 and box 2 respectively. Note that the above trace is
invariant under simultaneous similarity transformations of $R''$ and $S''$. It will however, change under general
similarity transformations so that this last result is dependent on our choice of basis.

\subsection{The General Construction}

In the previous section we have developed our discussion of the intertwiner using a system of two branes with strings
stretching between them. Our conclusion however, is completely general. For any system of branes with strings stretching
between the branes, the intertwiner is always given, up to normalization, by the product
(projection operator)$\times$(group element)$\times$(projection operator). The Gauss Law forces the net charge on any
given brane's worldvolume to vanish. This implies that for every string leaving a brane's worldvolume, there will
be a string ending on the worldvolume. Thus, starting with any particular brane with a stretched string attached, we can
follow the string to the next brane, switch to the stretched string leaving that brane, follow it and so on, until we again
reach the first brane. If we move along $k$ stretched strings before returning to the starting point, the group element is
$\Gamma_R(n,n-k+1)$. The normalization factor easily follows using the results of Appendix B.

\subsection{Example}

Consider the excited brane system described by the diagram (see Appendix E for a summary of our graphical notation)
{\small
$$ \young({\,}{\,}{\,}{\,}{\,},{\,}{\,}{\,}{\,}{\onetwo},{\,}{\,}{\,},{\,}{\,}{\twothree},{\,},{\threeone}) .$$
}
The boxes are labeled by the upper index in each box and the weight of box $i$ is denoted $c_i$.
The projector $P_{R\to R'''_1}$ projects through the following sequence of irreps
{\small
$$ \yng(5,5,3,3,1,1)\to\yng(5,4,3,3,1,1)\to\yng(5,4,3,2,1,1)\to\yng(5,4,3,2,1).$$
}
The projector $P_{R\to R'''_2}$ projects through the following sequence of irreps
{\small
$$ \yng(5,5,3,3,1,1)\to\yng(5,5,3,3,1)\to\yng(5,4,3,3,1)\to\yng(5,4,3,2,1) .$$
}
The intertwiner is now given by
$$ I_{12}={\cal N} P_{R\to R'''_2}\Gamma_R\left((n,n-2)\right)P_{R\to R'''_1},$$
where
$${\cal N}^{-1}={\Tr_{R'''_2,R'''_1}\left(\Gamma_R\left((n,n-2)\right)\right)\over {\rm dim}_{R'''_1}}=
{1\over c_2-c_3}\sqrt{1-{1\over (c_1-c_2)^2}}\sqrt{1-{1\over (c_1-c_3)^2}},
$$
is easily computed using the methods of appendix B. To understand the order of the projection operators, note that
\begin{eqnarray}
 \Tr _{R'''_1,R'''_2}\Big(\Gamma_R(\sigma )\Big)&=&\sum_{i}\langle i,R'''_1 |\Gamma_R(\sigma)|i,R'''_2\rangle
\nonumber\\
&=&\Tr ({\cal N}^{-1} P_{R\to R'''_2}\Gamma_R(n,n-2)P_{R\to R'''_1}\Gamma_R(\sigma) ),
\nonumber
\end{eqnarray}
so that the row (column) index of the trace is column (row) index of the intertwiner respectively.

\section{Restricted Characters}

Starting from $S_n$, define a chain of subgroups ${\cal G}_i$ $i=1,...,d$ as follows
\begin{equation}
{\cal G}_1=\{\sigma\in S_n|\sigma(n)=n\}
\label{sgroup1}
\end{equation}
\begin{equation}
{\cal G}_i=\{\sigma\in {\cal G}_{i-1}|\sigma(n-i+1)=n-i+1\},\qquad i=2,3,...,d.
\label{sgroup2}
\end{equation}
In this appendix we will give a simple algorithm for the computation of
$$\chi_{R_1,R_2}\Big( (p_1,p_2,...,p_m) \Big)\equiv\Tr_{R_1,R_2} \Big(\Gamma_R\big( (p_1,p_2,...,p_m)\big) \Big)$$
with $R_1$ and $R_2$ irreps of ${\cal G}_d$ subduced from $R$, $(p_1,p_2,...,p_m)$ is an element of $S_n$ specified using
the cycle notation and $n-d< p_i \le n$ $\forall i$. We call $\chi_{R_1,R_2}$ a {\it restricted character}. If $R_1=R_2$,
we will simply write $\chi_{R_1}$. We have already seen that
restricted characters determine the normalization of the intertwiners. Further, they are also needed in the derivation of
the hopping identities that determine the interactions between strings and the branes to which they are attached.

The first subsection of this appendix reviews the subgroup swap rule in the setting of a specific example.
In the next subsection we will derive the algorithm for the computation of the restricted character. The third subsection
of this appendix describes a graphical notation which considerably simplifies the computation. The remainder of the appendix
then develops this diagrammatic notation further.

\subsection{Review of the Subgroup Swap Rule}

In this appendix, we review the subgroup swap rule.  The reader requiring a more detailed explanation can consult Appendix D of \cite{de Mello Koch:2007uu}.
Consider the restricted Schur polynomial
\[\chi^{(2)}_{R,R''}\Big|_1\Big|_2={1\over (n-2)!}\sum_{\sigma\in S_n}\mathrm{Tr}_{R''}\left(\Gamma_R(\sigma )\right)Z^{i_1}_{i_{\sigma (1)}}\cdots
Z^{i_{n-2}}_{i_{\sigma (n-2)}}(W^{(2)})^{i_{n-1}}_{i_{\sigma (n-1)}}(W^{(1)})^{i_n}_{i_{\sigma (n)}}.\]

The labelling on the left hand side tells us to first restrict with respect to the subgroup that leaves the index of $W^{(1)}$ inert, and 
then with respect to the subgroup that leaves the index of $W^{(2)}$ inert.  In general, we will get a different polynomial if we were to 
restrict first with respect to the subgroup that leaves the index of $W^{(2)}$ inert, and then with respect to the subgroup that leaves 
the index of $W^{(1)}$ inert.
There is a relation between these two sets of polynomials, which is known as the ``subgroup swap rule''.  

We use the weights of the boxes of the Young diagrams in the subgroup swap rule.  All weights are defined by the Young diagram before the swap.  
The weight of the box labelled with upper index 1 is denoted by $c_1^U$ and the weight of the box labelled with lower index 1 is denoted by $c_1^L$.  
Similarly for index 2. The upper and lower no-swap factors are given by
\[
  N_U = \sqrt{1-\frac{1}{(c_1^U-c_2^U)^2}}, \qquad N_L = \sqrt{1-\frac{1}{(c_1^L-c_2^L)^2}} .
\]
The upper and lower swap factors are given by
\[
 S_U = \frac{1}{c_1^U - c_2^U}, \qquad S_L = \frac{1}{c_1^L - c_2^L} .
\]
Our example uses a restricted Schur with three strings attached.  Swapping strings $2$ and $3$, the subgroup swap rule gives
\begin{eqnarray*}
\chi_{\young({\,}{\,}{\onetwo},{\,}{\twothree},{{\threeone}})} (\sigma) \Big|_1\Big|_2\Big|_3 &=&\left [ N_L N_U \chi_{\young({\,}{\,}{\onetwo},{\,}{\twothree},{{\threeone}})} (\sigma) + S_U N_L \chi_{\young({\,}{\,}{\onetwo},{\,}{3},{{\twoone}})} (\sigma) \right ] \Big|_1\Big|_3\Big|_2 \cr
 &+&\left [ S_L N_U \chi_{\young({\,}{\,}{\onetwo},{\,}{\twoone},{{3}})} (\sigma) + S_U S_L \chi_{\young({\,}{\,}{\onetwo},{\,}{\threeone},{{\twothree}})} (\sigma) \right ] \Big|_1\Big|_3\Big|_2
\end{eqnarray*}
where
$$
  N_U = \sqrt{1-\frac{1}{(c_2^U-c_3^U)^2}} = \frac{\sqrt{3}}{2}
$$
$$
  N_L = \sqrt{1-\frac{1}{(c_2^L-c_3^L)^2}} = \frac{\sqrt{3}}{2},
$$
$$
 S_U = \frac{1}{c_2^U - c_3^U} = \frac{1}{2},
$$
$$
 S_L = \frac{1}{c_2^L - c_3^L} = \frac{1}{2}.
$$
Whenever an index is swapped, we include a swap factor, and whenever there is no swap, 
we include a no-swap factor.

\subsection{Computing Restricted Characters}

Consider an irrep $R$ of $S_n$ labeled by a Young diagram which has at least two boxes, either of which can be dropped to
leave a valid Young diagram. Label these two boxes by 1 and 2. Denote the weights of these boxes by $c_1$ and $c_2$. Denote
the irrep of $S_{n-2}$ obtained by dropping box 1 and then box 2 by $R_1''$. Denote the irrep of $S_{n-2}$ obtained by dropping
box 2 and then box 1 by $R_2''$. Our first task is to compute
$$ \Tr_{R''_1,R''_2}\left(\Gamma_R\left( (n,n-1)\right)\right).$$
Using the subgroup swap rule obtained in \cite{de Mello Koch:2007uu}, we can write
\begin{eqnarray}
\chi_{R_1''}\left((n,n-1)\right)&=&\left[1-{1\over (c_1-c_2)^2}\right]\chi_{R_2''}\left((n,n-1)\right)
+{1\over (c_1-c_2)^2}\chi_{R_1''}\left((n,n-1)\right)\label{ssr1}\\
&+&\sqrt{1-{1\over (c_1-c_2)^2}}{1\over c_1-c_2 }
\left[\chi_{R_1'',R_2''}\left((n,n-1)\right)+\chi_{R_2'',R_1''}\left((n,n-1)\right)\right].
\nonumber
\end{eqnarray}
A second application of the subgroup swap rule gives
\begin{eqnarray}
\chi_{R_2'',R_1''}\left((n,n-1)\right)&=&\left[1-{1\over (c_1-c_2)^2}\right]\chi_{R_1'',R_2''}\left((n,n-1)\right)
+{1\over (c_1-c_2)^2}\chi_{R_2'',R_1''}\left((n,n-1)\right)
\nonumber\\
&+&\sqrt{1-{1\over (c_1-c_2)^2}}{1\over c_1-c_2 }
\left[\chi_{R_2''}\left((n,n-1)\right)-\chi_{R_1''}\left((n,n-1)\right)\right].
\label{ssr2}
\end{eqnarray}
Now, substituting the results\cite{de Mello Koch:2007uu}
$$\chi_{R_1''}\left((n,n-1)\right)={1\over c_1-c_2}{\rm dim}_{R_1''},\qquad \chi_{R_2''}\left((n,n-1)\right)={1\over c_2-c_1}{\rm dim}_{R_2''},$$
into (\ref{ssr1}) and (\ref{ssr2}) and solving, we obtain
$$\chi_{R_1'',R_2''}\left((n,n-1)\right)=\sqrt{1-{1\over (c_1-c_2)^2}}{\rm dim}_{R_1''}=\chi_{R_2'',R_1''}\left((n,n-1)\right).$$

Next, consider an irrep of $S_n$ labeled by Young diagram $R$ . Choose three boxes in this Young diagram, and label them 1, 2 and 3
respectively. Choose the boxes so that dropping box 1 gives a legal Young diagram $R'$ labeling an irrep of $S_{n-1}$, dropping box 1
and then box 2 gives a legal Young diagram $R''$ labeling an irrep of $S_{n-2}$, and dropping box 1, then box 2 and then box 3 again
gives a legal Young diagram $R'''$ labeling an irrep of $S_{n-3}$. We will compute
$$\chi_{R'''}\left( (n,n-2)\right)=\Tr_{R'''}\left(\Gamma_R \left( (n,n-2)\right)\right).$$
In what follows, we will frequently need to refer to vectors belonging to the carrier spaces of specific representations subduced by
$R$ when boxes are dropped from $R$. A convenient notation is to list the labels of the boxes that must be dropped from $R$ in the
order in which they must be dropped. Thus, the ket $|i,123\rangle$ is the $i^{\rm th}$ ket belonging to the carrier space of the
$S_{n-3}$ irrep obtained by dropping box 1, then box 2 and then box 3 from $R$; the ket $|j,231\rangle$ is the $j^{\rm th}$ ket belonging
to the carrier space of the $S_{n-3}$ irrep obtained by dropping box 2, then box 3 and then box 1 from $R$ (assuming of course that the
boxes can be dropped from $R$ in this order, giving a legal Young diagram at each step). Start by writing
\begin{eqnarray}
\chi_{R'''} ((&n&,n-2))=\sum_{i=1}^{{\rm dim}_{R'''}}
\langle i,123|\Gamma_R \left( (n,n-2)\right) |i, 123\rangle
\nonumber\\
&=&\sum_{i=1}^{{\rm dim}_{R'''}}
\langle i,123|\Gamma_{R'} \left( (n-1,n-2)\right) \Gamma_R \left( (n,n-1)\right)\Gamma_{R'} \left( (n-1,n-2)\right)|i, 123\rangle .
\nonumber
\end{eqnarray}
Noting that $\Gamma_{R'} \left( (n-1,n-2)\right)|i, 123\rangle $ must belong to the carrier space of $R'$, and using the
completeness relation ($1_{R'}$ is the identity on the $R'$ carrier space)
$$ 1_{R'}=\sum_{k=1}^{{\rm dim}_{R'}} |k,1\rangle\langle k,1|,$$
we have
$$
\chi_{R'''}\left( (n,n-2)\right)=\sum_{i=1}^{{\rm dim}_{R'''}}\sum_{j,k=1}^{{\rm dim}_{R'}}
\langle i,123|\Gamma_{R'} \left( (n-1,n-2)\right)|k,1\rangle\langle k,1| \Gamma_R \left( (n,n-1)\right)
|j,1\rangle
$$
$$
\times \langle j,1|\Gamma_{R'} \left( (n-1,n-2)\right)|i, 123\rangle .
$$
Now, decompose $R'$ into a direct sum of $S_{n-2}$ irreps $ R'=\oplus R_\beta ''$. Use the label $\beta$ to denote the box that must be dropped
from $R'$ to obtain $R_\beta''$. Thus, we can write
$$ 1_{R'}=\sum_{k=1}^{{\rm dim}_{R'}} |k,1\rangle\langle k,1|=\sum_{\beta}\sum_{k=1}^{{\rm dim}_{R_\beta ''}}
|k,1\beta\rangle\langle k,1\beta|,$$
and hence
$$
\chi_{R'''}\left( (n,n-2)\right)=\sum_{i=1}^{{\rm dim}_{R'''}}\sum_{\beta_1,\beta_2}\sum_{k=1}^{{\rm dim}_{R_{\beta_1} ''}}
\sum_{j=1}^{{\rm dim}_{R_{\beta_2} ''}}
\langle i,123|\Gamma_{R'} \left( (n-1,n-2)\right)|k,1\beta_1 \rangle
$$
$$
\times \langle k,1\beta_1 | \Gamma_R \left( (n,n-1)\right)
|j,1\beta_2\rangle\langle j,1\beta_2 |\Gamma_{R'} \left( (n-1,n-2)\right)|i, 123\rangle .
$$
Now, introduce the operator $O(2)$ obtained by summing all two cycles of the $S_{n-2}$ subgroup of which the $R_\beta''$
are irreps. This operator is a Casimir of $S_{n-2}$. If the Young diagram $R_\beta''$ has $r_i$ boxes in the $i^{\rm th}$
row and $c_i$ boxes in the $i^{\rm th}$ column, then when acting on the carrier space of $R_\beta''$ we have\cite{chen}
$$O(2)|i,1\beta\rangle =\left[\sum_i {r_i (r_i-1)\over 2}-\sum_j {c_j(c_j-1)\over 2}\right]|i,1\beta\rangle\equiv
\lambda_\beta |i,1\beta\rangle .$$
Clearly, for the problem we study here,
$\lambda_{\beta_1}=\lambda_{\beta_2}$ if and only if $R_{\beta_1}$ and $R_{\beta_2}$ have the same shape as Young
diagrams. From the definition of the ${\cal G}_2$ subgroup given above, it is clear that
$$\big[ O(2),\Gamma_R\left( (n,n-1)\right)\big]=0.$$
It is now a simple matter to see that
\begin{eqnarray}
\lambda_{\beta_1}\langle k,1\beta_1 | \Gamma_R \left( (n,n-1)\right)|j,1\beta_2\rangle &=&
\langle k,1\beta_1 |O(2) \Gamma_R \left( (n,n-1)\right) |j,1\beta_2\rangle\nonumber\\
&=&\langle k,1\beta_1 | \Gamma_R \left( (n,n-1)\right)O(2)|j,1\beta_2\rangle\nonumber\\
&=&\lambda_{\beta_2}\langle k,1\beta_1 | \Gamma_R \left( (n,n-1)\right)|j,1\beta_2\rangle\nonumber
\end{eqnarray}
so that $\langle k,1\beta_1 | \Gamma_R \left( (n,n-1)\right)|j,1\beta_2\rangle$ vanishes if $R_{\beta_1}$ and $R_{\beta_2}$ do not
have the same shape. A completely parallel argument, using a Casimir of $S_{n-3}$, can be used to show that
$\langle j,1\alpha_1\alpha_2 |\Gamma_{R'} \left( (n-1,n-2)\right)|i, 123\rangle$ is only non-zero if $\alpha_1=2$, $\alpha_2=3$ or
$\alpha_1=3$, $\alpha_2=2$. Thus,
\begin{eqnarray}
\chi_{R'''}\left( (n,n-2)\right)&=&\sum_{i=1,j,k}^{{\rm dim}_{R'''}}\Big[
\langle i,123|\Gamma_{R'} \left( (n-1,n-2)\right)|k,123 \rangle
\langle k,123 | \Gamma_R \left( (n,n-1)\right)
|j,123\rangle\nonumber\\
&\times&\langle j,123 |\Gamma_{R'} \left( (n-1,n-2)\right)|i, 123\rangle+\langle i,123|\Gamma_{R'} \left( (n-1,n-2)\right)|k,132 \rangle
\nonumber\\
&\times&\langle k,132 | \Gamma_R \left( (n,n-1)\right)
|j,132\rangle\langle j,132 |\Gamma_{R'} \left( (n-1,n-2)\right)|i, 123\rangle\Big]
\nonumber\\
&=&\left[{1\over (c_2-c_3)^2}{1\over c_1-c_2}+\left(1-{1\over (c_2-c_3)^2}\right){1\over c_1-c_3}\right]{\rm dim}_{R'''}.
\nonumber
\end{eqnarray}

This example illustrates the general algorithm to be used to compute restricted characters:

\begin{itemize}

\item{} The group element whose trace is to be computed, can be decomposed into a product of two cycles of the form
          $\Gamma_R\left((i,i+1)\right)$. A complete set of states is inserted between each factor.

\item{} Using appropriately chosen Casimirs, one can argue that the only non-zero matrix elements of each factor,
          are obtained when the order of boxes dropped to obtain the carrier space of the bra matches the order of
          boxes dropped to obtain the carrier space of the ket, except for the $(n-i+1)^{\rm th}$ and $(n-i+2)^{\rm th}$
          boxes, whose order can be swapped.

\item{} We can plug in the known value of the restricted character, which we have computed for precisely the two cases
          arising in the previous point.

\end{itemize}

\subsection{Strand Diagrams}

Strand diagrams are a graphical notation designed to compute restricted characters. Strand diagrams keep track of two things:

\begin{itemize}

\item{} The order in which boxes are to be dropped and the identity (= position within the Young diagram) of the boxes.

\item{} The group element whose trace we are computing.

\end{itemize}

\noindent
If we are to drop $n$ boxes, we draw a picture with $n$ columns. The columns are populated by labeled strands - each strand
represents one of the boxes that are to be dropped. We label the strands by the upper index in the box. The reader is strongly
advised to read Appendix E for a summary of our graphical notation. Whatever appears in
the first column is to be dropped first; whatever appears in the second column is to be dropped second and so on.  The strands
are ordered at the top of the diagram, according to the order in which they must be dropped to get the row index. The strands
are ordered at the bottom of the diagram according to the column index. The strands move from the top of the diagram to the
bottom of the diagram, without breaking, so that strands ends at the top connect to the corresponding strand ends at the bottom.
To connect the strands (which in general are in a different order at the top and bottom of the diagram) we need to weave the
strands, thereby allowing them to swap columns. The allowed swaps depends on the specific group element whose trace we are
computing. To determine the allowed swaps, write the group element as a product of cycles of the form $(i,i+1)$. For example,
we would write
$$ (n,n-2)=(n,n-1)(n-1,n-2)(n,n-1). $$
Each time we drop a box, we are considering a new subgroup. The action of the permutation group can be visualized as
a permutation of $n$ indices. The subgroups are obtained by considering elements that hold
certain indices fixed (see (\ref{sgroup1}) and (\ref{sgroup2})). Choose the subgroups involved so that when box $i$ is dropped,
$n-i+1$ is held fixed. Clearly then, each column $j$ is associated with the index $n-j+1$. Each cycle $(i,i+1)$ is drawn as
a box which straddles the columns associated with indices $i$ and $i+1$. When the strands pass through a box, they may do so
without swapping or by swapping columns. Each box is associated with a factor. Imagine that the strands passing through
the box, reading from left to right, are labeled $n$ and $m$. The weights associated with these boxes are $c_n$ and $c_m$
respectively. If the strands do not swap inside the box the factor for the box is
$$ f_{\rm no\,\, swap}={1\over c_n-c_m}.$$
If the strands do swap inside the box, the factor is
$$ f_{\rm swap}=\sqrt{1-{1\over (c_n-c_m)^2}}.$$
Denote the product of the factors, one from each box, by $F$. We have
$$ \Tr_{R_1,R_2}\Big(\Gamma_R(\sigma )\Big) = \sum_i F_i{\rm dim}_{R_1}, $$
where the index $i$ runs over all possible paths consistent with the boundary conditions.
With a little thought, the astute reader should be able to convince herself that this graphical rule is nothing
but a convenient representation of the computation of the last subsection.

\subsection{Strand Diagram Examples}

In this section we will illustrate the use of strand diagrams in the computation of restricted characters. For our first example,
we consider the computation of
$$ \chi_1=\Tr_{\young({\,}{\,}{\onethree},{\,}{\twoone},{\threetwo})}\Big(\Gamma_{\yng(3,2,1)}\big((6,4)\big)\Big).$$
\myfig{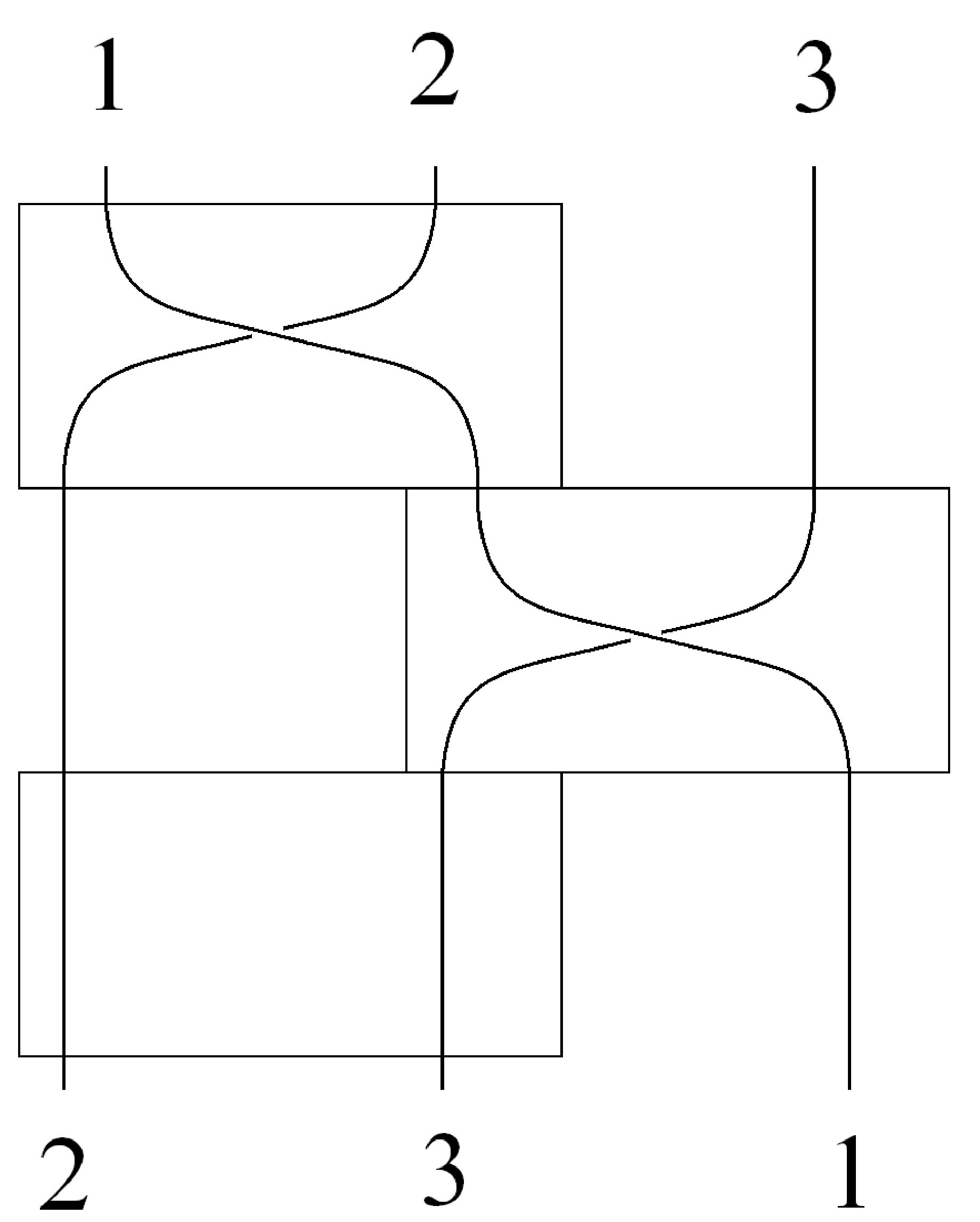}{4.0}{The strand diagram used in the computation of $\chi_1$.}
Writing $(6,4)=(6,5)(4,5)(6,5)$ we obtain the strand diagram shown in Figure \ref{fig:strand1.ps}. The factors for the upper most, middle
and lower most boxes are
$$\sqrt{1-{1\over (c_1-c_2)^2}},\qquad \sqrt{1-{1\over (c_1-c_3)^2}},\qquad {1\over c_2-c_3}$$
respectively. Thus,
\begin{eqnarray}
\chi_1&=&\sqrt{1-{1\over (c_1-c_2)^2}}\sqrt{1-{1\over (c_1-c_3)^2}}{1\over c_2-c_3}{\rm dim}_{\yng(2,1)}
\nonumber\\
&=&2\sqrt{1-{1\over (c_1-c_2)^2}}\sqrt{1-{1\over (c_1-c_3)^2}}{1\over c_2-c_3}.
\nonumber
\end{eqnarray}

The alert reader may worry that our recipe is not unique. Indeed we could also have written $(6,4)=(4,5)(6,5)(4,5)$. In this case,
we obtain the strand diagram given in Figure \ref{fig:strand2.ps}. In this case, the factors for the upper most, middle
and lower most boxes are
$${1\over c_2-c_3},\qquad \sqrt{1-{1\over (c_1-c_2)^2}},\qquad \sqrt{1-{1\over (c_1-c_3)^2}} $$
respectively. This gives exactly the same value for $\chi_1$.
\myfig{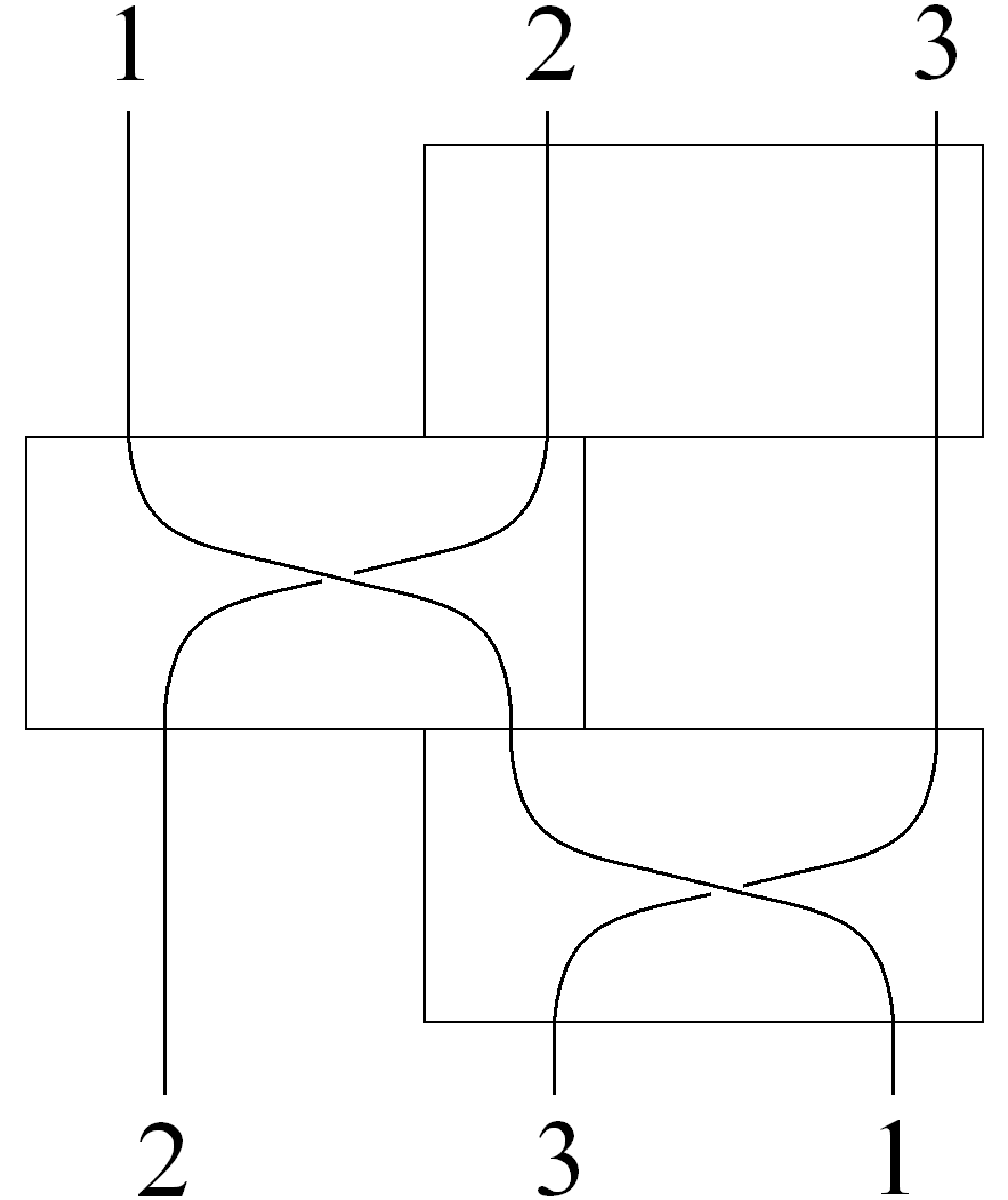}{4.0}{A second strand diagram that can be used in the computation of $\chi_1$.}

Next, we consider the computation of
$$ \chi_2=\Tr_{\young({\,}{\,}{1},{\,}{2},{3})}\Big(\Gamma_{\yng(3,2,1)}\big((6,4)\big)\Big).$$
This example is interesting as more than one path contributes. Writing $(6,4)=(4,5)(6,5)(4,5)$ we obtain the strand diagrams shown
in Figure \ref{fig:strand3.ps}.
The product of factors for the diagram on the left is
$${1\over c_1-c_3}\left[ 1-{1\over (c_2-c_3)^2}\right]. $$
The product of factors for the diagram on the right is
$${1\over c_1-c_2} {1\over (c_2-c_3)^2}. $$
Thus,
\begin{eqnarray}
\chi_2&=&\left({1\over c_1-c_3}\left[ 1-{1\over (c_2-c_3)^2}\right]+{1\over c_1-c_2} {1\over (c_2-c_3)^2}\right){\rm dim}_{\yng(2,1)}
\nonumber\\
&=&2\left({1\over c_1-c_3}\left[ 1-{1\over (c_2-c_3)^2}\right]+{1\over c_1-c_2} {1\over (c_2-c_3)^2}\right).
\nonumber
\end{eqnarray}
The reader can check that the same value for $\chi_2$ is obtained by decomposing $(6,4)=(6,5)(4,5)(6,5)$.
\myfig{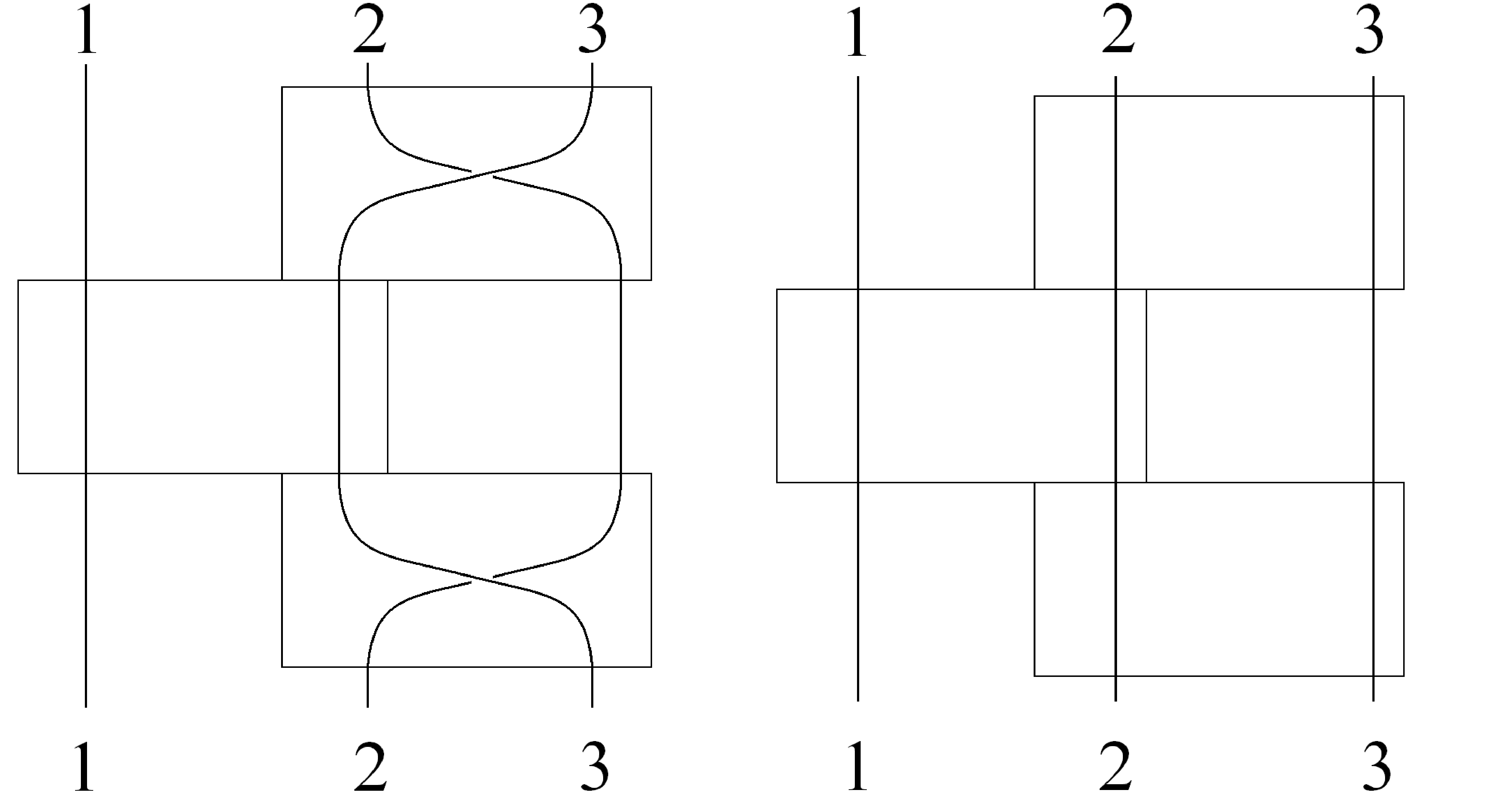}{9}{The strand diagrams used in the computation of $\chi_2$.}

Finally, consider
$$ \chi_3=\Tr_{\young({\,}{\,}{1},{\,}{2},{3})}\Big(\Gamma_{\yng(3,2,1)}\big( 1 \big)\Big).$$
Since we consider the identity element, the strand diagram has no boxes and hence $\chi_3={\rm dim}_{\tiny \yng(2,1)}=2$.
Since $(4,5)(4,5)=1$ we could also have written
$$ \chi_3=\Tr_{\young({\,}{\,}{1},{\,}{2},{3})}\Big(\Gamma_{\yng(3,2,1)}\big( (4,5)(4,5) \big)\Big).$$
In this case there are two strand diagrams given in Figure \ref{fig:strand4.ps}. Summing the contributions from these two
strand diagrams we obtain
$$ \chi_3={1\over (c_2-c_3)^2}{\rm dim}_{\tiny \yng(2,1)}+\left(1-{1\over (c_2-c_3)^2}\right){\rm dim}_{\tiny \yng(2,1)}
={\rm dim}_{\tiny \yng(2,1)}=2.$$
Once again, the two ways of writing the restricted character give the same result.
\myfig{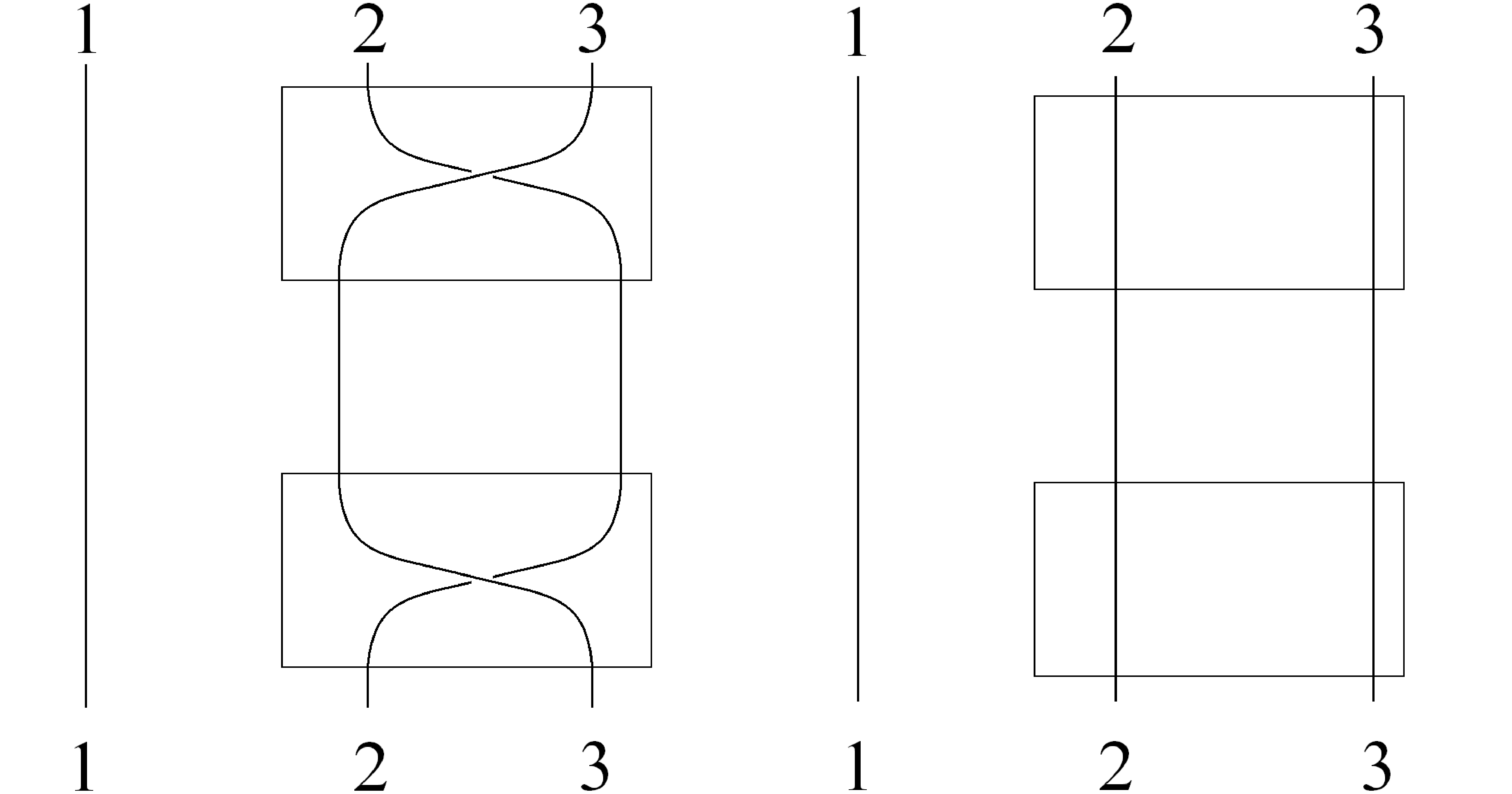}{9}{The strand diagrams used in the computation of $\chi_3$.}
Note that the trace
$$ \chi_3=\Tr_{\young({\,}{\,}{\onetwo},{\,}{\twoone},{3})}\Big(\Gamma_{\yng(3,2,1)}\big( 1 \big)\Big),$$
clearly vanishes because we are tracing the identity over an off the diagonal block. This is reflected graphically by the
fact that there is no strand diagram that can be drawn - the order of strands at the top of the diagram does not match the order
of strands at the bottom of the diagram and since we consider the identity element, the strand diagram has no boxes.

\subsection{Tests of the Restricted Character Results}

By summing well chosen restricted characters, one can recover the characters of $S_n$ which are known. This provides a number of tests that
our restricted character formulas pass. As an example, consider the computation of $\chi_R\left( (6,7)\right)$ for
$$ R=\yng(6,1). $$
From the character tables for $S_7$ we find $\chi_R\left( (6,7)\right)=4$. In terms of restricted characters
$$\chi_R\left( (6,7)\right)=\chi_{\young({\,}{\,}{\,}{\,}{2}{1},{\,})}\left( (6,7)\right)+
\chi_{\young({\,}{\,}{\,}{\,}{\,}{1},{2})}\left( (6,7)\right)+
\chi_{\young({\,}{\,}{\,}{\,}{\,}{2},{1})}\left( (6,7)\right).$$
Using the algorithm given above, it is straight forward to verify that
$$ \chi_{\young({\,}{\,}{\,}{\,}{2}{1},{\,})}\left( (6,7)\right)={\rm dim}_{\young({\,}{\,}{\,}{\,},{\,})}=4,$$
$$\chi_{\young({\,}{\,}{\,}{\,}{\,}{1},{2})}\left( (6,7)\right)={1\over 6},\quad
\chi_{\young({\,}{\,}{\,}{\,}{\,}{2},{1})}\left( (6,7)\right)=-{1\over 6},$$
which do indeed sum to give 4. The reader is invited to check some more examples herself.

As a further check of our methods, we have computed the restricted characters $\Tr_{R_1,R_2}\left(\Gamma_{R}\big[\sigma\big]\right)$
numerically. This was done by explicitly constructing the matrices $\Gamma_{R}\big[\sigma\big]$. Each representation used was
obtained by induction. One induces a reducible representation; the irreducible representation that participates was isolated using
projection operators built from the Casimir obtained by summing over all two cycles. See appendix B.2 of \cite{de Mello Koch:2007uu}
for more details. The resulting irreducible representations were tested by verifying the multiplication table of $S_n$. The
intertwiners were computed using the projection operators of \cite{de Mello Koch:2007uu} and the results of Appendix A; the
normalization of the intertwiner was computed numerically.

\subsection{Representations of $S_n$ from Strand Diagrams}

Using Strand diagrams, it is possible to write down the irreducible matrix representations of $S_n$. We will treat the simplest
nontrivial example of $S_3$. First consider the $\yng(2,1)$ irrep. Start by numbering the boxes in the Young diagram labeling the irrep, with
an ordering in which the boxes are to be removed, so that one is left with a legal Young diagram after each box is removed. These
labeled Young diagrams are in one-to-one correspondence with the matrix indices of the matrices in the irrep. For our example,
$$ i=1, \leftrightarrow \young({3}{1},{2})\qquad\qquad\qquad i=2, \leftrightarrow \young({3}{2},{1}).$$
Each matrix element of $\Gamma_{\tiny \yng(2,1)}\left((12)\right)$ is given by a single strand diagram
$$
\left[\Gamma_{\tiny \yng(2,1)}\left((12)\right)\right]_{11}=\Tr_{\young({3}{1},{2})}\left( (12)\right)={1\over c_1-c_2}={1\over 2},
$$
$$
\left[\Gamma_{\tiny \yng(2,1)}\left((12)\right)\right]_{12}=\Tr_{\young({3}{\onetwo},{\twoone})}\left( (12)\right)=
\sqrt{1-{1\over (c_1-c_2)^2}}={\sqrt{3}\over 2},
$$
$$
\left[\Gamma_{\tiny \yng(2,1)}\left((12)\right)\right]_{21}=\Tr_{\young({3}{\twoone},{\onetwo})}\left( (12)\right)=
\sqrt{1-{1\over (c_1-c_2)^2}}={\sqrt{3}\over 2},
$$
and
$$
\left[\Gamma_{\tiny \yng(2,1)}\left((12)\right)\right]_{22}=\Tr_{\young({3}{2},{1})}\left( (12)\right)={1\over c_1-c_2}=-{1\over 2},
$$
so that
$$
\Gamma_{\tiny \yng(2,1)}\left((12)\right)=\left[
\matrix{{1\over 2} &{\sqrt{3}\over 2}\cr {\sqrt{3}\over 2} &-{1\over 2}}\right].
$$
In exactly the same way we obtain
$$
\Gamma_{\tiny \yng(2,1)}\left((23)\right)=\left[
\matrix{-1 &0\cr 0 &1}\right].
$$
These two elements can now be used to generate the complete irrep.

Next consider $\yng(3)$. There is only one valid labeling $\young({3}{2}{1})$, so that the representation is one dimensional.
It is straight forward to obtain
$$ \Tr_{\young({3}{2}{1})}\left( (12)\right)={1\over c_1 -c_2}=1,\qquad
\Tr_{\young({3}{2}{1})}\left( (23)\right)={1\over c_2 -c_3}=1,$$
which are the correct results. Finally, consider $\yng(1,1,1)$. Again, there is only one valid labeling so
that the representation is again one dimensional. We find
$$ \Tr_{\young({3},{2},{1})}\left( (12)\right)={1\over c_1 -c_2}=-1,\qquad
\Tr_{\young({3},{2},{1})}\left( (23)\right)={1\over c_2 -c_3}=-1,$$
which are again the correct results.

\section{Hopping Identity}

In this appendix, we derive identities that can be used to obtain the Cuntz chain Hamiltonian that accounts for the $O(g_{YM}^2)$ correction
to the anomalous dimension of our operators. To construct the ``hop off'' process, we use the fact that whenever a $Z$ field hops past the
borders of the open string word $W$, the resulting restricted Schur polynomial decomposes into a sum of two types of systems, one is a giant
with a closed string and another is a string-giant system where the giant is now bigger. In the large $N$ limit only the second type needs to
be considered. The identities we derive in this appendix express this decomposition.
The irreps which play a role in the derivation of the identities are illustrated in Figure
\ref{fig:irreps.ps}. The basic structure of the derivation of these identities is very similar. For this reason, we explicitly derive an
identity in the next subsection and simply state the remaining identities. In contrast to the case of a single string
attached\cite{de Mello Koch:2007uv}, here it does make a difference if the first or last sites of the string participate in the hopping. The
identities needed in these two cases are listed separately. We have performed extensive numerical checks of the identities, which we describe
next. Finally, we explain how to express the leading large $N$ form of the identities, in terms of states of the Cuntz chain.
\myfig{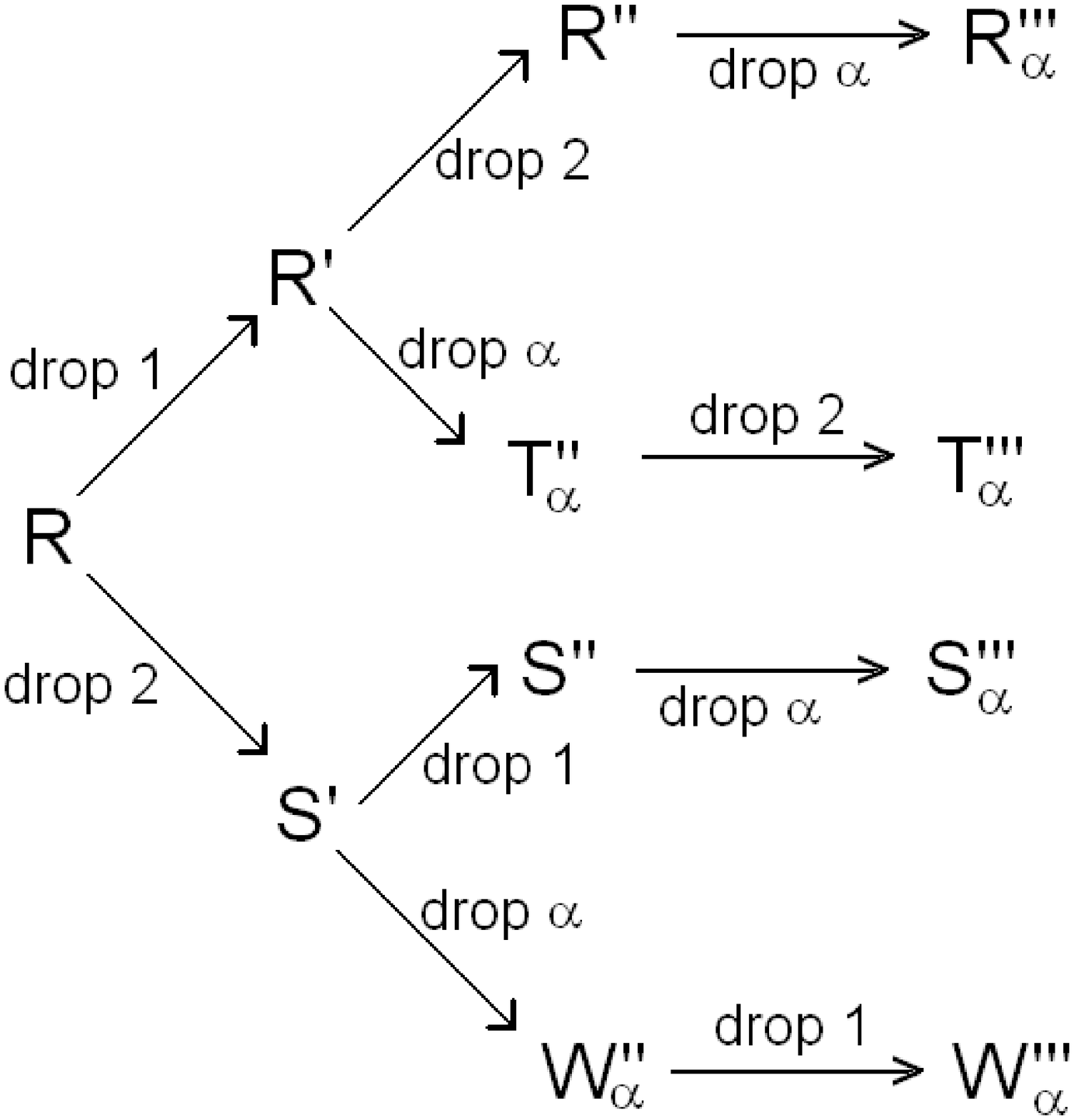}{12}{This figure shows the irreps that are used in the hopping identities. Starting from $R$, the figure shows which irrep
is obtained when boxes in $R$ are dropped.}

\subsection{Derivation of a Hopping Identity}

Our starting point is the restricted Schur polynomial
$$\chi^{(2)}_{R,R''}\Big|_1\Big|_2={1\over (n-2)!}\sum_{\sigma\in S_n}\Tr_{R''}\left(\Gamma_R(\sigma )\right)Z^{i_1}_{i_{\sigma (1)}}\cdots
Z^{i_{n-2}}_{i_{\sigma (n-2)}}(W^{(2)})^{i_{n-1}}_{i_{\sigma (n-1)}}(W^{(1)})^{i_n}_{i_{\sigma (n)}}.$$
There are two labeled boxes in $R$; dropping box 1 gives irrep $R'$; dropping box 2 gives irrep $R''$. Since $R'$ is an irrep of the
$S_{n-1}$ subgroup ${\cal G}_1=\{\sigma\in S_n|\sigma(n)=n\}$, we say that the open string described by the word $W^{(1)}$ is associated
to box 1. Since $R''$ is an irrep of the $S_{n-2}$ subgroup ${\cal G}_2=\{\sigma\in {\cal G}_1 |\sigma(n-1)=n-1\}$, we say that the open
string described by the word $W^{(2)}$ is associated with box 2. Notice that, in the chain of subductions used to define the restricted
Schur polynomial, the box associated with $W^{(1)}$ is dropped before the box associated to $W^{(2)}$. We have indicated this with the
notation $\Big|_1\Big|_2$. Rewrite the sum over $S_n$ as a sum over ${\cal G}_1$ and its cosets
\begin{eqnarray}
\chi^{(2)}_{R,R''}(Z,W^{(1)},W^{(2)})\Big|_1\Big|_2 &=& {1\over (n-2)!}\sum_{\sigma\in {\cal G}_1}\left[
\Tr_{R''}\left(\Gamma_{R'}(\sigma )\right)Z^{i_1}_{i_{\sigma (1)}}\cdots
Z^{i_{n-2}}_{i_{\sigma (n-2)}}(W^{(2)})^{i_{n-1}}_{i_{\sigma (n-1)}}\Tr (W^{(1)})\right.\nonumber\\
&+&\Tr_{R''}\left(\Gamma_{R}((1,n)\sigma )\right)(W^{(1)}Z)^{i_1}_{i_{\sigma (1)}}\cdots
Z^{i_{n-2}}_{i_{\sigma (n-2)}}(W^{(2)})^{i_{n-1}}_{i_{\sigma (n-1)}}+\cdots+\nonumber\\
&+&\Tr_{R''}\left(\Gamma_{R}((n-2,n)\sigma )\right)Z^{i_1}_{i_{\sigma (1)}}\cdots
(W^{(1)}Z)^{i_{n-2}}_{i_{\sigma (n-2)}}(W^{(2)})^{i_{n-1}}_{i_{\sigma (n-1)}}+\nonumber\\
&+&\left. \Tr_{R''}\left(\Gamma_{R}((n-1,n)\sigma )\right)Z^{i_1}_{i_{\sigma (1)}}\cdots
Z^{i_{n-2}}_{i_{\sigma (n-2)}}((W^{(1)}W^{(2)})^{i_{n-1}}_{i_{\sigma (n-1)}}\right].\nonumber
\end{eqnarray}
The first term on the right hand side is
$${1\over (n-2)!}\sum_{\sigma\in {\cal G}_1}
\Tr_{R''}\left(\Gamma_{R'}(\sigma )\right)Z^{i_1}_{i_{\sigma (1)}}\cdots
Z^{i_{n-2}}_{i_{\sigma (n-2)}}(W^{(2)})^{i_{n-1}}_{i_{\sigma (n-1)}}\Tr (W^{(1)})= \chi^{(1)}_{R',R''}(Z,W^{(2)})\Tr (W^{(1)}).$$
Using the methods of appendix B, we know that
$$\Tr_{R''}\left(\Gamma_{R}((n-1,n)\sigma )\right)={1\over c_1-c_2}\Tr_{R''}\left(\Gamma_{R'}(\sigma )\right),$$
so that the last term on the right hand side is
$${1\over (n-2)!}\sum_{\sigma\in {\cal G}_1}
\Tr_{R''}\left(\Gamma_{R}((n,n-1)\sigma )\right)Z^{i_1}_{i_{\sigma (1)}}\cdots
Z^{i_{n-2}}_{i_{\sigma (n-2)}}(W^{(1)}W^{(2)})^{i_{n-1}}_{i_{\sigma (n-1)}}$$
$$={1\over c_1-c_2}\chi^{(1)}_{R',R''}(Z,W^{(1)}W^{(2)}).$$
Focus on the remaining terms on the right hand side. Each of these terms makes the same contribution. We need to evaluate
$$\Tr_{R''}\left(\Gamma_{R}((j,n)\sigma \right)=\sum_{i=1}^{{\rm dim}_{R''}}
\langle i,12|\Gamma_{R}((j,n))\Gamma_{R'}(\sigma )|i,12\rangle .$$
Using the techniques of appendix B, it is straight forward to show that (the sum on $\alpha$ in the next equation is a sum
over all boxes that can be removed from $R''$ to leave a valid Young diagram; the relevant $S_{n-3}$ subgroup is given by
$\{\sigma\in {\cal G}_2|\sigma(j)=j\}$)
\begin{eqnarray}
\Tr_{R''}\left(\Gamma_{R}((j,n)\sigma \right)&=&\sum_\alpha\sum_{i,k=1}^{{\rm dim}_{R_\alpha'''}}
\langle i,12\alpha |\Gamma_R((j,n))|k,12\alpha\rangle\langle k,12\alpha |\Gamma_{R'} (\sigma )|i,12\alpha\rangle\nonumber\\
&+&\sum_\alpha\sum_{i,k=1}^{{\rm dim}_{R_\alpha'''}}
\langle i,12\alpha |\Gamma_R((j,n))|k,1\alpha 2\rangle\langle k,1\alpha 2|\Gamma_{R'} (\sigma )|i,12\alpha\rangle\nonumber\\
&=&\sum_\alpha {1\over c_1-c_\alpha}\left[1+{1\over (c_1-c_2)(c_2-c_\alpha)}\right]\Tr_{R_\alpha'''}(\Gamma_{R'}(\sigma ))\nonumber\\
&+&\sum_\alpha {1\over c_1-c_2} {1\over c_1-c_\alpha}\sqrt{1-{1\over (c_2-c_\alpha)^2}}
\Tr_{T_\alpha''',R_\alpha'''}(\Gamma_{R'}(\sigma )).\nonumber
\end{eqnarray}
Thus, summing the remaining $n-2$ terms we obtain
\begin{eqnarray}
&\sum_\alpha& {1\over c_1-c_\alpha}\left[1+{1\over (c_1-c_2)(c_2-c_\alpha)}\right]
\chi^{(2)}_{R',R_\alpha'''}(Z,W^{(1)}Z,W^{(2)})\Big|_2\Big|_1\nonumber\\
+&\sum_\alpha& {1\over c_1-c_2} {1\over c_1-c_\alpha}\sqrt{1-{1\over (c_2-c_\alpha)^2}}
\chi^{(2)}_{R'\to T_\alpha''',R_\alpha'''}(Z,W^{(1)}Z,W^{(2)})\Big|_2\Big|_1 .\nonumber
\end{eqnarray}
A straight forward application of the subgroup swap rule gives
$$ \chi^{(2)}_{R',R_\alpha'''}(Z,W^{(1)}Z,W^{(2)})\Big|_2\Big|_1 =
\left[\left(1-{1\over (c_2-c_\alpha)^2}\right)\chi^{(2)}_{R',T_\alpha'''}(Z,W^{(1)}Z,W^{(2)})\right.
$$
$$
+{1\over (c_2-c_\alpha)^2}\chi^{(2)}_{R',R_\alpha'''}(Z,W^{(1)}Z,W^{(2)})
+\sqrt{1-{1\over (c_2-c_\alpha)^2}}{1\over c_2-c_\alpha}\left(
\chi^{(2)}_{R'\to R_\alpha''' T_\alpha'''}(Z,W^{(1)}Z,W^{(2)})\right.
$$
$$
\left.\left. +\chi^{(2)}_{R'\to T_\alpha''' R_\alpha'''}(Z,W^{(1)}Z,W^{(2)})\right)\right]\Big|_1\Big|_2 ,
$$
$$ \chi^{(2)}_{R'\to T_\alpha''' R_\alpha'''}(Z,W^{(1)}Z,W^{(2)})\Big|_2\Big|_1 =
\left[\left(1-{1\over (c_2-c_\alpha)^2}\right)\chi^{(2)}_{R'\to R_\alpha''' T_\alpha'''}(Z,W^{(1)}Z,W^{(2)})\right.
$$
$$
-{1\over (c_2-c_\alpha)^2}\chi^{(2)}_{R'\to T_\alpha''' R_\alpha'''}(Z,W^{(1)}Z,W^{(2)})
+\sqrt{1-{1\over (c_2-c_\alpha)^2}}{1\over c_2-c_\alpha}\left(
\chi^{(2)}_{R', R_\alpha'''}(Z,W^{(1)}Z,W^{(2)})\right.
$$
$$
\left.\left. -\chi^{(2)}_{R', T_\alpha'''}(Z,W^{(1)}Z,W^{(2)})\right)\right]\Big|_1\Big|_2 .
$$
Thus, we finally obtain
\begin{eqnarray}
\chi^{(2)}_{R,R''}(Z,W^{(1)},W^{(2)})\Big|_1\Big|_2 &=&
\chi^{(1)}_{R',R''}(Z,W^{(2)})\Tr (W^{(1)})+{1\over c_1-c_2}\chi^{(1)}_{R',R''}(Z,W^{(1)}W^{(2)})\nonumber\\
&+&\sum_\alpha\left[{1\over c_1-c_\alpha}\left(1-{1\over (c_2-c_\alpha)^2}\right)\chi^{(2)}_{R',T_\alpha'''}(Z,W^{(1)}Z,W^{(2)})
\right.\nonumber\\
&+&{1\over c_1-c_2}{1\over (c_2-c_\alpha)^2}\chi^{(2)}_{R',R_\alpha'''}(Z,W^{(1)}Z,W^{(2)})\label{mikesexample}\\
&+&{1\over c_1-c_2}{1\over c_2-c_\alpha}\sqrt{1-{1\over (c_2-c_\alpha)^2}}
\chi^{(2)}_{R'\to R_\alpha'''T_\alpha'''}(Z,W^{(1)}Z,W^{(2)})\nonumber\\
&+&\left. {1\over c_1-c_\alpha}{1\over c_2-c_\alpha}\sqrt{1-{1\over (c_2-c_\alpha)^2}}
\chi^{(2)}_{R'\to T_\alpha'''R_\alpha'''}(Z,W^{(1)}Z,W^{(2)})\right]\Big|_1\Big|_2\nonumber \, .
\end{eqnarray}

The above identity is relevant for interactions in which the impurity hops out of the last site of the string. For the
hopping interaction in which the impurity hops out of the first site of the string, the right hand side of our identity should
be written in terms of $ZW^{(1)}$. This identity is easily derived by rewriting the sum over $S_n$ in terms of right
cosets of ${\cal G}_1$ instead of left cosets as we have done above.

The identity derived above is relevant for the description of interactions in which string 1 exchanges momentum with the
branes in the boundstate. To derive identities that allow string 2 to exchange momentum with the branes in the boundstate,
we first use the subgroup swap rule to swap strings 1 and 2. We then rewrite the sum over $S_n$ in terms of a sum over
$S_{n-1}$ and its cosets and then employ character identities as above. We give a complete set of identities in the next
two subsections.

On first inspection, our identity (\ref{mikesexample}) may appear intimidating. For this reason, we conclude this
section with a concrete example of the use of our identity.

\vspace{3mm}
\noindent
Consider for example $\chi^{(2)}_{R,R''} = $ {\small $ \young({\,}{\,}{1},{\,}{2},{\,})$}. In \ref{mikesexample} the sum on $\alpha$ now yields one term for $\chi^{(2)}_{R',T_\alpha'''}(Z,W^{(1)}Z,W^{(2)})$, two terms for $\chi^{(2)}_{R',R_\alpha'''}(Z,W^{(1)}Z,W^{(2)})$ and one term for $\chi^{(2)}_{R'\to R_\alpha'''T_\alpha'''}(Z,W^{(1)}Z,W^{(2)})$ and $\chi^{(2)}_{R'\to T_\alpha'''R_\alpha'''}(Z,W^{(1)}Z,W^{(2)})$. Explicitly:

\begin{eqnarray*}
    \chi^{(2)}_{R',T_\alpha'''}(Z,W^{(1)}Z,W^{(2)}) & = & \mbox{{\small $\young({\,}{\,},{\,}{2},{{\oneplus}})$}}, \\
    \chi^{(2)}_{R',R_\alpha'''}(Z,W^{(1)}Z,W^{(2)}) & = & \mbox{{\small $\young({\,}{\,},{\,}{{\oneplus}},{2})$, $\young({\,}{2},{\,}{{\oneplus}},{\,})$ }},\\
    \chi^{(2)}_{R'\to R_\alpha'''T_\alpha'''}(Z,W^{(1)}Z,W^{(2)}) & = & \mbox{{\small $\young({\,}{\,},{\,}{{\oneplustwo}},{\twooneplus})$}},\\
    \chi^{(2)}_{R'\to T_\alpha'''R_\alpha'''}(Z,W^{(1)}Z,W^{(2)}) & = & \mbox{{\small $\young({\,}{\,},{\,}{\twooneplus},{\oneplustwo})$}},
\end{eqnarray*}

\noindent
Now, $c_{1} = \mbox{N}+2$, $c_{2} = \mbox{N}$ and $c_{\alpha}$ (for a particular term in the identity) is equal to the weight of the labelled box in the restricted Schur polynomial of that term that does not correspond to either of the labelled boxes in the original Schur polynomial. The identity \ref{mikesexample} therefore becomes:

\begin{eqnarray*}
\mbox{{\small $ \young({\,}{\,}{1},{\,}{2},{\,})$}}\Big|_1\Big|_2 &=&
\young({\,}{\,},{\,}{2},{\,}) \; \Tr (W^{(1)})+\frac{1}{2} \; \mbox{{\small $ \young({\,}{\,},{\,}{{\w12}},{\,})$}} 
+\frac{3}{16} \; \mbox{{\small $\young({\,}{\,},{\,}{2},{{\oneplus}})$}} \Big|_{1^{+}}\Big|_2 \nonumber\\ \\
&+&\frac{1}{8} \; \mbox{{\small $\young({\,}{\,},{\,}{{\oneplus}},{2})$}} \Big|_{1^{+}}\Big|_2
+ \frac{1}{2} \; \mbox{{\small $\young({\,}{2},{\,}{{\oneplus}},{\,})$ }} \Big|_{1^{+}}\Big|_2
+ \frac{\sqrt{3}}{16} \; \mbox{{\small $\young({\,}{\,},{\,}{\twooneplus},{\oneplustwo})$}} \Big|_{1^{+}}\Big|_2
+ \frac{\sqrt{3}}{8} \; \mbox{{\small $\young({\,}{\,},{\,}{{\oneplustwo}},{\twooneplus})$}}  \Big|_{1^{+}}\Big|_2. 
\end{eqnarray*}

\subsection{Identities Relevant to Hopping off the first site of the string}

\begin{eqnarray}
\chi^{(2)}_{R,R''}(&Z&,W^{(1)},W^{(2)})\Big|_1\Big|_2 =
\chi^{(1)}_{R',R''}(Z,W^{(2)})\Tr (W^{(1)})+{1\over c_1-c_2}\chi^{(1)}_{R',R''}(Z,W^{(2)}W^{(1)})\nonumber\\
&+&\sum_\alpha\left[{1\over c_1-c_\alpha}\left(1-{1\over (c_2-c_\alpha)^2}\right)\chi^{(2)}_{R',T_\alpha'''}(Z,ZW^{(1)},W^{(2)})
\right.\nonumber\\
&+&{1\over c_1-c_2}{1\over (c_2-c_\alpha)^2}\chi^{(2)}_{R',R_\alpha'''}(Z,ZW^{(1)},W^{(2)})\label{firstone}\\
&+&{1\over c_1-c_2}{1\over c_2-c_\alpha}\sqrt{1-{1\over (c_2-c_\alpha)^2}}
\chi^{(2)}_{R'\to T_\alpha'''R_\alpha'''}(Z,ZW^{(1)},W^{(2)})\nonumber\\
&+&\left. {1\over c_1-c_\alpha}{1\over c_2-c_\alpha}\sqrt{1-{1\over (c_2-c_\alpha)^2}}
\chi^{(2)}_{R'\to R_\alpha'''T_\alpha'''}(Z,ZW^{(1)},W^{(2)})\right]\Big|_1\Big|_2\nonumber
\end{eqnarray}
The form of this identity is rather intuitive. The first term on the right hand side contributes to the process in which the bound state
emits string 1; the second term describes the process in which the two open strings join to form one long open string. In both of these
processes, the box which string 1 occupied on the left hand side does not appear on the right hand side. These two processes
will not contribute to our Cuntz chain Hamiltonian; they are relevant for the description of interactions which change the number of
open strings attached to the boundstate and do not contribute at the leading order of the large $N$ expansion.

It is instructive to consider the form of this identity for well separated branes. For well separated branes, we have $|c_1-c_2|\gg 1$.
For $|c_1-c_\alpha |\sim 1$, $|c_2-c_\alpha |\gg 1$ so that of the last four terms only the first one contributes, giving
$\approx {1\over c_1-c_\alpha}\chi^{(2)}_{R',T_\alpha'''}(Z,ZW^{(1)},W^{(2)}) .$
Thus, string 2 stays in box 2 and string 1 is close to where it started. Note that dropping terms of order $(c_1-c_2)^{-1}$ or
$(c_\alpha-c_2)^{-1}$ we obtain
$$\chi^{(2)}_{R,R''}(Z,W^{(1)},W^{(2)})\Big|_1\Big|_2 \approx \chi^{(1)}_{R',R''}(Z,W^{(2)})\Tr (W^{(1)})+\sum_\alpha
{1\over c_1-c_\alpha}\chi^{(2)}_{R',T_\alpha'''}(Z,ZW^{(1)},W^{(2)}),$$
which is the identity of \cite{de Mello Koch:2007uv}.

Next, consider the stretched string identities
\begin{eqnarray}
\chi^{(2)}_{R\to R'' S''}(&Z&,W^{(1)},W^{(2)})\Big|_1\Big|_2 =
\sqrt{1-{1\over (c_1-c_2)^2}}\chi^{(1)}_{R',R''}(Z,W^{(2)}W^{(1)})\nonumber\\
&+&\sum_\alpha\left[{1\over c_1-c_\alpha}
{1\over c_2-c_\alpha}\sqrt{1-{1\over (c_1-c_2)^2}}
\chi^{(2)}_{R',R_\alpha'''}(Z,ZW^{(1)},W^{(2)})\right.\label{firstwo}\\
&+&\left. {1\over c_1-c_\alpha}\sqrt{1-{1\over (c_2-c_\alpha)^2}}\sqrt{1-{1\over (c_1-c_2)^2}}
\chi^{(2)}_{R'\to T_\alpha'''R_\alpha'''}(Z,ZW^{(1)},W^{(2)})\right]\Big|_1\Big|_2\nonumber
\end{eqnarray}

\begin{eqnarray}
\chi^{(2)}_{R\to S'' R''}(&Z&,W^{(1)},W^{(2)})\Big|_1\Big|_2 =
\sqrt{1-{1\over (c_1-c_2)^2}}\chi^{(1)}_{S',S''}(Z,W^{(2)}W^{(1)})\nonumber\\
&+&\sum_\alpha\left[{1\over c_1-c_\alpha}
{1\over c_2-c_\alpha}\sqrt{1-{1\over (c_1-c_2)^2}}
\chi^{(2)}_{S',S_\alpha'''}(Z,ZW^{(1)},W^{(2)})\right.\label{firstthree}\\
&+&\left. {1\over c_2-c_\alpha}\sqrt{1-{1\over (c_1-c_\alpha)^2}}\sqrt{1-{1\over (c_1-c_2)^2}}
\chi^{(2)}_{S'\to W_\alpha'''S_\alpha'''}(Z,ZW^{(1)},W^{(2)})\right]\Big|_1\Big|_2\nonumber
\end{eqnarray}
Notice that in contrast to (\ref{firstone}), (\ref{firstwo}) and (\ref{firstthree}) do not have a term on the right hand
side corresponding to emission of string $1$. This is what we would expect for an operator dual to a state with two strings
stretching between branes, since if string 1 is emitted, it leaves a state with string 2 stretched between branes;
this state is not allowed as it violates the Gauss Law. The process in which the two open strings join at their endpoints
is allowed. In this process, it is the box with the upper 1 label that is removed. Thus, we can identify the Chan-Paton label
for the side of the string defining the first lattice site of the Cuntz chain with the upper label for the string, in our
diagrammatic notation. This corresponds to the first label of the restricted Schur polynomial. We will see further evidence
for this interpretation when we interpret the final form of the Hamiltonian.

If we again consider the limit of two well separated branes, we find that (\ref{firstwo}) becomes
$$
\chi^{(2)}_{R\to R'' S''}(Z,W^{(1)},W^{(2)})\Big|_1\Big|_2 \approx
\chi^{(1)}_{R',R''}(Z,W^{(2)} W^{(1)})
+\sum_\alpha {1\over c_1-c_\alpha}
\chi^{(2)}_{R'\to T_\alpha'''R_\alpha'''}(Z,ZW^{(1)},W^{(2)})\Big|_1\Big|_2 .
$$
In this case, the box with upper 1 label and lower 2 label moves from box 1 to box $\alpha$ (which are close to each other in
the Young diagram) and box with upper 2 label and lower 1 label stays where it is.

The first three identities that we have discussed corresponded to an interaction in which an impurity from the first site of string 1
interacts with the brane. The next three identities that we discuss correspond to an interaction in which an impurity from the first
site of string 2 interacts with the brane. The first three terms of the identity
\begin{eqnarray}
\chi^{(2)}_{R,R''}(&Z&,W^{(1)},W^{(2)})\Big|_1\Big|_2 =
\left(1-{1\over (c_1-c_2)^2}\right)\chi^{(1)}_{S',S''}(Z,W^{(1)})\Tr (W^{(2)})
\nonumber\\
&+&{1\over (c_1-c_2)^2}\chi^{(1)}_{R',R''}(Z,W^{(1)})\Tr(W^{(2)})
+{1\over c_1-c_2}\chi^{(1)}_{R',R''}(Z,W^{(1)}W^{(2)})\nonumber\\
&+&\sum_\alpha\left[{1\over c_2-c_\alpha}\left(1-{1\over (c_1-c_2)^2}\right)\chi^{(2)}_{S',S_\alpha'''}(Z,W^{(1)},ZW^{(2)})
\right.\nonumber\\
&+&{1\over c_2-c_\alpha}{1\over (c_1-c_2)^2}\chi^{(2)}_{R',R_\alpha'''}(Z,W^{(1)},ZW^{(2)})\label{firstfour}\\
&+&\left. {1\over c_1-c_2}{1\over c_1-c_\alpha}\sqrt{1-{1\over (c_2-c_\alpha)^2}}
\chi^{(2)}_{R'\to R_\alpha''' T_\alpha'''}(Z,W^{(1)},ZW^{(2)}) \right]\Big|_1\Big|_2\nonumber
\end{eqnarray}
change the number of open strings attached to the boundstate. The first two terms correspond to gravitational radiation; for both of
these terms, string 2 is emitted as a closed string. The third term corresponds to a process in which the two open strings join to give
a single open string. The order of the open string words in this term is not the same as the order in the corresponding term of
(\ref{firstone}). The term above is natural because it is the first site of string 2 that is interacting; the order in (\ref{firstone})
also looks natural because in that case it is the first site of string 1 that is interacting. Notice that the above identity is rather
different to (\ref{firstone}). Physically this is surprising - since in both cases it is the first site of the string interacting, these
identities should presumably look identical. This mismatch between the two identities is a consequence of the fact that we have treated
string 1 and string 2 differently when constructing the operator. See section 3 for further discussion of this point.

If we again consider the limit of two well separated branes, we find that (\ref{firstfour}) becomes
(take $|c_1-c_2|\gg 1$, $|c_1-c_\alpha|\gg 1$ and $|c_2-c_\alpha|\sim 1$)
$$\chi^{(2)}_{R,R''}(Z,W^{(1)},W^{(2)})\Big|_1\Big|_2\approx \chi^{(1)}_{S',S''}(Z,W^{(1)})\Tr(W^{(2)})
+\sum_\alpha {1\over c_2-c_\alpha}\chi^{(2)}_{S',S_\alpha'''}(Z,W^{(1)},ZW^{(2)}).$$
This again reproduces the identity of \cite{de Mello Koch:2007uv}. Thus, the content of the formula for well separated branes matches
the corresponding limit of (\ref{firstone}). This is satisfying, because in this limit the order in which the strings are attached does
not matter. This follows because the swap factor of \cite{de Mello Koch:2007uv} behaves as $|c_1-c_2|^{-1}$.

The remaining two identities are stretched string identities. In contrast to what we found above, there are terms corresponding to gravitational
radiation in these identities. We interpret this as a signal that there is some mixing between the operators we have defined (which as explained
above, made some arbitrary choices) to get to a ``physical basis''. See section 3 for more details. The first term in both identities
\begin{eqnarray}
\chi^{(2)}_{R\to R'' S''}(&Z&,W^{(1)},W^{(2)})\Big|_1\Big|_2 =
\sqrt{1-{1\over (c_1-c_2)^2}}\chi^{(1)}_{S',S''}(Z,W^{(1)}W^{(2)})
\nonumber\\
&+&{1\over c_1-c_2}\sqrt{1-{1\over (c_1-c_2)^2}}\left(
\chi^{(1)}_{R',R''}(Z,W^{(1)})-\chi^{(1)}_{S',S''}(Z,W^{(1)})\right)\Tr(W^{(2)})\nonumber\\
&+&\sum_\alpha\left[{1\over c_1-c_\alpha}{1\over c_2-c_1}\sqrt{1-{1\over (c_1-c_2)^2}}\chi^{(2)}_{S',S_\alpha'''}(Z,W^{(1)},ZW^{(2)})
\right.\label{firstfive}\\
&+&{1\over c_2-c_\alpha}\sqrt{1-{1\over (c_1-c_2)^2}}
\sqrt{1-{1\over (c_\alpha-c_1)^2}}\chi^{(2)}_{S'\to S_\alpha''' W_\alpha'''}(Z,W^{(1)},ZW^{(2)})\nonumber\\
&+&\left. {1\over c_1-c_2}{1\over c_1-c_\alpha}\sqrt{1-{1\over (c_1-c_2)^2}}
\chi^{(2)}_{R',R_\alpha'''}(Z,W^{(1)},ZW^{(2)}) \right]\Big|_1\Big|_2,\nonumber
\end{eqnarray}

\begin{eqnarray}
\chi^{(2)}_{R\to S'' R''}(&Z&,W^{(1)},W^{(2)})\Big|_1\Big|_2 =
\sqrt{1-{1\over (c_1-c_2)^2}}\chi^{(1)}_{R',R''}(Z,W^{(1)}W^{(2)})
\nonumber\\
&+&{1\over c_1-c_2}\sqrt{1-{1\over (c_1-c_2)^2}}\left(
\chi^{(1)}_{R',R''}(Z,W^{(1)})-\chi^{(1)}_{S',S''}(Z,W^{(1)})\right)\Tr(W^{(2)})\nonumber\\
&+&\sum_\alpha\left[{1\over c_2-c_\alpha}{1\over c_1-c_2}\sqrt{1-{1\over (c_1-c_2)^2}}\chi^{(2)}_{R',R_\alpha'''}(Z,W^{(1)},ZW^{(2)})
\right.\label{firstsix}\\
&+&{1\over c_1-c_\alpha}\sqrt{1-{1\over (c_1-c_2)^2}}
\sqrt{1-{1\over (c_\alpha-c_2)^2}}\chi^{(2)}_{R'\to R_\alpha''' T_\alpha'''}(Z,W^{(1)},ZW^{(2)})\nonumber\\
&+&\left. {1\over c_2-c_1}{1\over c_2-c_\alpha}\sqrt{1-{1\over (c_1-c_2)^2}}
\chi^{(2)}_{S',S_\alpha'''}(Z,W^{(1)},ZW^{(2)}) \right]\Big|_1\Big|_2,\nonumber
\end{eqnarray}
corresponds to two open strings joining to form one long open string. The order of the open string words in these terms again looks
natural given that it is the first site of string 2 that is interacting. They will again not contribute in the leading order of the
large $N$ expansion. It is satisfying that the content of the large distance limit of (\ref{firstfive})
$$\chi^{(2)}_{R\to R'' S''}(Z,W^{(1)},W^{(2)})\Big|_1\Big|_2 \approx
 \chi^{(1)}_{S',S''}(Z,W^{(1)}W^{(2)})+\sum_\alpha{1\over c_2-c_\alpha}\chi^{(2)}_{S'\to S_\alpha''' W_\alpha'''}(Z,W^{(1)},ZW^{(2)}),$$
is in complete agreement with the large distance limit of (\ref{firstwo}).

\subsection{Identities Relevant to Hopping off the last site of the string}

In this subsection, impurities hop between the last site of the strings and the threebrane. There are again six
possible identities that we could consider. The first three identities describe an interaction between the last
site of string 1 and the threebrane. The first identity
\begin{eqnarray}
\chi^{(2)}_{R,R''}(&Z&,W^{(1)},W^{(2)})\Big|_1\Big|_2 =
\chi^{(1)}_{R',R''}(Z,W^{(2)})\Tr (W^{(1)})+{1\over c_1-c_2}\chi^{(1)}_{R',R''}(Z,W^{(1)}W^{(2)})\nonumber\\
&+&\sum_\alpha\left[{1\over c_1-c_\alpha}\left(1-{1\over (c_2-c_\alpha)^2}\right)\chi^{(2)}_{R',T_\alpha'''}(Z,W^{(1)}Z,W^{(2)})
\right.\nonumber\\
&+&{1\over c_1-c_2}{1\over (c_2-c_\alpha)^2}\chi^{(2)}_{R',R_\alpha'''}(Z,W^{(1)}Z,W^{(2)})\label{lastone}\\
&+&{1\over c_1-c_2}{1\over c_2-c_\alpha}\sqrt{1-{1\over (c_2-c_\alpha)^2}}
\chi^{(2)}_{R'\to R_\alpha'''T_\alpha'''}(Z,W^{(1)}Z,W^{(2)})\nonumber\\
&+&\left. {1\over c_1-c_\alpha}{1\over c_2-c_\alpha}\sqrt{1-{1\over (c_2-c_\alpha)^2}}
\chi^{(2)}_{R'\to T_\alpha'''R_\alpha'''}(Z,W^{(1)}Z,W^{(2)})\right]\Big|_1\Big|_2\nonumber
\end{eqnarray}
can be obtained from (\ref{firstone}) by (i) swapping the labels on the twisted string states on the right hand side
and (ii) swapping the order
of the open string words in the second term on the right hand side. This is exactly what we would expect - it is now the
last site of the string that is interacting; to swap the first and last sites, we must swap Chan-Paton indices i.e. we
must swap the labels on the twisted string states. The discussion of this identity now parallels the discussion of (\ref{firstone})
and is not repeated.

Consider next the stretched string identities
\begin{eqnarray}
\chi^{(2)}_{R\to S'' R''}(&Z&,W^{(1)},W^{(2)})\Big|_1\Big|_2 =
\sqrt{1-{1\over (c_1-c_2)^2}}\chi^{(1)}_{R',R''}(Z,W^{(1)}W^{(2)})\nonumber\\
&+&\sum_\alpha\left[{1\over c_1-c_\alpha}
{1\over c_2-c_\alpha}\sqrt{1-{1\over (c_1-c_2)^2}}
\chi^{(2)}_{R',R_\alpha'''}(Z,W^{(1)}Z,W^{(2)})\right.\label{lasttwo}\\
&+&\left. {1\over c_1-c_\alpha}\sqrt{1-{1\over (c_2-c_\alpha)^2}}\sqrt{1-{1\over (c_1-c_2)^2}}
\chi^{(2)}_{R'\to R_\alpha''' T_\alpha'''}(Z,W^{(1)}Z,W^{(2)})\right]\Big|_1\Big|_2,\nonumber
\end{eqnarray}

\begin{eqnarray}
\chi^{(2)}_{R\to R'' S''}(&Z&,W^{(1)},W^{(2)})\Big|_1\Big|_2 =
\sqrt{1-{1\over (c_1-c_2)^2}}\chi^{(1)}_{S',S''}(Z,W^{(1)}W^{(2)})\nonumber\\
&+&\sum_\alpha\left[{1\over c_1-c_\alpha}
{1\over c_2-c_\alpha}\sqrt{1-{1\over (c_1-c_2)^2}}
\chi^{(2)}_{S',S_\alpha'''}(Z,W^{(1)}Z,W^{(2)})\right.\label{lastthree}\\
&+&\left. {1\over c_2-c_\alpha}\sqrt{1-{1\over (c_1-c_\alpha)^2}}\sqrt{1-{1\over (c_1-c_2)^2}}
\chi^{(2)}_{S'\to S_\alpha''' W_\alpha'''}(Z,W^{(1)}Z,W^{(2)})\right]\Big|_1\Big|_2 .\nonumber
\end{eqnarray}
It is satisfying that
identity (\ref{lasttwo}) can be obtained from (\ref{firstwo}) and (\ref{lastthree}) from (\ref{firstthree}) by
swapping the labels for stretched string states on both sides, and reversing the order of the open string words
in the first term on the right hand side. The discussion of these identities now parallel the discussion of
(\ref{firstwo}) and (\ref{firstthree}) and is not repeated.

The remaining three identities describe an interaction between the last site of string 2 and the threebrane. The identity
\begin{eqnarray}
\chi^{(2)}_{R,R''}(&Z&,W^{(1)},W^{(2)})\Big|_1\Big|_2 =
\left(1-{1\over (c_1-c_2)^2}\right)\chi^{(1)}_{S',S''}(Z,W^{(1)})\Tr (W^{(2)})
\nonumber\\
&+&{1\over (c_1-c_2)^2}\chi^{(1)}_{R',R''}(Z,W^{(1)})\Tr(W^{(2)})
+{1\over c_1-c_2}\chi^{(1)}_{R',R''}(Z,W^{(2)}W^{(1)})\nonumber\\
&+&\sum_\alpha\left[{1\over c_2-c_\alpha}\left(1-{1\over (c_1-c_2)^2}\right)\chi^{(2)}_{S',S_\alpha'''}(Z,W^{(1)},W^{(2)}Z)
\right.\nonumber\\
&+&{1\over c_2-c_\alpha}{1\over (c_1-c_2)^2}\chi^{(2)}_{R',R_\alpha'''}(Z,W^{(1)},W^{(2)}Z)\label{lastfour}\\
&+&\left. {1\over c_1-c_2}{1\over c_1-c_\alpha}\sqrt{1-{1\over (c_2-c_\alpha)^2}}
\chi^{(2)}_{R'\to T_\alpha'''R_\alpha'''}(Z,W^{(1)},W^{(2)}Z) \right]\Big|_1\Big|_2\nonumber
\end{eqnarray}
can be obtained from (\ref{firstfour}) by (i) swapping the labels on the twisted string states on the right hand side
and (ii) swapping the order of the open string words in the second term on the right hand side. Finally, the stretched string
identities
\begin{eqnarray}
\chi^{(2)}_{R\to R'' S''}(&Z&,W^{(1)},W^{(2)})\Big|_1\Big|_2 =
\sqrt{1-{1\over (c_1-c_2)^2}}\chi^{(1)}_{R',R''}(Z,W^{(2)}W^{(1)})
\nonumber\\
&+&{1\over c_1-c_2}\sqrt{1-{1\over (c_1-c_2)^2}}\left(
\chi^{(1)}_{R',R''}(Z,W^{(1)})-\chi^{(1)}_{S',S''}(Z,W^{(1)})\right)\Tr(W^{(2)})\nonumber\\
&+&\sum_\alpha\left[{1\over c_2-c_\alpha}{1\over c_1-c_2}\sqrt{1-{1\over (c_1-c_2)^2}}\chi^{(2)}_{R',R_\alpha'''}(Z,W^{(1)},W^{(2)}Z)
\right.\label{lastfive}\\
&+&{1\over c_1-c_\alpha}\sqrt{1-{1\over (c_1-c_2)^2}}
\sqrt{1-{1\over (c_\alpha-c_2)^2}}\chi^{(2)}_{R'\to T_\alpha''' R_\alpha'''}(Z,W^{(1)},W^{(2)}Z)\nonumber\\
&-&\left. {1\over c_1-c_2}{1\over c_2-c_\alpha}\sqrt{1-{1\over (c_1-c_2)^2}}
\chi^{(2)}_{S',S_\alpha'''}(Z,W^{(1)},W^{(2)}Z) \right]\Big|_1\Big|_2\nonumber
\end{eqnarray}

\begin{eqnarray}
\chi^{(2)}_{R\to S'' R''}(&Z&,W^{(1)},W^{(2)})\Big|_1\Big|_2 =
\sqrt{1-{1\over (c_1-c_2)^2}}\chi^{(1)}_{S',S''}(Z,W^{(2)}W^{(1)})
\nonumber\\
&-&{1\over c_1-c_2}\sqrt{1-{1\over (c_1-c_2)^2}}\left(
\chi^{(1)}_{S',S''}(Z,W^{(1)})-\chi^{(1)}_{R',R''}(Z,W^{(1)})\right)\Tr(W^{(2)})\nonumber\\
&+&\sum_\alpha\left[{1\over c_1-c_\alpha}{1\over c_2-c_1}\sqrt{1-{1\over (c_1-c_2)^2}}\chi^{(2)}_{S',S_\alpha'''}(Z,W^{(1)},W^{(2)}Z)
\right.\label{lastsix}\\
&+&{1\over c_2-c_\alpha}\sqrt{1-{1\over (c_1-c_2)^2}}
\sqrt{1-{1\over (c_\alpha-c_1)^2}}\chi^{(2)}_{S'\to W_\alpha''' S_\alpha'''}(Z,W^{(1)},W^{(2)}Z)\nonumber\\
&+&\left. {1\over c_1-c_2}{1\over c_1-c_\alpha}\sqrt{1-{1\over (c_1-c_2)^2}}
\chi^{(2)}_{R',R_\alpha'''}(Z,W^{(1)},W^{(2)}Z) \right]\Big|_1\Big|_2\nonumber
\end{eqnarray}
can be obtained from (\ref{firstfour}) and (\ref{firstfive}) by swapping the labels for stretched string states on both sides, and
reversing the order of the open string words in the first term on the right hand side.

\subsection{Numerical Test}

An important result of this article are the identities presented in the previous two subsections, since they determine the hop
off interaction. The hop on interaction follows from the hop off interaction by Hermitian conjugation
and the kissing interaction by composing the hop on and the hop off interactions. Thus, the complete boundary
interaction and the corresponding back reaction on the brane are determined by these identities. For this reason,
we have tested the identities numerically. In this subsection we will explain the check we have performed.

Our formulas are identities between restricted Schur polynomials. They must hold if we evaluate them
for {\it any}\footnote{In particular, not necessarily Hermitian.} numerical value of the matrices $Z$ and $W$.
Our check entails evaluating our identities for randomly generated matrices $W^{(1)}$, $W^{(2)}$ and $Z$, to
check their validity. Evaluating a restricted Schur polynomial entails evaluating a restricted character as well
as a product of traces of a product of the matrices $W^{(1)}$, $W^{(2)}$ and $Z$.

The restricted character $\Tr_{R'',S''}\left(\Gamma_{R}\big[\sigma\big]\right)$ or
$\Tr_{R''}\left(\Gamma_{R}\big[\sigma\big]\right)$ was computed by explicitly
constructing the matrices $\Gamma_{R}\big[\sigma\big]$. Each representation used was obtained by induction.
One induces a reducible representation; the irreducible representation that participates was isolated using
projection operators built from the Casimir obtained by summing over all two cycles. See appendix B.2 of
\cite{de Mello Koch:2007uu} for more details. The resulting irreducible representations were tested by verifying the
multiplication table of $S_n$. The restricted trace is then evaluated with the help of a projection operator
or an intertwiner. The intertwiner was computed using the results of appendix A.

The trace $\Tr (\sigma Z^{\otimes n-1} W^{(1)}W^{(2)})=
Z^{i_1}_{i_{\sigma(1)}}Z^{i_2}_{i_{\sigma(2)}}\cdots
Z^{i_{n-2}}_{i_{\sigma(n-2)}}(W^{(2)})^{i_{n-1}}_{i_{\sigma(n-1)}} (W^{(1)})^{i_n}_{i_{\sigma(n)}}$ for any given $\sigma\in S_n$ is
easily expressed as a product of traces of powers of $Z$, $W^{(1)}$ and $W^{(2)}$.

In total we verified over 50 specific instances of our identities, which provides a significant check of each identity.

\subsection{Identities in terms of Cuntz Chain States}

The state-operator correspondence is available for any conformal field theory. Using this correspondence, we can trade our (local)
operators for a set of states. Concretely, this involves quantizing with respect to radial time. Considering a fixed ``radial time"
slice we obtain a round sphere. The states dual to the restricted Schur polynomial operators are the states of our Cuntz chain.
Thus, we need to rewrite the identities obtained in this appendix as statements in terms of the states of the Cuntz oscillator chain.
The states of the Cuntz oscillator chain are normalized. Normalized states correspond to operators whose two point function is normalized.
Using the technology of \cite{de Mello Koch:2007uu} it is a simple task to compute the free equal time correlators of the restricted
Schur polynomials. After making use of the free field correlators to write our identities in terms operators with unit two point functions,
we find that not all terms are of the same order in $N$. We drop all terms which are subleading in $N$. These
terms are naturally interpreted in terms of string splitting and joining processes, so that they will be important when interactions
that change the number of open strings are considered.

The discussion for all of the identities above is rather similar, so we will be content to discuss a specific example which illustrates
the general features. Consider the right hand side of (\ref{firstone}). From the equal time correlator (there are a total of $h_i$ fields
in open string word $W^{(i)}$; $f_R$ is the product of the weights of the Young diagram $R$; $d_R$ is the dimension of $R$ as an irrep
of the symmetric group; $n_R$ is the number of boxes in Young diagram $R$)
$$ \langle \chi^{(1)}_{R',R''}(Z,W^{(2)})\Tr (W^{(1)})\chi^{(1)}_{R',R''}(Z,W^{(2)})^\dagger\Tr (W^{(1)})^\dagger\rangle $$
\begin{equation}
=\left( {4\pi\lambda\over N}\right)^{h_1+h_2+n_{R''}}h_1 N^{h_1+h_2-1} n_{R''} f_{R'}{d_{R''}\over d_{R'}}
\label{twopoint}
\end{equation}
we know that the operator $\chi^{(1)}_{R',R''}(Z,W^{(2)})\Tr (W^{(1)})$ corresponds to the state (all Cuntz chain states are normalized to 1)
$$ \sqrt{\left( {4\pi\lambda\over N}\right)^{h_1+h_2+n_{R''}}h_1 N^{h_1+h_2-1} n_{R''} f_{R'}{d_{R''}\over d_{R'}}}
|R',R'',W^{(2)};W^{(1)}\rangle .$$
The result (\ref{twopoint}) is not exact. When computing $\langle \Tr (W^{(1)})\Tr (W^{(1)})^\dagger\rangle$ we have only summed the
leading planar contribution. When computing $\langle \chi^{(1)}_{R',R''}(Z,W^{(2)})\chi^{(1)}_{R',R''}(Z,W^{(2)})^\dagger\rangle $ we
have only kept the $F_0$ contribution in the language of \cite{de Mello Koch:2007uu}. We have also factorized
$\langle \chi^{(1)}_{R',R''}(Z,W^{(2)})\Tr (W^{(1)})\chi^{(1)}_{R',R''}(Z,W^{(2)})^\dagger\Tr (W^{(1)})^\dagger\rangle$ as
$\langle \chi^{(1)}_{R',R''}(Z,W^{(2)})\chi^{(1)}_{R',R''}(Z,W^{(2)})^\dagger\rangle$ $\times \langle \Tr (W^{(1)})\Tr (W^{(1)})^\dagger\rangle$
which is valid at large $N$. Similarly, (again we sum only the leading order at large $N$)
$$\langle\chi^{(1)}_{R',R''}(Z,W^{(2)}W^{(1)})\chi^{(1)}_{R',R''}(Z,W^{(2)}W^{(1)})^\dagger\rangle=
\left( {4\pi\lambda\over N}\right)^{h_1+h_2+n_{R''}} N^{h_1+h_2-1} n_{R''} f_{R'}{d_{R''}\over d_{R'}}
$$
implies that $\chi^{(1)}_{R',R''}(Z,W^{(2)}W^{(1)})$ corresponds to the state
$$ \sqrt{\left( {4\pi\lambda\over N}\right)^{h_1+h_2+n_{R''}} N^{h_1+h_2-1} n_{R''} f_{R'}{d_{R''}\over d_{R'}}} |R',R'',W^{(2)}W^{(1)})\rangle .$$
Finally, the correlators (again we sum only the leading order at large $N$)
$$ \langle\chi^{(2)}_{R',T_\alpha'''}(Z,ZW^{(1)},W^{(2)})\chi^{(2)}_{R',T_\alpha'''}(Z,ZW^{(1)},W^{(2)})^\dagger \rangle
= \left( {4\pi\lambda\over N}\right)^{h_1+h_2+1+n_{T_\alpha'''}} N^{h_1+h_2-1}n_{R'}^2 {d_{T_\alpha'''}\over d_{R'}}f_{R'},$$
$$ \langle\chi^{(2)}_{R',R_\alpha'''}(Z,ZW^{(1)},W^{(2)})\chi^{(2)}_{R',R_\alpha'''}(Z,ZW^{(1)},W^{(2)})^\dagger \rangle
= \left( {4\pi\lambda\over N}\right)^{h_1+h_2+1+n_{T_\alpha'''}} N^{h_1+h_2-1}n_{R'}^2 {d_{R_\alpha'''}\over d_{R'}}f_{R'},$$
$$ \langle\chi^{(2)}_{R'\to T_\alpha'''R_\alpha'''}(Z,ZW^{(1)},W^{(2)})\chi^{(2)}_{R'\to T_\alpha'''R_\alpha'''}(Z,ZW^{(1)},W^{(2)})^\dagger\rangle $$
$$ = \left( {4\pi\lambda\over N}\right)^{h_1+h_2+1+n_{T_\alpha'''}} N^{h_1+h_2-1}n_{R'}^2 {d_{T_\alpha'''}\over d_{R'}}f_{R'},$$
$$ \langle\chi^{(2)}_{R'\to R_\alpha'''T_\alpha'''}(Z,ZW^{(1)},W^{(2)})\chi^{(2)}_{R'\to R_\alpha'''T_\alpha'''}(Z,ZW^{(1)},W^{(2)})^\dagger\rangle $$
$$ = \left( {4\pi\lambda\over N}\right)^{h_1+h_2+1+n_{T_\alpha'''}} N^{h_1+h_2-1}n_{R'}^2 {d_{T_\alpha'''}\over d_{R'}}f_{R'}$$
imply the correspondences
$$ \chi^{(2)}_{R',T_\alpha'''}(Z,ZW^{(1)},W^{(2)}) \longleftrightarrow
\sqrt{\left( {4\pi\lambda\over N}\right)^{h_1+h_2+1+n_{T_\alpha'''}} N^{h_1+h_2-1}n_{R'}^2 {d_{T_\alpha'''}\over d_{R'}}f_{R'}}
|R',T_{\alpha'''},ZW^{(1)},W^{(2)}\rangle , $$
$$ \chi^{(2)}_{R',R_\alpha'''}(Z,ZW^{(1)},W^{(2)}) \longleftrightarrow
\sqrt{\left( {4\pi\lambda\over N}\right)^{h_1+h_2+1+n_{T_\alpha'''}} N^{h_1+h_2-1}n_{R'}^2 {d_{R_\alpha'''}\over d_{R'}}f_{R'}}
|R',R_\alpha''',ZW^{(1)},W^{(2)}\rangle , $$
$$ \chi^{(2)}_{R'\to T_\alpha'''R_\alpha'''}(Z,ZW^{(1)},W^{(2)}) \longleftrightarrow
\sqrt{\left( {4\pi\lambda\over N}\right)^{h_1+h_2+1+n_{T_\alpha'''}} N^{h_1+h_2-1}n_{R'}^2 {d_{T_\alpha'''}\over d_{R'}}f_{R'}}
|R',T_\alpha'''R_\alpha''',ZW^{(1)},W^{(2)}\rangle
$$
$$ \chi^{(2)}_{R'\to R_\alpha'''T_\alpha'''}(Z,ZW^{(1)},W^{(2)}) \longleftrightarrow
\sqrt{\left( {4\pi\lambda\over N}\right)^{h_1+h_2+1+n_{T_\alpha'''}} N^{h_1+h_2-1}n_{R'}^2 {d_{T_\alpha'''}\over d_{R'}}f_{R'}}
|R',R_\alpha'''T_\alpha''',ZW^{(1)},W^{(2)}\rangle
$$
Consider the factor
$$ n_{R'}^2 {d_{R_\alpha'''}\over d_{R'}}={({\rm hooks})_{R'}\over ({\rm hooks})_{R_\alpha'''}},$$
where (hooks)$_R$ is the product of the hook lengths of Young diagram $R$. It is straight forward to compute this ratio of
hook lengths, which is generically of order $N^2$ implying that ${d_{R_\alpha'''}\over d_{R'}}$ is of order 1. Using this
observation, it is equally easy to verify that ${d_{T_\alpha'''}\over d_{R'}}$ and ${d_{R''}\over d_{R'}}$ are also both $O(1)$.
Given these results, it is simple to see that the sum of operators
\begin{eqnarray}
&&\chi^{(1)}_{R',R''}(Z,W^{(2)})\Tr (W^{(1)})+{1\over c_1-c_2}\chi^{(1)}_{R',R''}(Z,W^{(2)}W^{(1)})\nonumber\\
&+&\sum_\alpha\left[{1\over c_1-c_\alpha}\left(1-{1\over (c_2-c_\alpha)^2}\right)\chi^{(2)}_{R',T_\alpha'''}(Z,ZW^{(1)},W^{(2)})
\right.\nonumber\\
&+&{1\over c_1-c_2}{1\over (c_2-c_\alpha)^2}\chi^{(2)}_{R',R_\alpha'''}(Z,ZW^{(1)},W^{(2)})\nonumber\\
&+&{1\over c_1-c_2}{1\over c_2-c_\alpha}\sqrt{1-{1\over (c_2-c_\alpha)^2}}
\chi^{(2)}_{R'\to T_\alpha'''R_\alpha'''}(Z,ZW^{(1)},W^{(2)})\nonumber\\
&+&\left. {1\over c_1-c_\alpha}{1\over c_2-c_\alpha}\sqrt{1-{1\over (c_2-c_\alpha)^2}}
\chi^{(2)}_{R'\to R_\alpha'''T_\alpha'''}(Z,ZW^{(1)},W^{(2)})\right]\Big|_1\Big|_2\nonumber
\end{eqnarray}
corresponds to the following sum of normalized states
\begin{eqnarray}
&&\sqrt{\left( {4\pi\lambda\over N}\right)^{h_1+h_2+n_{R''}} N^{h_1+h_2-1}n_{R'}^2 f_{R'}}
\left[
\sqrt{h_1 d_{R''}\over n_{R'}d_{R'}}|R',R'',W^{(2)};W^{(1)}\rangle\right.\nonumber\\
&+&{1\over c_1-c_2}\sqrt{d_{R''}\over n_{R'}d_{R'}}|R',R'',W^{(2)}W^{(1)}\rangle\nonumber\\
&+&\sum_\alpha\left[{1\over c_1-c_\alpha}\left(1-{1\over (c_2-c_\alpha)^2}\right)
\sqrt{d_{T_\alpha'''}\over d_{R'}}|R',T_\alpha''',ZW^{(1)},W^{(2)}\rangle
\right.\nonumber\\
&+&{1\over c_1-c_2}{1\over (c_2-c_\alpha)^2}
\sqrt{d_{R_\alpha'''}\over d_{R'}}|R',R_\alpha''',ZW^{(1)},W^{(2)}\rangle\nonumber\\
&+&{1\over c_1-c_2}{1\over c_2-c_\alpha}\sqrt{1-{1\over (c_2-c_\alpha)^2}}
\sqrt{d_{T_\alpha'''}\over d_{R'}}|R',T_\alpha'''R_\alpha''',ZW^{(1)},W^{(2)}\rangle\nonumber\\
&+&\left.\left. {1\over c_1-c_\alpha}{1\over c_2-c_\alpha}\sqrt{1-{1\over (c_2-c_\alpha)^2}}
\sqrt{d_{T_\alpha'''}\over d_{R'}}|R',R_\alpha''' T_\alpha''',ZW^{(1)},W^{(2)}\rangle\right]\right].\nonumber
\end{eqnarray}
Recalling that $h_1=O(\sqrt{N})$ and $n_{R'}=O(N)$, it is clear that the first two terms are subleading.
These two terms correspond to gravitational radiation (first term) and string joining (second
term); they are the only terms that correspond to an interaction that changes the number of open strings attached to the excited giant
system. Although we have illustrated things with an example, this conclusion is general - for all of the identities obtained
in this appendix, terms that do not correspond to two strings attached to the giant system can be dropped in the leading large $N$ limit.

\section{State/Operator Map}

In this section we will simply quote the six normalization factors that enter the relation between the restricted Schur polynomials
and the normalized Cuntz chain states relevant for the excited two giant graviton bound state\footnote{See the introduction for the restricted
Schur polynomials corresponding to these states.}. The normalization factors are not exact - we simply quote the leading large $N$
value of these normalizations. These factors are determined completely by the $F_0^{(1)}F_0^{(2)}$ contribution in the language of
\cite{de Mello Koch:2007uu}. The factor $f_R$ is the product of weights of the Young diagram $R$. The open string word $W^{(1)}$
contains a total number of $h_1$ Higgs fields; the open string word $W^{(2)}$ contains a total number of $h_2$ Higgs fields.
\begin{center}
\begin{tabular}{|c|c|}
\hline
State & Normalization \\
\hline
$|b_0-1,b_1,11,22\rangle$ & $\left({4\pi\lambda\over N}\right)^{2b_0+b_1+h_1+h_2-2\over 2}b_0\sqrt{f_R} \sqrt{N^{h_1+h_2-2}}$\\
\hline
$|b_0-1,b_1,22,11\rangle$ & $\left({4\pi\lambda\over N}\right)^{2b_0+b_1+h_1+h_2-2\over 2}b_0\sqrt{f_R} \sqrt{N^{h_1+h_2-2}}$\\
\hline
$|b_0-1,b_1,12,21\rangle$ & $\left({4\pi\lambda\over N}\right)^{2b_0+b_1+h_1+h_2-2\over 2}b_0\sqrt{f_R} \sqrt{N^{h_1+h_2-2}}$\\
\hline
$|b_0-1,b_1,21,12\rangle$ & $\left({4\pi\lambda\over N}\right)^{2b_0+b_1+h_1+h_2-2\over 2}b_0\sqrt{f_R} \sqrt{N^{h_1+h_2-2}}$\\
\hline
$|b_0-2,b_1+2,22,22\rangle$ & $\left({4\pi\lambda\over N}\right)^{2b_0+b_1+h_1+h_2-2\over 2}b_0\sqrt{f_R} \sqrt{N^{h_1+h_2-2}}\sqrt{b_1+3\over b_1+1}$\\
\hline
$|b_0,b_1-2,11,11\rangle$ & $\left({4\pi\lambda\over N}\right)^{2b_0+b_1+h_1+h_2-2\over 2}b_0\sqrt{f_R} \sqrt{N^{h_1+h_2-2}}\sqrt{b_1-1\over b_1+1}$\\
\hline
\end{tabular}
\end{center}

\section{Review of the Restricted Schur Polynomial Notation}

In this appendix, we review the definition of the restricted Schur polynomial. The reader requiring more details can consult
\cite{Balasubramanian:2004nb,de Mello Koch:2007uu,de Mello Koch:2007uv}.

There is by now convincing evidence that the dual of a giant graviton is a Schur polynomial.
Schur polynomials are labeled by Young diagrams.
Excitations of giant gravitons can be described by attaching open strings to the giant graviton.
Operators dual to excitations of giant gravitons are obtained by inserting words $(W^{(a)})^j_i$ describing the open strings
(one word for each open string) into the operator describing the system of giant gravitons
\begin{equation}
\chi_{R,R_1}^{(k)}(Z,W^{(1)},...,W^{(k)})={1\over (n-k)!}
\sum_{\sigma\in S_n}\Tr_{R_1}(\Gamma_R(\sigma))\Tr(\sigma Z^{\otimes n-k}W^{(1)}\cdots W^{(k)}),
\label{restrictedschur}
\end{equation}
$$\Tr (\sigma Z^{\otimes n-k}W^{(1)}\cdots W^{(k)})= Z^{i_1}_{i_{\sigma (1)}}Z^{i_2}_{i_{\sigma (2)}}\cdots
Z^{i_{n-k}}_{i_{\sigma (n-k)}}(W^{(1)})^{i_{n-k+1}}_{i_{\sigma (n-k+1)}}\cdots
(W^{(k)})^{i_{n}}_{i_{\sigma (n)}}.$$
The representation $R$ of the giant graviton system is a Young diagram with $n$ boxes, i.e. it is a representation of $S_n$.
$\Gamma_R(\sigma )$ is the matrix representing $\sigma$ in irreducible representation $R$ of the symmetric group $S_n$.
The representation $R_1$ is a Young diagram with $n-k$ boxes, i.e. it is a representation of $S_{n-k}$.
Imagine that the $k$ words above are all distinct, corresponding to the case that the open strings are distinguishable.
Consider an $S_{n-k}\otimes (S_1)^k$ subgroup of $S_n$. The representation $R$ of $S_n$ will subduce a (generically)
reducible representation of the $S_{n-k}\otimes (S_1)^k$ subgroup. One of the irreducible representations appearing in this
subduced representation is $R_1$. $\Tr_{R_1}$ is an instruction to trace only over the indices belonging to this irreducible
component. If the representation $R_1$ appears more than once, things are more interesting. The example discussed in
\cite{Balasubramanian:2004nb} illustrates this point nicely. Suppose $R\to R_1\oplus R_2\oplus R_2$ under restricting $S_n$ to
$S_{n-2}\times S_1\times S_1$. Choose a basis so that
$$\Gamma_R (\sigma )=\left[
\matrix{\Gamma_{R_1}(\sigma)_{i_1 j_1} &0 &0\cr 0 &\Gamma_{R_2}(\sigma)_{i_2 j_2} &0\cr 0 &0 &\Gamma_{R_2}(\sigma)_{i_3 j_3}}
\right],\qquad \forall \sigma\in S_{n-2}\times S_1\times S_1 ,$$
%
%
$$\Gamma_R (\sigma )=\left[
\matrix{A^{(1,1)}_{i_1 j_1} &A^{(1,2)}_{i_1 j_2} &A^{(1,3)}_{i_1 j_3}\cr
        A^{(2,1)}_{i_2 j_1} &A^{(2,2)}_{i_2 j_2} &A^{(2,3)}_{i_2 j_3}\cr
        A^{(3,1)}_{i_3 j_1} &A^{(3,2)}_{i_3 j_2} &A^{(3,3)}_{i_3 j_3}}
\right],\qquad \sigma\notin S_{n-2}\times S_1\times S_1 .$$
There are four suitable definitions for $\Tr_{R_2}(\Gamma_R(\sigma ))$: $\Tr (A^{(2,2)})$, $\Tr (A^{(2,3)})$, $\Tr (A^{(3,2)})$ or $\Tr (A^{(3,3)})$.
Interpret the operator obtained using $\Tr (A^{(2,3)})$ or $\Tr (A^{(3,2)})$ as dual to the system with the open
strings stretching between the giants and the operator obtained using $\Tr (A^{(2,2)})$ or $\Tr (A^{(3,3)})$ as dual to the system
with one open string on each giant. In general, identify the ``on the diagonal" blocks with states in which the
two open strings are each on a specific giant and the ``off the diagonal" blocks as states in which the open strings stretch between two
giants. As a consequence of the fact that the representation $R_2$ appears with a multiplicity two, there is no unique way to extract
two $R_2$ representations out of $R$. The specific representations obtained will depend on the details of the subgroups used in
performing the restriction. These subgroups are the set of elements of the permutation group that leave an index invariant,
$\sigma(i)=i$. Choosing the index to be the index of an open string, we can associate the subgroups participating with specific open strings.
The subgroups are specified by dropping boxes from $R$, so that we can now associate boxes in $R$ with specific open strings. This
leads to a convenient graphical notation which has been developed in \cite{de Mello Koch:2007uu,de Mello Koch:2007uv}.
There is an obvious generalization to the case that a representation $R_1$ appears $n$ times after restricting
to the subgroup.

If any of the strings are identical, one needs to decompose with respect to a larger subgroup and to pick a representation
for the strings which are indistinguishable. Thus, for example, if we consider a bound state of a giant system with three identical
strings attached, we would consider an $S_{n-3}\otimes S_3$ subgroup of $S_n$. The restricted Schur polynomial would be given by
$\chi^{(3)}_{R,R_1}$ with $R$ an irrep of $S_n$ and $R_1$ an irrep of $S_{n-3}\otimes S_3$. The $S_3$ subgroup would act by
permuting the indices of the three identical strings; the $S_{n-3}$ subgroup would act by permuting the indices of the $Z$s out of
which the giant is composed. Write $R_1=r_1\times r_2$ with $r_1$ are irrep of $S_{n-3}$ and
$r_2$ an irrep of $S_3$. As an example, if we take $R$ to be an irrep of $S_9$
$$ R=\yng(4,4,1),\qquad {\rm dim}_R= 84 $$
then we can have
$$R_1 = \yng(3,3)\otimes\yng(1,1,1),\quad {\rm dim}_{R_1}=5,\qquad R_1 = \yng(3,3)\otimes\yng(2,1),\quad {\rm dim}_{R_1}=10,$$
$$R_1 = \yng(4,2)\otimes\yng(3),\quad {\rm dim}_{R_1}=9,\qquad R_1 = \yng(4,2)\otimes\yng(2,1),\quad {\rm dim}_{R_1}=18,$$
$$R_1 = \yng(3,2,1)\otimes \yng(2,1),\qquad {\rm dim}_{R_1} = 32,$$
or
$$R_1=\yng(4,1,1)\otimes\yng(3),\qquad {\rm dim}_{R_1}=10.$$
By summing the dimensions of these representations, it is easy to see that we have indeed listed all of the representations
that are subduced by $R$.

The giant graviton system is dual to an operator containing a
product of order $N$ fields; the open strings are dual to an operator containing a product of order $\sqrt{N}$ fields.
We have in mind the case that $k$ is $O(1)$, $n$ is $O(N)$ and the words $(W^{(a)})^j_i$ are a product of $O(\sqrt{N})$
fields.

We call the operator (\ref{restrictedschur}) a {\it restricted} Schur polynomial of representation $R$ with representation $R_1$
for the restriction. We end this appendix with a summary of the graphical notation of \cite{de Mello Koch:2007uu}, which is used
heavily in this article. An operator dual to an excited giant graviton takes the form
$$
\chi_{R,R_1}^{(k)}(Z,W^{(1)},...,W^{(k)})
={1\over (n-k)!}\sum_{\sigma\in S_n}\Tr (\Pi \Gamma_R(\sigma))\Tr(\sigma Z^{\otimes n-k}W^{(1)}\cdots W^{(k)}),
$$
where $\Pi$ is a product of projection operators and/or intertwiners, used to implement the restricted trace. $\Pi$ is defined by
the sequence of irreducible representations used to subduce $R_1$ from $R$, as well as the chain of subgroups
to which these representations belong. Since the row and column indices of the block that we trace over
(denoted by $R_1$ in the above formula) need not coincide, we need to specify this data separately for both indices.
The graphical notation summarizes this information.

For the case that we have $k$ strings, we label the words describing the open strings $1,2,...,k$. Denote the chain of
subgroups involved in the reduction by ${\cal G}_k\subset {\cal G}_{k-1}\subset\cdots\subset {\cal G}_2\subset {\cal G}_1\subset S_n$.
${\cal G}_m$ is obtained by taking all elements $S_n$ that leave the indices of the strings $W^{(i)}$ with $i\le m$ inert.
To specify the sequence of irreducible representations employed in subducing $R_1$, place a pair of labels into each box,
a lower label and an upper label.
The representations needed to subduce the row label of $R_1$ are obtained by starting with $R$. The second representation
is obtained by dropping the box with upper label equal to 1; the third representation is obtained from the
second by dropping the box with upper label equal to 2 and so on until the box with label $k$ is dropped. The
representations needed to subduce the column label are obtained in exactly the same way except that instead of using the
upper label, we now use the lower label. For further details and explicit examples, we refer the reader to
\cite{de Mello Koch:2007uu}.

\section{Boundstate of three Sphere Giants}

In this appendix, we will compute the $+1\to 1$ interaction for two strings attached to a bound state of three sphere giants.
This example is interesting because, firstly, it does partially illustrate our claim that the methods we have developed
apply to any bound state of giants and secondly, in this situation, we expect an emergent $U(3)$ gauge theory. The three sphere
giant boundstate is described by a Young diagram with three columns. When labeling the open string endpoints we will use the
labels `b', `m' and `l' for the first column (`b' for big brane), second column (`m' for medium brane) and third column
(`l' for little brane) respectively. The relevant Cuntz chain states together with their normalizations are shown in
the table below.
\begin{center}
\begin{tabular}{|c|c|}
\hline
State & Normalization \\
\hline
$|b_0,b_1-1,b_2,bb,mm\rangle$ &
$\left({4\pi\lambda\over N}\right)^{3b_0+2b_1+b_2+h_1+h_2-2\over 2}b_0\sqrt{f_R} \sqrt{N^{h_1+h_2-2}}\sqrt{(b_1+b_2+1)b_1\over (b_1+b_2+2)(b_1+1)}$\\
\hline
$|b_0-1,b_1+1,b_2-1,bb,ll\rangle$ &
$\left({4\pi\lambda\over N}\right)^{3b_0+2b_1+b_2+h_1+h_2-2\over 2}b_0\sqrt{f_R} \sqrt{N^{h_1+h_2-2}}\sqrt{(b_1+2)b_2\over (b_2+1)(b_1+1)}$\\
\hline
$|b_0,b_1-1,b_2,mm,bb\rangle$ &
$\left({4\pi\lambda\over N}\right)^{3b_0+2b_1+b_2+h_1+h_2-2\over 2}b_0\sqrt{f_R} \sqrt{N^{h_1+h_2-2}}\sqrt{(b_1+b_2+1)b_1\over (b_1+b_2+2)(b_1+1)}$\\
\hline
$|b_0-1,b_1,b_2+1,mm,ll\rangle$ &
$\left({4\pi\lambda\over N}\right)^{3b_0+2b_1+b_2+h_1+h_2-2\over 2}b_0\sqrt{f_R} \sqrt{N^{h_1+h_2-2}}\sqrt{(b_2+2)(b_1+b_2+3)\over (b_2+1)(b_1+b_2+2)}$\\
\hline
$|b_0-1,b_1+1,b_2-1,ll,bb\rangle$ &
$\left({4\pi\lambda\over N}\right)^{3b_0+2b_1+b_2+h_1+h_2-2\over 2}b_0\sqrt{f_R} \sqrt{N^{h_1+h_2-2}}\sqrt{(b_1+2)b_2\over (b_2+1)(b_1+1)}$\\
\hline
$|b_0-1,b_1,b_2+1,ll,mm\rangle$ &
$\left({4\pi\lambda\over N}\right)^{3b_0+2b_1+b_2+h_1+h_2-2\over 2}b_0\sqrt{f_R} \sqrt{N^{h_1+h_2-2}}\sqrt{(b_2+2)(b_1+b_2+3)\over (b_2+1)(b_1+b_2+2)}$\\
\hline
$|b_0,b_1,b_2-2,bb,bb\rangle$ &
$\left({4\pi\lambda\over N}\right)^{3b_0+2b_1+b_2+h_1+h_2-2\over 2}b_0\sqrt{f_R} \sqrt{N^{h_1+h_2-2}}\sqrt{(b_2-1)(b_1+b_2)\over (b_2+1)(b_1+b_2+2)}$\\
\hline
$|b_0,b_1-2,b_2+2,mm,mm\rangle$ &
$\left({4\pi\lambda\over N}\right)^{3b_0+2b_1+b_2+h_1+h_2-2\over 2}b_0\sqrt{f_R} \sqrt{N^{h_1+h_2-2}}\sqrt{(b_2+3)(b_1-1)\over (b_2+1)(b_1+1)}$\\
\hline
$|b_0-2,b_1+2,b_2,ll,ll\rangle$ &
$\left({4\pi\lambda\over N}\right)^{3b_0+2b_1+b_2+h_1+h_2-2\over 2}b_0\sqrt{f_R} \sqrt{N^{h_1+h_2-2}}\sqrt{(b_1+3)(b_1+b_2+4)\over (b_1+1)(b_1+b_2+2)}$\\
\hline
$|b_0,b_1-1,b_2,bm,mb\rangle$ &
$\left({4\pi\lambda\over N}\right)^{3b_0+2b_1+b_2+h_1+h_2-2\over 2}b_0\sqrt{f_R} \sqrt{N^{h_1+h_2-2}}\sqrt{(b_1+b_2+1)b_1\over (b_1+b_2+2)(b_1+1)}$\\
\hline
$|b_0,b_1-1,b_2,mb,bm\rangle$ &
$\left({4\pi\lambda\over N}\right)^{3b_0+2b_1+b_2+h_1+h_2-2\over 2}b_0\sqrt{f_R} \sqrt{N^{h_1+h_2-2}}\sqrt{(b_1+b_2+1)b_1\over (b_1+b_2+2)(b_1+1)}$\\
\hline
$|b_0-1,b_1+1,b_2-1,bl,lb\rangle$ &
$\left({4\pi\lambda\over N}\right)^{3b_0+2b_1+b_2+h_1+h_2-2\over 2}b_0\sqrt{f_R} \sqrt{N^{h_1+h_2-2}}\sqrt{(b_1+2)b_2\over (b_2+1)(b_1+1)}$\\
\hline
$|b_0-1,b_1+1,b_2-1,lb,bl\rangle$ &
$\left({4\pi\lambda\over N}\right)^{3b_0+2b_1+b_2+h_1+h_2-2\over 2}b_0\sqrt{f_R} \sqrt{N^{h_1+h_2-2}}\sqrt{(b_1+2)b_2\over (b_2+1)(b_1+1)}$\\
\hline
$|b_0-1,b_1,b_2+1,ml,lm\rangle$ &
$\left({4\pi\lambda\over N}\right)^{3b_0+2b_1+b_2+h_1+h_2-2\over 2}b_0\sqrt{f_R} \sqrt{N^{h_1+h_2-2}}\sqrt{(b_2+2)(b_1+b_2+3)\over (b_2+1)(b_1+b_2+2)}$\\
\hline
$|b_0-1,b_1,b_2+1,lm,ml\rangle$ &
$\left({4\pi\lambda\over N}\right)^{3b_0+2b_1+b_2+h_1+h_2-2\over 2}b_0\sqrt{f_R} \sqrt{N^{h_1+h_2-2}}\sqrt{(b_2+2)(b_1+b_2+3)\over (b_2+1)(b_1+b_2+2)}$\\
\hline
\end{tabular}
\end{center}

The labels $b_0$, $b_1$ and $b_2$ again determine the momenta of the giants. The giant corresponding to the first column has a momentum
of $b_0+b_1+b_2$, the giant corresponding to the second column has a momentum of $b_0+b_1$ and the giant corresponding to the third column
has a momentum of $b_0$. We take $b_0$ to be $O(N)$ and $b_1,b_2$ to be $O(1)$.

To determine the boundary interactions, we start by rewriting the identities of Appendix C for the case that we have a Young
diagram with three columns. To obtain the boundary interaction terms in the Hamiltonian, these identities are then inverted and rewritten in
terms of normalized Cuntz chain states.

The term in the Hamiltonian describing the process in which a $Z$ hops out of the first site of string 1 is given by
$$
H_{+1\to 1}\left[
\matrix{
|b_0,b_1-1,b_2,bb,mm\rangle\cr
|b_0-1,b_1+1,b_2-1,bb,ll\rangle\cr
|b_0,b_1-1,b_2,mm,bb\rangle\cr
|b_0-1,b_1,b_2+1,mm,ll\rangle\cr
|b_0-1,b_1+1,b_2-1,ll,bb\rangle\cr
|b_0-1,b_1,b_2+1,ll,mm\rangle\cr
|b_0,b_1,b_2-2,bb,bb\rangle\cr
|b_0,b_1-2,b_2+2,mm,mm\rangle\cr
|b_0-2,b_1+2,b_2,ll,ll\rangle\cr
|b_0,b_1-1,b_2,bm,mb\rangle\cr
|b_0,b_1-1,b_2,mb,bm\rangle\cr
|b_0-1,b_1+1,b_2-1,bl,lb\rangle\cr
|b_0-1,b_1+1,b_2-1,lb,bl\rangle\cr
|b_0-1,b_1,b_2+1,ml,lm\rangle\cr
|b_0-1,b_1,b_2+1,lm,ml\rangle}
\right]=-\lambda\sqrt{1-{b_0\over N}}M
\left[
\matrix{
|b_0,b_1-1,b_2+1,bb,mm\rangle\cr
|b_0-1,b_1+1,b_2,bb,ll\rangle\cr
|b_0,b_1,b_2-1,mm,bb\rangle\cr
|b_0-1,b_1+1,b_2,mm,ll\rangle\cr
|b_0,b_1,b_2-1,ll,bb\rangle\cr
|b_0,b_1-1,b_2+1,ll,mm\rangle\cr
|b_0,b_1,b_2-1,bb,bb\rangle\cr
|b_0,b_1-1,b_2+1,mm,mm\rangle\cr
|b_0-1,b_1+1,b_2,ll,ll\rangle\cr
|b_0,b_1-1,b_2+1,bm,mb\rangle\cr
|b_0,b_1,b_2-1,mb,bm\rangle\cr
|b_0-1,b_1+1,b_2,bl,lb\rangle\cr
|b_0,b_1,b_2-1,lb,bl\rangle\cr
|b_0-1,b_1+1,b_2,ml,lm\rangle\cr
|b_0,b_1-1,b_2+1,lm,ml\rangle}
\right],$$
where the non-zero elements of $M$ are given by

{\small
$$ M_{1\, 1}=-(b_2)_1^2 (b_1+b_2)_2 ,\qquad M_{3\, 1}=-{(b_1+b_2)_2\over (b_2+1)^2 (b_2+2)},\qquad
M_{6\, 1}=-{\frac {\left (b_1+2\right )\sqrt {b_2+2}\sqrt {b_1}}{\sqrt {b_2+1}\left (b_1+1\right )^{3/2}\left (b_1+b_2+2\right )}},$$
$$ M_{4\, 1}={\frac {-b_1-b_2-3}{\sqrt {b_2+1}\left (b_1+1\right )^{3/2}\left (b_1+b_2+2\right )\sqrt {b_1}\sqrt {b_2+2}}},\qquad
M_{8\, 1}=-{\frac {\sqrt {b_1-1}\sqrt {b_2+3}\sqrt {b_1+b_2+3}}{\sqrt {b_2+1}\sqrt {b_1+b_2+2}\left (b_2+2\right )\sqrt {b_1}}},
$$
$$ M_{10\, 1}= -{(b_1+b_2)_2 (b_2)_1\over (b_2+1)(b_2 +2)},
\qquad
  M_{11\, 1}=-{(b_1+b_2)_2 (b_2)_1\over (b_2+1)},
\qquad
M_{14\, 1}=
{\frac {\sqrt {b_2+2}\sqrt {b_1+2}}{\left (b_1+1\right )^{3/2}\sqrt {b_2+1}\left (b_1+b_2+2\right )}},
$$
$$ M_{15\, 1}=
{\frac {\left (b_1+b_2+3\right )\sqrt {b_1+2}}{\left (b_1+1\right )^{3/2}\sqrt {b_2+1}\left (b_1+b_2+2\right )\sqrt {b_2+2}}},
\qquad M_{2\, 2}=-(b_1+b_2)_2^2 (b_2)_1,$$
$$
 M_{5\, 2}={(b_2)_1\over (b_1+b_2+2)^2 (b_1+b_2+3)},\qquad
 M_{4\, 2}=
-{\frac {b_1\,\sqrt {b_1+b_2+3}\sqrt {b_1+2}}{\left (b_1+1\right )^{3/2}\left (b_2+1\right )\sqrt {b_1+b_2+2}}},$$
$$ M_{6\, 2}=
{\frac {-b_2-2}{\left (b_2+1\right )\left (b_1+1\right )^{3/2}\sqrt {b_1+b_2+2}\sqrt {b_1+2}\sqrt {b_1+b_2+3}}},\qquad
M_{13\, 2}=-{(b_2)_1(b_1+b_2)_2\over (b_1+b_2+2)},$$
$$ M_{9\, 2}=
-{\frac {\sqrt {b_1+b_2+4}\sqrt {b_1+3}\sqrt {b_2+2}}{\sqrt {b_2+1}\sqrt {b_1+b_2+2}\sqrt {b_1+2}\left (b_1+b_2+3\right )}},
\qquad
M_{12\, 2}=-{(b_2)_1 (b_1+b_2)_2\over (b_1+b_2+2)(b_1+b_2+3)}$$
$$ M_{14\, 2}=
-{\frac {\left (b_2+2\right )\sqrt {b_1}}{\left (b_1+1\right )^{3/2}\left (b_2+1\right )\sqrt {b_1+b_2+2}\sqrt {b_1+b_2+3}}},
\qquad
M_{1\, 3}={(b_1)_1\over b_2 (b_2+1)^2},$$
$$ M_{15\, 2}=
-{\frac {\sqrt {b_1+b_2+3}\sqrt {b_1}}{\left (b_1+1\right )^{3/2}\left (b_2+1\right )\sqrt {b_1+b_2+2}}},
\qquad
M_{3\, 3}=-(b_2)_1^2 (b_1)_1 ,$$
$$ M_{2\, 3}=
{\frac {b_1+2}{\sqrt {b_2+1}\left (b_1+1\right )\left (b_1+b_2+2\right )^{3/2}\sqrt {b_1+b_2+1}\sqrt {b_2}}},
\qquad
M_{10\, 3}={(b_1)_1 (b_2)_1\over (b_2+1)},$$
$$ M_{5\, 3}=
-{\frac {\left (b_1+b_2+3\right )\sqrt {b_2}\sqrt {b_1+b_2+1}}{\sqrt {b_2+1}\left (b_1+1\right )\left (b_1+b_2+2\right )^{3/2}}},
\qquad
M_{11\, 3}={(b_1)_1 (b_2)_1\over b_2(b_2+1)},$$
$$ M_{7\, 3}=
{\frac {\sqrt {b_2 -1}\sqrt{b_1+b_2}\sqrt {b_1+2}}{\sqrt {b_2+1}\sqrt {b_1+1}b_2\,\sqrt {b_1+b_2+1}}},
\qquad
M_{12\, 3}=
{\frac {\sqrt {b_2}\sqrt {b_1+b_2+3}}{\left (b_1+b_2+2\right )^{3/2}\left (b_1+1\right )\sqrt {b_2+1}}},$$
$$ M_{13\, 3}=
-{\frac {\left (b_1+2\right )\sqrt {b_1+b_2+3}}{\left (b_1+b_2+2\right )^{3/2}\left (b_1+1\right )\sqrt {b_2+1}\sqrt {b_2}}},
\qquad
M_{2\, 4}=
{\frac {\sqrt {b_1+2}\left (b_1+b_2+1\right )\sqrt {b_1+b_2+3}}{\left (b_2+1\right )\sqrt {b_1+1}\left (b_1+b_2+2\right )^{3/2}}},$$
$$M_{4\, 4}=-(b_1)_1^2 (b_2)_1,
\qquad
M_{5\, 4}=
-{\frac {b_2}{\sqrt {b_1+1}\left (b_2+1\right )\left (b_1+b_2+2\right )^{3/2}\sqrt {b_1+b_2+3}\sqrt {b_1+2}}},$$
$$ M_{6\, 4}=
-{\frac {\sqrt {b_2+2}\sqrt {b_2}}{\left (b_1+1\right )^{2}\left (b_2+1\right )\left (b_1+2\right )}},
\qquad
M_{9\, 4}=
-{\frac {\sqrt {b_1+b_2+4}\sqrt {b_1+3}\sqrt {b_2}}{\sqrt{b_2+1}\sqrt {b_1+1}\left (b_1+2\right )\sqrt{b_1+b_2+3} }},$$
$$ M_{12\, 4}=
-{\frac {b_2\,\sqrt {b_1+b_2+1}}{\left (b_1+b_2+2\right )^{3/2}\left (b_2+1\right )\sqrt {b_1+1}\sqrt {b_1+2}}},
  \qquad
M_{13\, 4}=
{\frac {\sqrt {b_1+2}\sqrt {b_1+b_2+1}}{\left (b_1+b_2+2\right )^{3/2}\left (b_2+1\right )\sqrt {b_1+1}}},$$
$$M_{14\, 4}=-{(b_2)_1(b_1)_1\over (b_1+1)(b_1+2)},\qquad
M_{1\, 5}=
{\frac {b_1}{\left (b_2+1\right )^{3/2}\left (b_1+1\right )\sqrt {b_1+b_2+2}\sqrt {b_2}\sqrt {b_1+b_2+1}}},$$
$$ M_{15\, 4}=-{(b_2)_1(b_1)_1\over (b_1+1)},
\qquad
M_{2\, 5}={(b_1)_1\over (b_1+b_2+2)^2 (b_1+b_2+1)},$$
$$ M_{3\, 5}=
{\frac {\sqrt {b_1+b_2+1}\left (b_2+2\right )\sqrt {b_2}}{\left (b_2+1\right )^{3/2}\left (b_1+1\right )\sqrt {b_1+b_2+2}}},
\qquad
M_{5\, 5}=-(b_1)_1 (b_1+b_2)_2^2,$$
$$
M_{7\, 5}=
{\frac {\sqrt{b_2-1}\sqrt{b_1+b_2}\sqrt {b_1}}{\sqrt {b_1+1}\sqrt {b_1+b_2+2}\left (b_1+b_2+1\right )\sqrt {b_2}}},
\qquad
M_{10\, 5}=
-{\frac {\sqrt {b_1+b_2+1}\sqrt {b_2+2}}{\left (b_2+1\right )^{3/2}\left (b_1+1\right )\sqrt {b_1+b_2+2}}},
$$
$$ M_{11\, 5}=
-{\frac {b_1\,\sqrt {b_2+2}}{\left (b_2+1\right )^{3/2}\left (b_1+1\right )\sqrt {b_1+b_2+2}\sqrt {b_1+b_2+1}}},
\qquad
M_{12\, 5}={(b_1)_1(b_1+b_2)_2\over b_1+b_2+2},$$
$$ M_{13\, 5}=-{(b_1)_1(b_1+b_2)_2\over (b_1+b_2+2)(b_1+b_2+1)},
\qquad
M_{1\, 6}=
{\frac {\sqrt {b_1}b_2\,\sqrt {b_2+2}}{\sqrt {b_1+1}\left (b_2+1\right )^{3/2}\left (b_1+b_2+2\right )}},$$
$$M_{3\, 6}=
{\frac {b_1+b_2+1}{\sqrt {b_1+1}\left (b_2+1\right )^{3/2}\left (b_1+b_2+2\right )\sqrt {b_2+2}\sqrt {b_1}}},
\qquad
M_{4\, 6}={(b_1+b_2)_2\over b_1 (b_1+1)^2},$$
$$ M_{6\, 6}=-(b_1)_1^2(b_1+b_2)_2,\qquad
M_{8\, 6}=
{\frac {\sqrt {b_1-1}\sqrt {b_2+3}\sqrt {b_1+b_2+1}}{\sqrt {b_1+1}\sqrt{b_1+b_2+2}\sqrt{b_2+2}b_1}},$$
$$ M_{10\, 6}=
{\frac {\left (b_1+b_2+1\right )\sqrt {b_2}}{\left (b_2+1\right )^{3/2}\sqrt {b_1+1}\left (b_1+b_2+2\right )\sqrt {b_1}}},
  \qquad
M_{11\, 6}=
{\frac {\sqrt {b_1}\sqrt {b_2}}{\left (b_2+1\right )^{3/2}\sqrt {b_1+1}\left (b_1+b_2+2\right )}},$$
$$M_{14\, 6}={(b_1)_1 (b_1+b_2)_2\over b_1+1},
\qquad
M_{1\,7}=
-{\frac {\sqrt {b_1}\sqrt{b_2+2}\sqrt{b_1+b_2+3}}{\left (b_2+1\right )^{2}\sqrt {b_1+1}\sqrt {b_1+b_2+2}\sqrt {b_2}}},$$

$$M_{15\, 6}=-{(b_1)_1 (b_1+b_2)_2\over b_1(b_1+1)},
\qquad
M_{2\, 7}=
-{\frac {\sqrt {b_1+2}\sqrt{b_2+2}\sqrt{b_1+b_2+3}}{\sqrt {b_2+1}\sqrt {b_1+1}\left (b_1+b_2+2\right )^{2}\sqrt {b_1+b_2+1}}},$$
$$
M_{3\, 7}=
-{\frac {\sqrt {b_1}\sqrt{b_2+2}\sqrt{b_1+b_2+3}\sqrt {b_2}}{\sqrt {b_1+1}\left (b_2+1\right )^{2}\sqrt {b_1+b_2+2}}},
\qquad
M_{5\, 7}=
-{\frac {\sqrt {b_1+2}\sqrt{b_2+2}(b_1+b_2)_2}{\sqrt {b_1+1}\sqrt {b_2+1}\left (b_1+b_2+2\right )}},$$
$$ M_{7\, 7}=-{\sqrt{b_1+b_2+3}\sqrt{b_2+2}\sqrt{b_2-1}\sqrt{b_1+b_2}\over\sqrt{b_2}\sqrt{b_1+b_2+1}\sqrt{b_2+1}\sqrt{b_1+b_2+2}},
\qquad
M_{10\, 7}=
{\frac {\sqrt {b_1}\sqrt {b_1+b_2+3}}{\left (b_2+1\right )^{2}\sqrt {b_1+1}\sqrt {b_1+b_2+2}}},$$
$$ M_{11\, 7}=
{\frac {\sqrt {b_1}\left (b_2+2\right )\sqrt {b_1+b_2+3}}{\left (b_2+1\right )^{2}\sqrt {b_1+1}\sqrt {b_1+b_2+2}}},
  \qquad
M_{12\, 7}=
{\frac {\sqrt {b_1+2}\sqrt {b_2+2}}{\left (b_1+b_2+2\right )^{2}\sqrt {b_2+1}\sqrt {b_1+1}}},$$
$$ M_{13\, 7}=
{\frac {\sqrt {b_1+2}\sqrt {b_2+2}\left (b_1+b_2+3\right )}{\left (b_1+b_2+2\right )^{2}\sqrt {b_2+1}\sqrt {b_1+1}}},
\qquad
  M_{1\, 8}=
{\frac {\sqrt {b_1+2}\sqrt {b_1+b_2+1}(b_2)_1}{\left (b_2+1\right )\sqrt {b_1+1}\sqrt {b_1+b_2+2}}},$$
$$ M_{3\, 8}=
-{\frac {\sqrt {b_1+2}\sqrt {b_1+b_2+1}\sqrt {b_2}}{\left (b_2+1\right )^{2}\sqrt {b_1+1}\sqrt {b_1+b_2+2}\sqrt {b_2+2}}},
\qquad
M_{4\, 8}=
-{\frac {\sqrt {b_1+2}\sqrt {b_1+b_2+3}\sqrt {b_2}}{\left (b_1+1\right )^{2}\sqrt{b_2+1}\sqrt{b_1+b_2+2}\sqrt {b_1}}},$$
$$
M_{6\, 8}=
-{\frac {\sqrt {b_1+2}\sqrt {b_1+b_2+3}\sqrt {b_2}\sqrt {b_1}}{\sqrt{b_2+1}\left (b_1+1\right )^{2}\sqrt{b_1+b_2+2}}},
\qquad
M_{8\, 8}=
-{\frac {\sqrt {b_1-1}\sqrt {b_2+3}\sqrt {b_1+2}\sqrt {b_2}}{\sqrt {b_1+1}\sqrt {b_2+1}\sqrt {b_1}\sqrt {b_2+2}}},$$
$$ M_{10\, 8}=
-{\frac {\sqrt {b_1+2}\sqrt {b_1+b_2+1}b_2}{\left (b_2+1\right )^{2}\sqrt {b_1+1}\sqrt {b_1+b_2+2}}},
  \qquad
M_{11\, 8}=
{\frac {\sqrt {b_1+2}\sqrt {b_1+b_2+1}}{\left (b_2+1\right )^{2}\sqrt {b_1+1}\sqrt {b_1+b_2+2}}},$$
$$ M_{14\, 8}=
{\frac {\sqrt {b_1+b_2+3}\sqrt {b_2}}{\left (b_1+1\right )^{2}\sqrt{b_2+1}\sqrt{b_1+b_2+2}}},
\qquad
M_{15\, 8}=
{\frac {\left (b_1+2\right )\sqrt {b_1+b_2+3}\sqrt {b_2}}{\left (b_1+1\right )^{2}\sqrt{b_2+1}\sqrt{b_1+b_2+2}}},
$$
$$ M_{2\, 9}=
{\frac {\sqrt {b_1}\sqrt {b_2}(b_1+b_2)_2}{\sqrt {b_1+1}\sqrt {b_2+1}\left (b_1+b_2+2\right )}},
\qquad
M_{4\, 9}=
{\frac {\sqrt {b_1}\sqrt {b_1+b_2+1}\sqrt {b_2+2}\sqrt {b_1+2}}{\sqrt{b_1+b_2+2}\sqrt{b_2+1}\left (b_1+1\right )^{2}}}
$$
$$
M_{5\, 9}=
-{\frac {\sqrt {b_1}\sqrt {b_1+b_2+1}\sqrt {b_2}}{\sqrt {b_1+1}\sqrt {b_2+1}\left (b_1+b_2+2\right )^{2}\sqrt {b_1+b_2+3}}},
\qquad
M_{6\, 9}=
-{\frac {\sqrt {b_1}\sqrt {b_1+b_2+1}\sqrt {b_2+2}}{\sqrt{b_2+1}\left (b_1+1\right )^{2}\sqrt{b_1+b_2+2}\sqrt {b_1+2}}},
$$
$$ M_{9\, 9}=
-{\frac {\sqrt {b_1+b_2+4}\sqrt {b_1+3}\sqrt {b_1}\sqrt {b_1+b_2+1}}{\sqrt {b_1+1}\sqrt {b_1+b_2+2}\sqrt {b_1+b_2+3}\sqrt {b_1+2}}},
\qquad
M_{12\, 9}=
-{\frac {\sqrt {b_1}\left (b_1+b_2+1\right )\sqrt {b_2}}{\left (b_1+b_2+2\right )^{2}\sqrt {b_1+1}\sqrt {b_2+1}}},$$
$$ M_{13\, 9}=
{\frac {\sqrt {b_1}\sqrt {b_2}}{\left (b_1+b_2+2\right )^{2}\sqrt {b_1+1}\sqrt {b_2+1}}},
  \qquad
M_{14\, 9}=
-{\frac {b_1\,\sqrt {b_1+b_2+1}\sqrt {b_2+2}}{(b_1+1 )^{2}\sqrt{b_2+1}\sqrt{b_1+b_2+2}}},$$
$$ M_{15\, 9}=
{\frac {\sqrt {b_1+b_2+1}\sqrt {b_2+2}}{\left (b_1+1\right )^{2}\sqrt{b_2+1}\sqrt{b_1+b_2+2}}},
\qquad
M_{3\, 10}=-{\sqrt{b_2+3}(b_1+b_2)_2\over (b_2+2)\sqrt{b_2+1}},$$
$$
M_{4\, 10}=
{\frac {\sqrt {b_2+3}}{\left (b_1+b_2+2\right )\sqrt {b_2+2}\sqrt {b_1}\sqrt {b_1+1}}},
\qquad
M_{8\, 10}=
{\frac {\sqrt {b_1-1}\sqrt {b_1+b_2+3}}{\sqrt {b_1+b_2+2}\left (b_2+2\right )\sqrt {b_1}}},$$
$$
M_{10\, 10}=
-{\frac {\sqrt {b_2+3}\sqrt {b_1+b_2+1}\sqrt {b_1+b_2+3}\sqrt {b_2}}{\sqrt {b_2+1}\sqrt {b_2+2}\left (b_1+b_2+2\right )}},
\qquad
M_{15\, 10}=
-{\frac {\sqrt {b_2+3}\sqrt {b_1+2}}{\sqrt {b_1+1}\left (b_1+b_2+2\right )\sqrt {b_2+2}}},$$
$$ M_{1\, 11}=
{\frac {\left (b_2-1\right )\sqrt {b_1}\sqrt {b_1+2}}{\left (b_1+1\right )\sqrt {{b_2}^{2}-1}b_2}},
  \qquad
M_{2\, 11}=
{\frac {\sqrt {b_2-1}}{\left (b_1+1\right )\sqrt {b_2}\sqrt {b_1+b_2+1}\sqrt {b_1+b_2+2}}},$$
$$ M_{7\, 11}=
-{\frac {\sqrt {b_1+b_2}\sqrt {b_1+2}}{\sqrt {b_1+1}b_2\,\sqrt {b_1+b_2+1}}},
\qquad
M_{11\, 11}=
-{\frac {\sqrt {b_2-1}\sqrt {b_1}\sqrt {b_1+2}\sqrt {b_2+2}}{\sqrt {b_2+1}\sqrt {b_2}\left (b_1+1\right )}},$$
$$ M_{13\, 11}=
-{\frac {\sqrt {b_2-1}\sqrt {b_1+b_2+3}}{\sqrt {b_1+b_2+2}\left (b_1+1\right )\sqrt {b_2}}},
\qquad
M_{5\, 12}=
-{\frac {\sqrt{b_1+b_2+4}\sqrt {b_2+2}\sqrt {b_2}}{(b_2+1)\sqrt {b_1+b_2+2}\left (b_1+b_2+3\right )}},
$$
$$ M_{6\, 12}=
-{\frac {\sqrt {b_1+b_2+4}}{\left (b_2+1\right )\sqrt {b_1+b_2+3}\sqrt {b_1+2}\sqrt {b_1+1}}},
\qquad
  M_{9\, 12}=
{\frac {\sqrt {b_1+3}\sqrt {b_2+2}}{\sqrt {b_2+1}\left (b_1+b_2+3\right )\sqrt {b_1+2}}},$$
$$ M_{12\, 12}=
-{\frac {\sqrt {b_1+b_2+4}\sqrt {b_2+2}\sqrt {b_2}\sqrt {b_1+b_2+1}}{\sqrt {b_1+b_2+2}\sqrt {b_1+b_2+3}\left (b_2+1\right )}},
\qquad
M_{14\, 12}=
-{\frac {\sqrt {b_1+b_2+4}\sqrt {b_1}}{\sqrt {b_1+1}\left (b_2+1\right )\sqrt {b_1+b_2+3}}},$$
$$ M_{1\, 13}=
-{\frac {\sqrt {b_1+b_2}}{\left (b_1+1\right )\sqrt {b_1+b_2+1}\sqrt {b_2}\sqrt {b_2+1}}},
\qquad
M_{2\, 13}=
{\frac {\sqrt{b_1+b_2}\sqrt {b_1}\sqrt {b_1+2}}{(b_1+1)\sqrt {b_1+b_2+2}\left (b_1+b_2+1\right )}},$$
$$ M_{7\, 13}=
-{\frac {\sqrt {b_2-1}\sqrt {b_1}}{\sqrt {b_1+1}\left (b_1+b_2+1\right )\sqrt {b_2}}},
\qquad
M_{11\, 13}=
{\frac {\sqrt {b_1+b_2}\sqrt {b_2+2}}{\sqrt {b_2+1}\left (b_1+1\right )\sqrt {b_1+b_2+1}}},$$
$$ M_{13\, 13}=
-{\frac {\sqrt {b_1+b_2}\sqrt {b_1}\sqrt {b_1+2}\sqrt {b_1+b_2+3}}{\sqrt {b_1+b_2+1}\sqrt {b_1+b_2+2}\left (b_1+1\right )}},
\qquad
M_{5\, 14}=
{\frac {\sqrt {b_1+3}}{\left (b_2+1\right )\sqrt {b_1+2}\sqrt {b_1+b_2+3}\sqrt {b_1+b_2+2}}},
$$
$$ M_{6\, 14}=
-{\frac {\left (b_1+3\right )\sqrt {b_2+2}\sqrt {b_2}}{\left (b_2+1\right )\sqrt {{b_1}^{2}+4\,b_1+3}\left (b_1+2\right )}},
  \qquad
  M_{9\, 14}=
{\frac {\sqrt {b_1+b_2+4}\sqrt {b_2}}{\sqrt{b_2+1}\left (b_1+2\right )\sqrt{b_1+b_2+3} }},
$$
$$ M_{12\, 14}=
{\frac {\sqrt {b_1+3}\sqrt {b_1+b_2+1}}{\sqrt {b_1+b_2+2}\left (b_2+1\right )\sqrt {b_1+2}}},
\qquad
M_{14\, 14}=
-{\frac {\sqrt {b_1+3}\sqrt {b_2+2}\sqrt {b_2}\sqrt {b_1}}{\sqrt {b_1+1}\sqrt {b_1+2}\left (b_2+1\right )}},
$$
$$ M_{3\, 15}=
{\frac {\sqrt {b_1-1}}{\left (b_1+b_2+2\right )\sqrt {b_1}\sqrt {b_2+2}\sqrt {b_2+1}}},
  \qquad
 M_{4\, 15}=
{\frac {\left (b_1-1\right )(b_1+b_2)_2}{\sqrt {{b_1}^{2}-1}b_1}},
$$
$$ M_{8\, 15}=
-{\frac {\sqrt {b_2+3}\sqrt {b_1+b_2+1}}{\sqrt{b_1+b_2+2} b_1\,\sqrt{b_2+2} }},
  \qquad
M_{10\, 15}=
{\frac {\sqrt {b_1-1}\sqrt {b_2}}{\sqrt {b_2+1}\left (b_1+b_2+2\right )\sqrt {b_1}}},$$
$$ M_{15\, 15}=
-{\frac {\sqrt {b_1-1}\sqrt {b_1+b_2+1}\sqrt {b_1+b_2+3}\sqrt {b_1+2}}{\sqrt {b_1+1}\sqrt {b_1}\left (b_1+b_2+2\right )}}.$$
}

For large $b_1$ and $b_2$, we find that $M=-{\bf 1}$ with ${\bf 1}$ the $15\times 15$ identity matrix. We can also
identify terms in $M$ that behave as $b_1^{-1}$
$$M_1={1\over b_1}\left[
\matrix{
0 &0 &0 &0 &0 &0 &0 &0 &0 &0 &0 &0 &0 &0 &0\cr
0 &0 &0 &0 &0 &0 &0 &0 &0 &0 &0 &0 &0 &0 &0\cr
0 &0 &0 &0 &1 &0 &0 &0 &0 &0 &0 &0 &0 &0 &0\cr
0 &0 &0 &0 &0 &0 &0 &0 &1 &0 &0 &0 &0 &0 &1\cr
0 &0 &-1 &0 &0 &0 &0 &0 &0 &0 &0 &0 &0 &0 &0\cr
0 &0 &0 &0 &0 &0 &0 &-1 &0 &0 &0 &0 &0 &-1 &0\cr
0 &0 &0 &0 &0 &0 &0 &0 &0 &0 &0 &0 &0 &0 &0\cr
0 &0 &0 &0 &0 &1 &0 &0 &0 &0 &0 &0 &0 &0 &-1\cr
0 &0 &0 &-1 &0 &0 &0 &0 &0 &0 &0 &0 &0 &1 &0\cr
0 &0 &0 &0 &0 &0 &0 &0 &0 &0 &0 &0 &0 &0 &0\cr
0 &0 &0 &0 &0 &0 &0 &0 &0 &0 &0 &0 &1 &0 &0\cr
0 &0 &0 &0 &0 &0 &0 &0 &0 &0 &0 &0 &0 &0 &0\cr
0 &0 &0 &0 &0 &0 &0 &0 &0 &0 &-1 &0 &0 &0 &0\cr
0 &0 &0 &0 &0 &1 &0 &0 &-1 &0 &0 &0 &0 &0 &0\cr
0 &0 &0 &-1 &0 &0 &0 &1 &0 &0 &0 &0 &0 &0 &0}
\right],$$
terms that behave as $b_2^{-1}$
$$M_2={1\over b_2}\left[
\matrix{
0 &0 &0 &0 &0 &0 &0 &1 &0 &0 &1 &0 &0 &0 &0\cr
0 &0 &0 &1 &0 &0 &0 &0 &0 &0 &0 &0 &0 &0 &0\cr
0 &0 &0 &0 &0 &0 &-1 &0 &0 &-1 &0 &0 &0 &0 &0\cr
0 &-1 &0 &0 &0 &0 &0 &0 &0 &0 &0 &0 &0 &0 &0\cr
0 &0 &0 &0 &0 &0 &0 &0 &0 &0 &0 &0 &0 &0 &0\cr
0 &0 &0 &0 &0 &0 &0 &0 &0 &0 &0 &0 &0 &0 &0\cr
0 &0 &1 &0 &0 &0 &0 &0 &0 &0 &-1 &0 &0 &0 &0\cr
-1 &0 &0 &0 &0 &0 &0 &0 &0 &1 &0 &0 &0 &0 &0\cr
0 &0 &0 &0 &0 &0 &0 &0 &0 &0 &0 &0 &0 &0 &0\cr
0 &0 &1 &0 &0 &0 &0 &-1 &0 &0 &0 &0 &0 &0 &0\cr
-1 &0 &0 &0 &0 &0 &1 &0 &0 &0 &0 &0 &0 &0 &0\cr
0 &0 &0 &0 &0 &0 &0 &0 &0 &0 &0 &0 &0 &1 &0\cr
0 &0 &0 &0 &0 &0 &0 &0 &0 &0 &0 &0 &0 &0 &0\cr
0 &0 &0 &0 &0 &0 &0 &0 &0 &0 &0 &-1 &0 &0 &0\cr
0 &0 &0 &0 &0 &0 &0 &0 &0 &0 &0 &0 &0 &0 &0}
\right],$$
and terms that behave as $(b_1+b_2)^{-1}$
$$M_3={1\over b_1+b_2}\left[
\matrix{
0 &0 &0 &0 &0 &1 &0 &0 &0 &0 &0 &0 &0 &0 &0\cr
0 &0 &0 &0 &0 &0 &0 &0 &1 &0 &0 &0 &1 &0 &0\cr
0 &0 &0 &0 &0 &0 &0 &0 &0 &0 &0 &0 &0 &0 &0\cr
0 &0 &0 &0 &0 &0 &0 &0 &0 &0 &0 &0 &0 &0 &0\cr
0 &0 &0 &0 &0 &0 &-1 &0 &0 &0 &0 &-1 &0 &0 &0\cr
-1 &0 &0 &0 &0 &0 &0 &0 &0 &0 &0 &0 &0 &0 &0\cr
0 &0 &0 &0 &1 &0 &0 &0 &0 &0 &0 &0 &-1 &0 &0\cr
0 &0 &0 &0 &0 &0 &0 &0 &0 &0 &0 &0 &0 &0 &0\cr
0 &-1 &0 &0 &0 &0 &0 &0 &0 &0 &0 &1 &0 &0 &0\cr
0 &0 &0 &0 &0 &0 &0 &0 &0 &0 &0 &0 &0 &0 &1\cr
0 &0 &0 &0 &0 &0 &0 &0 &0 &0 &0 &0 &0 &0 &0\cr
0 &0 &0 &0 &1 &0 &0 &0 &-1 &0 &0 &0 &0 &0 &0\cr
0 &-1 &0 &0 &0 &0 &1 &0 &0 &0 &0 &0 &0 &0 &0\cr
0 &0 &0 &0 &0 &0 &0 &0 &0 &0 &0 &0 &0 &0 &0\cr
0 &0 &0 &0 &0 &0 &0 &0 &0 &-1 &0 &0 &0 &0 &0}
\right].$$
By looking at the Cuntz chain states, it is straight forward to see that $M_1$ is reproduced by ribbon diagrams in which a
pair of labels undergoes a $l\leftrightarrow m$ transition, that $M_2$ is reproduced by ribbon diagrams in which a
pair of labels undergoes a $b\leftrightarrow m$ transition and that $M_3$ is reproduced by ribbon diagrams in which a
pair of labels undergoes a $l\leftrightarrow b$ transition. This is exactly the structure expected from an emergent
$U(3)$ gauge theory.

\end{document}